\documentclass[epj]{svjour}

\usepackage{amssymb,amsmath,amsfonts,graphicx,color}
\usepackage{bm}

\begin{document}

\title{Supernovae: an example of complexity in the physics of compressible fluids}

\author{Yves Pomeau\inst{1} \and Martine Le Berre\inst{2} \and Pierre-Henri Chavanis \inst{3} \and Bruno Denet \inst{4}}

\institute{Department of Mathematics, University of Arizona,
Tucson, AZ 85721, USA. \and  Institut des Sciences Mol\'eculaires d'Orsay ISMO
- CNRS, Universit\'e Paris-Sud, Bat. 210, 91405 Orsay Cedex, France.
\and Laboratoire de Physique Th\'{e}orique (UMR 5152 du CNRS), Universit\'{e} Paul Sabatier, 118 route de Narbonne, 31062 Toulouse Cedex 4, France.
 \and  Universit\'e Aix-Marseille, IRPHE, UMR 7342 CNRS, et Centrale Marseille,
Technopole de Ch\^{a}teau-Gombert, 49 rue Joliot-Curie, 13384 Marseille Cedex 13, France.}

\date{\today }

\abstract{Because the collapse of massive stars occurs in a few seconds, while the stars evolve on billions of years, the supernovae are typical complex phenomena  in fluid mechanics with multiple time scales. We describe them in the light of catastrophe theory, assuming that successive equilibria between pressure and gravity present a saddle-node bifurcation. In the early stage we show that the loss of equilibrium may be described by a generic equation of the Painlev\'e I form. This is confirmed by two approaches, first by the full numerical solutions of the Euler-Poisson equations for a particular pressure-density relation, secondly by a derivation of the normal form of the solutions close to the saddle-node.
In the final stage of the collapse, just before the divergence of the central density, we show that the existence of a self-similar collapsing solution compatible with the numerical observations imposes that the gravity forces are stronger than the pressure ones. This situation differs drastically in its principle from the one generally admitted where pressure and gravity forces are assumed to be of the same order. Moreover it leads to different scaling laws for the density and the velocity of the collapsing material. The new self-similar solution (based on the hypothesis of dominant gravity forces) which matches the smooth solution of the outer core solution, agrees globally well with our numerical results, except a delay in the very central part of the star, as discussed. Whereas some differences with the earlier self-similar solutions are minor, others are very important. For example, we find that the velocity field becomes singular at the collapse time, diverging at the center, and decreasing slowly outside the core, whereas previous works described a finite velocity field in the core which tends to a supersonic constant value at large distances. This discrepancy should be important for explaining the emission of remnants in the post-collapse regime. Finally we describe the post-collapse dynamics, when mass begins to accumulate in the center, also within the hypothesis that gravity forces are dominant.
} 
\maketitle

\section{Introduction}

It is a great pleasure to write this contribution in honor of Paul Manneville.
We present below work belonging to the general field where he contributed so
eminently, nonlinear effects in fluid mechanics.
 However, our topic is perhaps
slightly unusual in this respect because it has to do with fluid mechanics
on a grand scale, namely the scale of
the Universe.

 We all know that Astrophysics has to tackle a huge variety of
phenomena, mixing widely different scales of space and time. Our contribution
below is perhaps the closest one can imagine of a problem of nonlinear and
highly non trivial fluid mechanics in Astrophysics, the explosion of supernovae.
In this fascinating field, many basic questions remain to be answered. The most
basic one can be formulated as follows: stars evolve on very long time scales,
in the billions years range, so why is it that some stars abruptly collapse (the
word collapse is used here in a loose sense, without implying for the moment an
inward fall of the star material) in a matter of days or even of seconds (the
ten seconds duration of the neutrino burst observed in 1987A, the only case
where neutrino emission of a supernova was recorded)? This huge difference of
time scales is described here in the light of catastrophe theory.
The basic mechanism for star collapse is by the loss of equilibrium between pressure and self-gravity. The theory of this equilibrium with the relevant equations is well-known.
We consider the case where the star is in equilibrium during a long period, then the series of equilibria presents a
saddle-node bifurcation. We expose in section \ref{Scaling laws} the hypothesis that the early stage
of the loss of equilibrium at the saddle-node should follow a kind of
universal equation of the Painlev\'e I form. Using  a
particular equation of state, we show in section
\ref{EulerPoisson}  that by a slow decrease of a given parameter (here the temperature), the series of equilibria do show a saddle-node bifurcation.  In section \ref{sec_dynA} we study the approach towards the saddle-node. We show that the full Euler-Poisson equations can be reduced to a normal form of the Painlev\'e I form valid at the first stage of the catastrophe, then we compare the
numerical solution of the full Euler-Poisson equations with the
solution of this universal equation. Section
\ref{sec:singular} is devoted to the final stage of the collapse, just
before the appearance of the singularity (divergence of the density
and velocity). We show that the existence of a self-similar collapsing
solution which agrees with the numerical simulations imposes that the
gravity forces are stronger than the pressure ones, a situation which
was not understood before. Usually the self-similar collapse, also
called ``homologous'' collapse, is treated by assuming that pressure
and gravity forces are of the same order that leads to scaling laws
such as $\rho\sim r^{-\alpha}$ for the density with parameter $\alpha$
equal to $2$. This corresponds to the Penston-Larson
solution \cite{Penston,larson}. Assuming that the gravity forces are
larger than the pressure ones inside the core, we show first that a
collapsing solution with $\alpha$ larger than $2$ displays relevant
asymptotic behavior in the outer part of the core, then we prove that
it requires that $\alpha$ takes the value $24/11$, which is larger
than $2$. We show that this result is actually in agreement with
the numerical works of Penston (see Fig. 1 in
\cite{Penston}) and Larson (see Fig. 1 in \cite{larson}) and many
others\footnote{Our initial condition (a star undergoing a loss of equilibrium at a saddle node) differs drastically from the initial conditions taken in \cite{Penston,larson,Brenner}. These studies assume an initial constant density over the whole star, $\rho(r)=\rho_0$, that seems very far from any physical situation. Note that, in this context, Brenner and Witelski \cite{Brenner} point out the existence of solutions which do \textit{not} behave as the theoretical Penston-Larson self-similar solution with $\alpha=2$. The numerical study presented here corresponds to a  parameter value $N=50$ in the notation of \cite{Brenner}. Note that despite the very different initial conditions, their Figs. 9 and 10 which are for $N=50$  show an asymptotic behavior with $\alpha$ larger than $2$ and a velocity  diverging in the core, in agreement with our results (see below).}
(see Figs. 4, 9, 10 and the first stage of Fig. 8  in
 \cite{Brenner})
 and that this small discrepancy between $\alpha=2$
and $\alpha=24/11$ leads to non negligible consequences for the
collapse characteristics. Contrary to the $\alpha=2$ case for which
the velocity remains finite close to the center and tends to a
constant supersonic value at large distances, our self-similar
solution (in the sense of Zel'dovich) displays a velocity diverging at
the center, and slowly vanishing as the boundary of the star is
approached. The latter property could be important for helping the
output of material in the post-collapse regime, see the next
paragraph. Finally, in section \ref{sec:beyond}, we describe the
post-collapse dynamics without introducing any new ingredient in the
physics.  We point out that just at the collapse time, there is no
mass in the center of the star, as in the case of the Bose-Einstein
condensation \cite{BoseE,bosesopik}; the mass begins to accumulate in
the inner core just after the singularity. Within the same frame as
before (gravity forces dominant with respect to pressure ones), we
derive the self-similar equations for the post-collapse regime and
compare the solutions with a generalized version of the parametric
free-fall solution proposed by Penston \cite{Penston}.

Let us discuss now some ideas concerning the difficulty of interpretation of what happens after the collapse. Indeed  the understanding of the pre-collapse stage does not help as much as one
would like to explain the observations: besides the neutrino burst of 1987 A,
supernovae are sources of intense radiation in the visible range or
nearby, this occurring days if not weeks after the more energetic part of the
collapse. Although this does not seem to follow from general principles, the
collapse is a true collapse because it shows a {\it{centripetal}} motion of the
material in the star, at least in its early stage. Instead what is observed is
the {\it{centrifugal}} motion of a dilute glowing gas (with usually a complex nuclear
chemistry) called the remnant, something believed to follow a centripetal
collapse. Such a change of sign (from centripetal to centrifugal) occurring in
the course of time has to be explained. It has long been a topic of active
research, relying on increasingly complex equations of state of nuclear matter
with high resolution numerical simulations of the fluid equations. Without
attempting to review the literature on this topic, one can say that no clear-cut
conclusion seems to have emerged on this. In particular there remains a
sensitivity of the results to a poorly understood production of neutrinos.  In
short, one has to explain how an inward motion to the center of the star
reverses itself into an outward motion, something requiring a large
acceleration. To understand how this reverse is possible, one may think  to the
classical Saint-Venant analysis of bouncing of a vertical rod
\cite{saintvenant}: at the end of its free-fall this rod hits the ground and
then reverses its motion to lift off the ground. This reversal is possible
because the initial kinetic energy is stored first in the compression elastic
energy when the lower end of the rod is in contact with the ground and then the
energy is released to feed an upward motion. Even though Saint-Venant dealt with
solid mechanics, it is not so different of fluid mechanics. Somehow, the
comparison with Saint-Venant brings two things to the fore: what could be the
equivalent of the solid ground in a collapsing star? Then how much time will it
take to trigger an outward motion out of the compression of the star? In
particular, thanks to the well defined initial value problem derived  in section
\ref{Scaling laws}, we can have a fair picture of what happens until the elastic
wave due to the bouncing reaches the outer edge of the star and starts the
emission of matter, as does Saint-Venant's rod. But, compared to this classical
problem, there is something different (among many other things of course) in
supernova explosion. To explain the emission of remnants, one has to do more
than to reverse the speed from inward to outward:  the outward speed must be
above the escape velocity to counteract the gravitational attraction of what
remains of the star (this excluding cases where the core of the star becomes a
black hole).  This requires some kind of explosion and, somehow, an explosion
requires an explosive, particularly because an additional supply of energy has
to be injected into the fluid to increase the outward velocity beyond the escape
value. This source of energy was long identified by Hoyle and Fowler \cite{HF}
in the nuclear reactions taking place in the compressed star material. This
explains type I supernovae. In this model, the pressure
increase in the  motionless material
left behind the outgoing shock should be due to a nuclear reaction triggered by
the shock, defining a detonation wave. Such a wave could be triggered by the
infalling material on the center, which has a very large (even diverging) speed
in the model of singularity developed here in section \ref{sec:singular}.

In the other type of supernova, called type II, the infall on the center is believed to yield a neutron core, observed in few cases as a neutron star emitting radio waves near the center of the cloud of remnants. The increase of pressure in the shocked gas would be due there to the neutrinos. They are emitted, by the reaction making one neutron from one proton $+$ one electron, out of the dense neutron core in formation which is bombarded by infalling nuclear matter.
Such a boosting of the pressure is likely localized in the neighborhood of the interface between the neutron core (at the center of the star) and the collapsing nuclear matter, and can hardly increase the pressure far away from this interface. As observed in the numerics, it is hard to maintain a shock wave far from the surface of the neutron core, and so it could well be that nuclear reactions behind the propagating shock are necessary to increase the pressure sufficiently to reach the escape velocity when the shock reaches the outer edge of the star. This is also a consequence of our discussion of the initial conditions for the collapse of the star: the singularity at the center of the star occurs at a time where the star has collapsed by a finite amount and keeps a radius of the same order of magnitude as its initial radius, making it order of magnitude bigger than the radius of its neutron core. Therefore the emission of neutrinos from the boundary of this neutron core cannot increase the pressure far from the core. The observation of neutrinos in supernova explosion could be due to the nuclear reactions taking place in the detonation wave, not to the nuclear reaction due to the growth of the neutron core.

Our approach of the phenomenon of supernova explosion is not to try to describe quantitatively this immensely complex phenomenon, something which could well be beyond reach because it depends on so many uncontrolled and poorly known physical phenomena, like equations of state of matter in conditions not realizable in laboratory experiments, the definition of the initial conditions for the star collapse, the distribution of various nuclei in the star, etc. Therefore we try instead to solve a simple model in a, what we believe, completely correct way.  The interest of our model and analysis is that we fully explain the transition from the slow evolution before the collapse to the fast collapse itself. Continuing the evolution we observe and explain the occurrence of a finite time singularity at the center, a singularity where the velocity field diverges. This singularity is not the standard homologous Penston-Larson collapse where all terms in the fluid equations are of the same order of magnitude. Instead this is a singularity of free-fall dynamics, that is such that the pressure force becomes (locally) negligible compared to gravitational attraction\footnote{Of course, the free-fall solution of a self-gravitating gas is well-known \cite{Penston}. However, it has been studied assuming either a purely homogeneous distribution of matter or an inhomogeneous distribution of matter behaving as $\rho(r,t)-\rho(0,t)\sim r^2$ for $r\rightarrow 0$, leading to a large distance decay $\rho\sim r^{-\alpha}$ with an exponent $\alpha=12/7$. We show that these assumptions are not relevant to our problem, and we consider for the first time a  behavior $\rho(r,t)-\rho(0,t)\sim r^4$ for $r\rightarrow 0$, leading to the large distance decay with the exponent $\alpha=24/11$.}. This point is more than a mathematical nicety, because the laws for this collapse, contrary to the ones of the homologous Penston-Larson collapse, are such that the velocity of infall tends to zero far from the center instead of tending to a constant supersonic value. This makes possible that the shock wave generated by the collapse escapes the center without the additional help of neutrinos as needed in models where the initial conditions are a homologous Penston-Larson collapse far from the center.

\section{The Painlev\'e equation and the scaling laws}
\label{Scaling laws}

A supernova explosion lasts about ten seconds, when measured by the duration of the neutrino burst in SN1987A, and this follows a ``slow" evolution over billions of years, giving  an impressive $10^{13}$ to  $10^{14}$ ratio of the slow to fast time scale. Such hugely different time scales make it a priori impossible to have the same numerical method for the slow and the fast dynamics. More generally it is a challenge to put in the same mathematical picture a dynamics with so widely different time scales. On the other hand the existence of such huge dimensionless numbers in a problem is an incentive to analyze it by using asymptotic methods. Recently it has been shown \cite{catastrophe} that such a slow-to-fast transition can be described as resulting from a slow sweeping across a saddle-node bifurcation. In such a bifurcation, if it has constant parameters, two fixed points, one stable the other unstable, merge and disappear when a parameter changes, but not as a function of time.
We have to consider here a dynamical transition, occurring when a parameter changes slowly as a function of time. It means that the relevant parameter drifts in time until it crosses a critical value at the time of the catastrophe, this critical time being at the onset of saddle-node bifurcation for the dynamical system. Such a slow-to-fast  transition is well known to show up in the van der Pol equation in the relaxation limit \cite{dor}. Interestingly, the analysis shows that this slow-to-fast transition occurs on a time scale intermediate between the slow and long time scale, and that it is described by a universal equation solvable by the Riccati method. This concerns dynamical systems with dissipation, where the ``universal equation" is first order in time. The supernovae likely belong to the class of dynamical catastrophes in our sense, because of the huge difference of time scales, but, if one assumes that the early post-bifurcation dynamics is described by inviscid fluid dynamics, one must turn to a model of non dissipative dynamics.

Such a dynamical model of catastrophes without dissipation and with time dependent sweeping across a saddle-node bifurcation is developed below and applied to supernovae. We deal mostly with the early stage of the collapse, which we assume to be described by compressible fluid mechanics, without viscosity. Indeed the slow evolution of a star before the transition is a highly complex process not modeled in this approach because of the large difference in time scales: it is enough to assume that this slow evolution makes a parameter cross a critical value where a pair of equilibria merge by a saddle-node bifurcation. The universal equation describing the transition is the Painlev\'e I equation, valid for the dissipationless case. We explain how to derive it from the fluid mechanical equations in the inviscid case, assumed to be valid for the interior of the star. Although applications of the ideas developed below could be found in more earthly situations like in subcritical bifurcation of Euler's Elastica with broken symmetry or the venerable Archimedes problem of (loss of) stability of floating bodies in an inviscid fluid \cite{Coullet}, we shall refer below explicitly to the supernova case only. Our starting point is the following equation of Newtonian dynamics,
\begin{equation}
\frac{{\mathrm{d}}^2 r_0}{{\mathrm{d}} t^2} = - \frac{\partial V}{\partial r_0}
\mathrm{,}
\label{eq:1}
\end{equation}
where $r_0$ can be seen as the radius of the star and $V(r_0,t)$ a time dependent potential. No mass multiplies the acceleration, which is always possible by rescaling the potential $V(.)$. We shall derive later this equation for an inviscid compressible fluid
with gravitation and an equation of state changing slowly as a function of time, for a radially symmetric geometry and with a finite mass. Contrary to the case studied in \cite{catastrophe}, this equation is second order in time because one neglects dissipation compared to inertia.
 The potential $V(.)$ on the right-hand side represents the potential energy of the star,
with the contributions of gravity and of internal energy \cite{LL}. At equilibrium the right-hand side is zero. Given the potential $V(.)$ this depends on two parameters (linked to the total mass and energy), $r_0$ and another physical parameter which may be seen as the temperature. Because of the long term evolution of the star interior by nuclear reactions and radiation to the outside, its temperature changes slowly.
We shall assume that this slow change of parameter makes  the equilibrium solution disappear by a saddle-node bifurcation when the temperature $T$ crosses a critical value.

A saddle-node bifurcation is sometimes called turning, or tipping point
instability, whereas the word ``saddle-node" (noeud-col in french) was coined by H. Poincar\'e in his Ph.D. thesis. Such a bifurcation is a fairly standard problem treated by Emden \cite{emden} for a
self-gravitating gas at finite (and  changing, but not as function of time)
temperature in a spherical
box. It was also discussed
by Ebert \cite{ebert}, Bonnor \cite{bonnor}, and McCrea \cite{crea} by varying
the pressure, and by Antonov \cite{antonov} and Lynden-Bell and Wood \cite{lbw}
by varying the energy. See
Chavanis
\cite{aaiso,grand} for recent studies. A saddle node also occurs in the
mass-radius relation of neutron stars determined by Oppenheimer and
Volkoff \cite{ov} when the mass crosses a critical value $M_{OV}$ (see
also section 109 of \cite{LL}, figure 52) and in the mass-radius
relation of boson stars \cite{colpi,prd1,ch}. A saddle node is also
present in the caloric curve of self-gravitating fermions at finite
temperature which has the form of a ``dinosaur's neck'' \cite{dino}.

As we do not solve the energy equation, the parameter $T$ could be any parameter describing the smooth changes of the star interior prior to the fast transition. Following the ideas of reference \cite{catastrophe}  we look for a finite change in the system on a time scale much shorter than the time scale of the control parameter (here the temperature $T$). Two time scales are involved: the long time scale of evolution of $T$, denoted as $\theta$ below, and the short time scale $\tau$ which is the fundamental period of a pressure oscillation in the star. Our approach will show that the early stage of the collapse is on a time scale intermediate between the fast and slow scale and give a precise definition of the initial conditions for the fast process.

\begin{figure}[htbp]
\centerline{
 \includegraphics[height=2.0in]{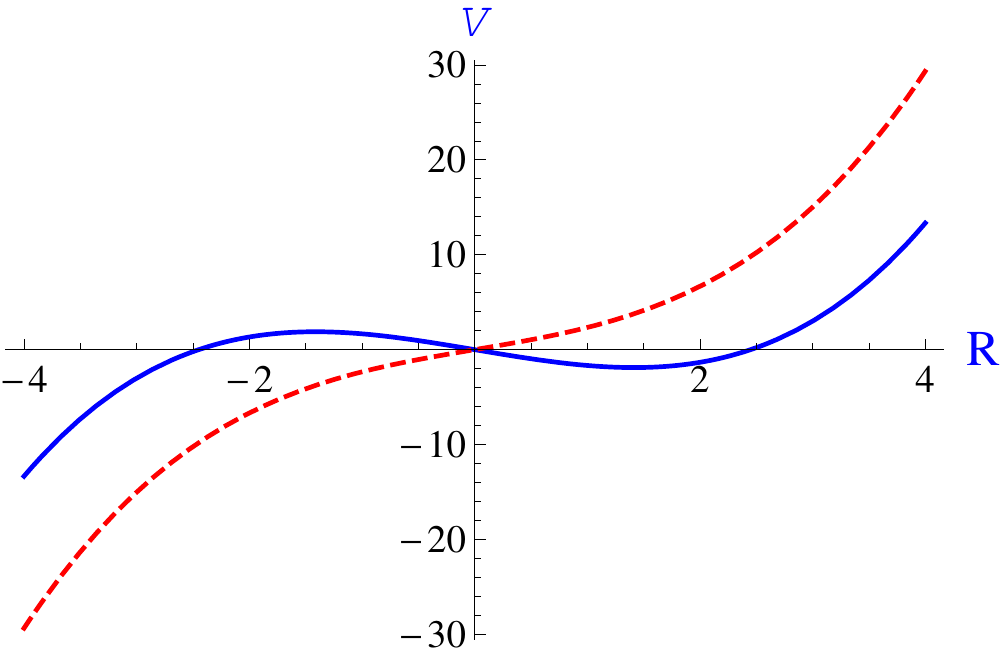}
}
\caption{Potential evolution close to a saddle-node, equation (\ref{eq:2}) with $b=c=1$ and two values of $a=-ct$; $t=-2$ for the blue curve, $t=2$ for the red dashed curve.
}
\label{Fig:data1Ra}
\end{figure}

Let us expand the potential $V(.)$ in Poincar\'e normal form near the saddle-node bifurcation:
 \begin{equation}
V = -a R + \frac{b}{3} R^3 +...
\mathrm{,}
\label{eq:2}
\end{equation}
In the expression above, $R$, a relative displacement, can be seen as
the difference between $r_c$, the value of the radius of the star at
the saddle-node bifurcation and its actual value, $R=(r_0-r_c)/r_c$, a
quantity which decreases as time increases, because we describe the
collapse of the star. Actually the quantity $R$ will be seen later as
the Lagrangian radial coordinate, a function depending on $r$, the
radial distance. The saddle-node bifurcation is when the - now time
dependent - coefficient $a$ of equation (\ref{eq:2}) crosses
$0$. Setting to zero the time of this crossing, one writes $a = - ct$,
where $c$, a constant, is small because the evolution of $V$ is
slow. This linear time dependence is an approximation because $a(t)$
is, in general, a more complex function of $t$ than a simple
ramp. However, near the transition, one can limit oneself to this
first term in the Taylor expansion of $a(t)$ with respect to $t$,
because the transition one is interested in takes place on time scales
much shorter than the typical time of change of $a(t)$. Limiting
oneself to displacements small compared to $r_c$, one can keep in
$V(R)$ terms which are linear and cubic (the coefficient $b$ is
assumed positive) with respect to $R$ because the quadratic term
vanishes at the saddle-node transition (the formal statement
equivalent to this lack of quadratic term in this Taylor expansion of
$V(R)$ is the existence of a non trivial solution of the linearized
equation at the bifurcation).  Moreover higher order terms in the
Taylor expansion of $V(.)$ near $R = 0$ are neglected in this analysis
because they are negligible with the scaling law to be found for the
magnitude of $R$ near the transition. This is true at least until a
well defined time where the solution has to be matched with the one of
another dynamical problem, valid for finite $R$.  At $t = 0$, the
potential $V(.)$ is a cubic function of $R$, exactly the local shape
of a potential in a metastable state. For $a$ and $b$ positive, the
potential has two extrema, one corresponding to a stable equilibrium
point at $R = \sqrt{a/b}$ and one unstable at $R =-\sqrt{a/b}$. In the
time dependent case, the potential evolves as shown in
Fig. \ref{Fig:data1Ra} and the equations (\ref{eq:1})-(\ref{eq:2})
become
\begin{equation}
\frac{d^2R}{dt^2} = \ddot{R} = - ct - bR^2
\mathrm{,}
 \label{eq:ab}
 \end{equation}
where the parameter $c$ is supposed to be positive, so that the solution at large negative time is close to equilibrium and positive, crosses zero at a time close to zero and diverges at finite positive time.

 To show that the time scale for the dynamical saddle-node bifurcation is intermediate between the long time scale of the evolution of the potential $V(.)$ and the short time scale of the pressure wave in the star, let us derive explicitly these two relevant short and long time scales.
For large negative time the solution of equation (\ref{eq:ab}) is assumed to evolve very slowly such that the left-hand side can be set to zero. It gives
\begin{equation}
R(t)\simeq \sqrt{\frac{c}{b}(-t)}
 \label{eq:depart}
 \end{equation}
which defines the long time scale as $\theta= {b}/{c}$
(recall that $R$, a relative displacement scaled to the star radius $r_c$, has no physical scale).

As for the short time scale, it appears
close to the time  $t=t_*$ where the solution of equation (\ref{eq:ab}) tends to minus infinity. In this domain the first term in the right-hand side is negligible with respect to the second one, the equation reduces to $\ddot{R} = - bR^2$, which has the characteristic time $\tau= {1}/{\sqrt{b}}$.

Let us scale out the two parameters $b,c$ of equation (\ref{eq:ab}).
  Defining $ \hat{ R}={R}/{r_s} $ and $\hat{ t}={t}/{t_0}$
the original equation takes the scaled form
\begin{equation}
\frac{d^2\hat{R}}{d\hat{t}^2}= -\hat{t} - \hat{R}^2
\mathrm{,}
 \label{eq:eqfin}
 \end{equation}
when setting $c ={r_s}/{t_0^3}$ and $b={1}/({r_st_0^2})$. Inversely, $t_0=1/(bc)^{1/5}$ and $r_s=c^{2/5}/b^{3/5}$. The solution of equation (\ref{eq:eqfin}) is called the first Painlev\'e transcendent, and cannot be reduced to elementary functions \cite{Ince}.

The writing of the Painlev\'e equation in its parameter free form yields the characteristic time scale $t_0$ of equation (\ref{eq:ab}) in terms of the short and long times,
\begin{equation}
t_0=(\theta \tau ^4 )^{1/5}
\mathrm{.}
 \label{eq:to}
\end{equation}
This intermediate time is  such that  $\tau \ll t_0 \ll \theta$; it could be of the order of  several hours when taking $\theta \sim$ one  billion years, $\tau \sim 10$ sec. The corresponding spatial extension $R$ is of order
\begin{equation}
 r_s=  \left (\frac{\tau}{\theta}\right )^{2/5}
 \mathrm{,}
 \label{eq:rs}
\end{equation}
much smaller than unity. The one-fifth power in equations (\ref{eq:to}) and  (\ref{eq:rs}) is ``typical" of the Painlev\'e I equation, which has a symmetry expressed in terms of the complex fifth root of unity.

\begin{figure}[htbp]
\centerline{
 \includegraphics[height=2.0in]{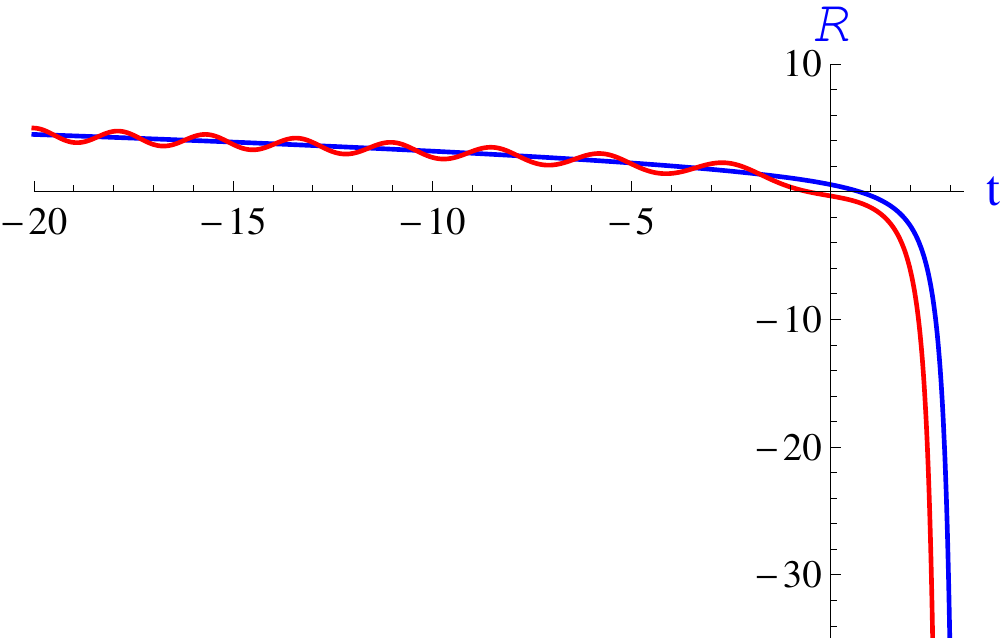}
}
\caption{
Numerical solution of equation (\ref{eq:eqfin}), or equation (\ref{eq:ab}) with
$b=c=1$, for two different initial conditions taken at time $t_i=-20$; (i) relation (\ref{eq:ci2}) for the blue curve without any oscillation; (ii) $R(t_i)
=\sqrt{-t_i}+0.5$  and $R'(t_i)=-\frac{1}{2\sqrt{-t_i}}$ for the red oscillating curve.
}
\label{Fig:data1R}
\end{figure}

To solve  equation (\ref{eq:eqfin}) we have to define the initial conditions. Choosing the initial conditions at large negative time $t_i$, we may assume that  the asymptotic relation (\ref{eq:depart}) is fulfilled at this time, that gives,

  \begin{equation}
 \left \{ \begin{array}{l}
\hat{R}(\hat{t}_i)=\sqrt{-\hat{t}_i}, \\
\dot{\hat{R}}(\hat{t}_i)=-\frac{1}{2\sqrt{-\hat{t}_i}}
\mathrm{.}
\end{array}
\right. \label{eq:ci2}
\end{equation}
The numerical solution of equation (\ref{eq:eqfin}) is drawn in Fig. \ref{Fig:data1R} leading to a finite time singularity. With  the initial conditions (\ref{eq:ci2}) the solution is a non oscillating function (blue curve) diverging at a finite time $\hat{t}_*\simeq 3.4$ (note that the divergence is not yet reached in Fig. \ref{Fig:data1R}).

But we may assume that, at very large negative time, the initial conditions slightly differ from the asymptotic quasi-equilibrium value (\ref{eq:ci2}). In that case the solution displays oscillations of increasing amplitude and period as time increases, in agreement with a WKB solution of the linearized problem.  Let us  put $\hat{R}(\hat{t}) \approx \sqrt{-\hat{t}} + \delta \hat{R}$, $\delta \hat{R}$ small which satisfies the linear equation
\begin{equation}
\delta \ddot{\hat{R}} = -2 \sqrt{-\hat{t}}  { \delta \hat{R}} \mathrm{.}
\end{equation}
A WKB solution, valid for $(-\hat{t})$ very large  is
\begin{equation}
\delta \hat{R} = \sum_{\pm} c_{\pm} (-\hat{t})^{-1/4} e^{\pm i \frac{4\sqrt{2}}{5} (-\hat{t})^{5/4}} \mathrm{.}
\end{equation}
It represents oscillations in the bottom of the potential $V(\hat{R},\hat{t}) =  \hat{t} \hat{R} + {\hat{R}^3}/{3}$ near $ \hat{R} = \sqrt{-\hat{t}}$. The two complex conjugate coefficients $c_{\pm}$ defining the amplitudes are arbitrary and depend on two real numbers.  Therefore, the cancelation of the oscillations defines uniquely a solution of the Painlev\'e I equation. This is illustrated in Fig. \ref{Fig:data1R} where the blue curve has no oscillation (see above) while the red curve displays oscillations of increasing period and a shift of the divergence time.

Near the singularity, namely just before time $\hat{t} = \hat{t}_*$, the dominant term on the right-hand side of equation (\ref{eq:eqfin}) is $\hat{R}^2$ so that
$\hat{R}$ becomes  approximately $\hat{R}(t)\simeq - {6}/{(\hat{t}_* - \hat{t})^2}$, or in terms of the original variables $R$ and $t$,
\begin{equation}
R(t)\simeq -6 r_s\left(\frac{t_0}{t_* - t}\right)^2.
\label{eq:t-2}
\end{equation}
This behavior will be compared later to the full Euler-Poisson model (see Fig. \ref{Fig:rhoc} and relative discussion).
Note that this divergence is completely due to the nonlinearity, and
has little to do with a {\it{linear}} instability. The applicability
of this theory requires $R\ll 1$, because it relies on the Taylor
expansion of $V(.)$ in equation (\ref{eq:1}) near $r_0=r_c$.  It is
valid if $|t - t_*| \gg \tau$. Therefore the collapse (we mean by
collapse the very fast dynamics following the saddle-node bifurcation)
can be defined within a time interval of order $\tau$, the center of
this interval being the time where the solution of equation
(\ref{eq:eqfin}) diverges, not the time where the linear term in the
same equation changes sign.  Moreover the duration of the early stage
of the collapse is, physically, of order $(\theta
\tau^{4})^{1/5}$, much shorter than the time scale of evolution of the
temperature, but much longer than the elastic reaction of the star interior.

The blow-up of the solution of equation (\ref{eq:ab}) at finite time does {\emph{not}} imply a physical singularity at this instant. It only shows that, when $t$ approaches $t_*$ by negative values, $R(t)$ grows enough to reach an order of magnitude, here the radius of the star, such that the approximation of $V$ by the first two terms (linear and cubic with respect to $R$) of its Taylor expansion is no longer valid, imposing to switch to a theory valid for finite displacements.  In this case, it means that one has to solve, one way or another, the full equations of inviscid hydrodynamics, something considered in section \ref{EulerPoisson}.  A warning at this stage is necessary: we have to consider more than one type of finite time singularity in this problem. Here we have met first a singularity of the solution of the Painlev\'e I equation, a singularity due to various approximations made for the full equations which disappear when the full system of Euler-Poisson equations is considered. But, as we shall see, the solution of this Euler-Poisson set of dynamical equations shows a finite time singularity also, which is studied in section \ref{sec:singular} and which is related directly to the supernova explosion.

Below we assume exact spherical symmetry, although non spherical stars could be quite different. A given star being likely not exactly spherically symmetric, the exact time $t_*$ is not so well defined at the accuracy of the short time scale $\tau$ because it depends on small oscillations of the star interior prior to the singularity (the amplitude of those oscillations depends on the constants $c_{\pm}$ in the WKB part of the solution, and the time $t_*$ of the singularity depends on this amplitude). One can expect those oscillations to have some randomness in space and so not to be purely radial. The induced loss of sphericity at the time of the collapse could explain the observed expulsion of the central core of supernovae  with large velocities, up to $500$ km per second \cite{coreexpulsion} a very large speed which requires large deviations to sphericity. However there is an argument against a too large loss of sphericity: the time scale $t_0$ for the part of the collapse described by the Painlev\'e equation is much longer than $\tau$, the typical time scale for the evolution of the inside of the star. Therefore one may expect that during a time of order $t_0$, the azimuthal heterogeneities are averaged, restoring spherical symmetry on average on the longer time scale $t_0$. However this does not apply if the star is intrinsically non spherically symmetric because of its rotation.

Within this assumption of given slow dependence with respect to a parameter called $T$, we shall derive the dynamical equation (\ref{eq:ab}) from the fluid equations with a general pressure-density relation and the gravity included.  To streamline equations and explanations, we shall not consider the constraint of conservation of energy (relevant on the fast time scale).

\section{Euler-Poisson system for a barotropic star presenting a saddle-node}
\label{EulerPoisson}

\subsection{Barotropic Euler-Poisson system}

We shall assume that the star can be  described as a compressible inviscid fluid with a barotropic equation of state $p=p(\rho)$. The relevant set of hydrodynamic equations are the barotropic Euler-Poisson system. These are dynamical equations for a compressible inviscid fluid with a pressure-density relation, including the gravitational interaction via Poisson equation. Note that there is no dynamical equation for the transport of energy. They read
\begin{equation}
\frac{\partial\rho}{\partial t}+\nabla\cdot (\rho {\bf u})=0,
\label{iso1}
\end{equation}
\begin{equation}
\rho\left\lbrack \frac{\partial {\bf u}}{\partial t}+({\bf u}\cdot \nabla){\bf u}\right\rbrack=-\nabla p-\rho\nabla\Phi,
\label{iso2}
\end{equation}
\begin{equation}
\Delta\Phi=4\pi G\rho
\mathrm{,}
\label{iso3}
\end{equation}
where
${\bf u}$ is the fluid velocity vector, $\rho$ the mass density, and $G$ Newton's constant.
Using the equation of continuity (\ref{iso1}), the momentum equation (\ref{iso2}) may be rewritten as
\begin{equation}
\frac{\partial}{\partial t}(\rho {\bf u})+\nabla (\rho {\bf u}\otimes {\bf u})=-\nabla p-\rho\nabla\Phi.
\label{iso4}
\end{equation}
The potential energy of this self-gravitating  fluid is $V=U+W$ where
\begin{eqnarray}
U=\int\rho\int^{\rho}\frac{p(\rho')}{{\rho'}^2}\, d\rho' d{\bf r},
\label{eos1}
\end{eqnarray}
is the internal energy and
\begin{eqnarray}
W=\frac{1}{2}\int\rho\Phi\, d{\bf r},
\label{eos2}
\end{eqnarray}
is the gravitational energy. The internal energy can be written as $U=\int \lbrack \rho h(\rho)-p(\rho)\rbrack\, d{\bf r}=\int H(\rho)\, d{\bf r}$ where we have introduced the  enthalpy $h(\rho)$, satisfying $dh(\rho)=dp(\rho)/\rho$, and its primitive $H(\rho) = \int_0^{\rho} h(\rho){\mathrm{d}}\rho $.

\subsection{Hydrostatic equilibrium and neutral mode}
\label{sec_henm}

In this section we briefly recall different formulations of the equilibrium state of a self-gravitating gas. From equation (\ref{iso2}), the condition of hydrostatic equilibrium writes
\begin{eqnarray}
\nabla p+\rho\nabla\Phi={\bf 0}.
\label{he1}
\end{eqnarray}
Dividing this equation by $\rho$, taking the divergence of the resulting expression, using Poisson equation (\ref{iso3}), and recalling that $p=p(\rho)$ for a barotropic gas, we obtain a differential equation for $\rho$ that is
\begin{eqnarray}
\nabla\cdot\left \lbrack\frac{p'(\rho)}{\rho}\nabla\rho\right \rbrack+4\pi G\rho=0.
\label{he2}
\end{eqnarray}
For a barotropic equation of state by definition $p=p(\rho)$. The condition of hydrostatic equilibrium (\ref{he1}) implies $\rho=\rho(\Phi)$. Substituting this relation in Poisson equation (\ref{iso3}), we obtain a differential equation for $\Phi$ that is
\begin{eqnarray}
\Delta\Phi=4\pi G\rho(\Phi).
\label{he3}
\end{eqnarray}

Introducing the enthalpy, satisfying $\nabla h=\nabla p/\rho$, the  condition of hydrostatic equilibrium (\ref{he1}) can be rewritten as
\begin{eqnarray}
\nabla h+\nabla\Phi={\bf 0}.
\label{he4}
\end{eqnarray}
Therefore, at equilibrium, $h({\bf r})=-\Phi({\bf r})+C$ where $C$ is a constant. Since the gas is barotropic, we also have $\rho=\rho(h)$. Taking the divergence of equation (\ref{he4}) and using Poisson equation (\ref{iso3}), we obtain a differential equation for $h$ that is
\begin{eqnarray}
\Delta h+4\pi G\rho(h)=0.
\label{he5}
\end{eqnarray}
These different formulations are equivalent. In the following, we will solve the differential equation (\ref{he5}).

To determine the dynamical stability of a steady state of the Euler-Poisson system (\ref{iso1})-(\ref{iso3}), we consider a small perturbation about that state and write $f({\bf r},t)=f({\bf r})+\delta f({\bf r},t)$ for $f=(\rho,{\bf u},\Phi)$ with $\delta f({\bf r},t)\ll f({\bf r})$. Linearizing the Euler-Poisson system about that state, and writing the perturbation as $\delta f({\bf r},t)\propto e^{\lambda t}$, we obtain the eigenvalue equation
\begin{equation}
\lambda^2\delta\rho=\nabla\cdot \left\lbrack \rho(\nabla\delta h+\nabla\delta\Phi)\right \rbrack.
\label{he6}
\end{equation}
The neutral mode ($\lambda=0$) which usually signals the change of stability is the solution of the differential equation
\begin{equation}
\nabla\delta h+\nabla\delta\Phi={\bf 0}.
\label{he7}
\end{equation}
Taking the divergence of this equation and using Poisson equation (\ref{iso3}), it can be rewritten as
\begin{equation}
\Delta\delta h+4\pi G\rho'(h)\delta h=0.
\label{he8}
\end{equation}
This equation may also be written in terms of $\delta\rho$ by using $\delta h=p'(\rho)\delta\rho/\rho$. We get
\begin{equation}
\Delta\left (\frac{p'(\rho)}{\rho}\delta \rho\right )+4\pi G\delta \rho=0.
\label{he9}
\end{equation}
In the following, we will solve the differential equation (\ref{he8}).

\subsection{An isothermal equation of state with a polytropic envelope implying a saddle node}
\label{sec_eos}

The series of equilibria of an isothermal self-gravitating
gas with $p=\rho T$ is known to present a saddle node \cite{emden,aaiso}. Therefore a self-gravitating isothermal
gas is a good candidate for our investigation. However, it
has the undesirable feature to possess an infinite mass because its
density decreases too slowly (as $r^{-2}$) at large distances. Therefore, to have a
finite mass, it must be confined artificially into a ``box''. In order to
skip this difficulty, we propose to use here an equation of state
that is isothermal at high densities and polytropic at low densities,
the polytropic equation of state serving as an envelope that confines
the system in a finite region of space without artificial container.
Specifically, we consider the
equation of state\footnote{This equation of state is
inspired by the study of self-gravitating boson stars in general
relativity
\cite{colpi,prd1,ch}. Such an equation of state could hold in the core of
neutron stars because of its superfluid properties \cite{ch}. The neutrons
(fermions) could form Cooper pairs and behave as bosons. In this
context
$\rho c^2$ represents the energy density and the
parameter $T$ has an interpretation different from the temperature (in
the core of neutron stars $T$ is much less than the Fermi temperature
or than the Bose-Einstein condensation temperature so it can be taken as
$T=0$). We use here this equation of state with a different
interpretation.}
\begin{eqnarray}
 p(\rho)=\rho_* T\left(\sqrt{1+\rho/\rho_*}-1\right )^2.
\label{eos3}
\end{eqnarray}
For $\rho\rightarrow +\infty$, it reduces to the isothermal equation
of state $p=\rho T$. For $\rho\rightarrow 0$, it reduces to the
polytropic equation of state $p=K\rho^2$ with polytropic index
$\gamma=2$ and polytropic constant $K=T/(4\rho_*)$.

The enthalpy function $h(\rho)$ defined by  $dh={dp}/{\rho}$ is explicitly given by
\begin{eqnarray}
h(\rho)=2 T \ln \left ( 1+\sqrt{1+\rho/\rho_*}\right )-2T \ln (2),
\label{eos5}
\end{eqnarray}
where the constant of integration has been determined such that
$h(\rho=0)=0$. With this choice, the enthalpy vanishes at the edge
of the star. The inverse relation writes
\begin{equation}
{\rho}({h})=4\rho_*\left (e^{{h}/T}-e^{{h}/2T}\right )
 \mathrm{.}
\label{eq:M4}
\end{equation}

In the following, it will be convenient to use dimensionless variables. The parameters regarded as fixed are $\rho_*$, $M$, and $G$. From $\rho_*$ and $M$ we can construct a length $L_*=(M/\rho_*)^{1/3}$. Then, we introduce the dimensionless quantities
\begin{equation}
{\tilde\rho}=\frac{\rho}{\rho_*},\quad {\tilde r}=\frac{r}{L},\quad
{\tilde\Phi}=\frac{\Phi}{G\rho_* L^2}.
\end{equation}
and
\begin{equation}
{\tilde T}=\frac{T}{G\rho_* L^2},\quad {\tilde p}=\frac{p}{GL^2\rho_*^2},\quad {\tilde t}=t \sqrt{G\rho_*}.
\end{equation}
Working with the dimensionless variables with tildes amounts to taking $G=\rho_*=M=1$ in the initial equations, a choice that we shall make in the following.

\subsection{Equilibrium solution and temperature-radius relation}
\label{sec_tr}

The equilibrium solution is obtained by solving equation (\ref{he5}) with equation (\ref{eq:M4}). Using the dimensionless variables defined in Sec. \ref{sec_eos}, assuming
spherical symmetry, and setting $\hat{r}=r/\sqrt{T}$, $\hat{h}=h/T$, $\hat{\Phi}=\Phi/T$, $\hat{\rho}=\rho$, and $\hat{M}=M/T^{3/2}$,
we obtain
\begin{equation}
 \hat{h}_{,\hat{r}^2}  + \frac{2}{\hat{r}} \hat{h}_{,\hat{r}} + 4 \pi  \hat{\rho}(\hat{h}) = 0
\mathrm{,}
\label{eq:5.2}
\end{equation}
where
\begin{equation}
\hat{\rho}(\hat{h}) =4\left (e^{\hat{h}}-e^{\hat{h}/2}\right )
\mathrm{.}
\label{eq:5.2b}
\end{equation}
Using Gauss
theorem $\Phi_{,r}={M(r)}/{r^2}$, where
\begin{equation}
M(r)=\int_0^r \rho(r') 4\pi {r'}^2\, dr'
 \mathrm{,}
\label{eq:massr}
\end{equation}
is the mass profile,  and the equilibrium relation $\Phi_{,r}=-h_{,r}$ , we obtain $\hat{\Phi}_{,\hat{r}}=-\hat{h}_{,\hat{r}}={\hat{M}(\hat{r})}/{\hat{r}^2}$ that allows us to determine the mass profile from the enthalpy profile using\footnote{Equation (\ref{fr}) may also be obtained by multiplying equation (\ref{eq:5.2}) by $\hat{r}^2$ and integrating between $0$ and $\hat{r}$.}
\begin{equation}
\hat{M}(\hat{r})=-\hat{r}^2 \hat{h}_{,\hat{r}}.
\label{fr}
\end{equation}

\begin{figure}[htbp]
\centerline{
\includegraphics[height=2.1in]{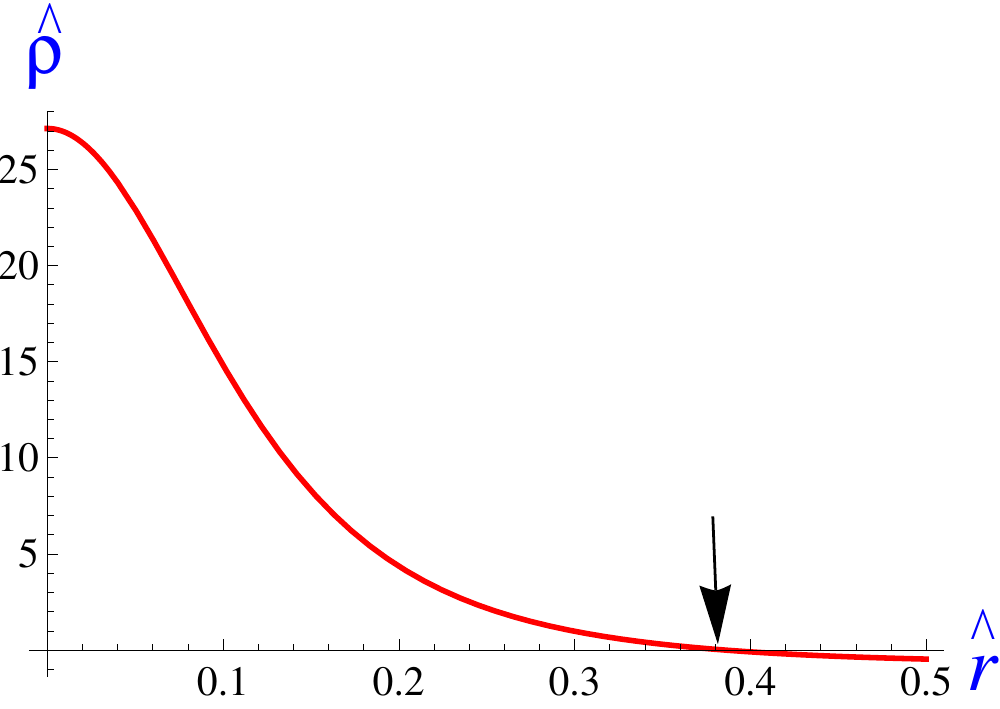}
}
\caption{Density $\hat{\rho}(\hat{r})$  versus the radial variable at the
saddle-node ($T=T_c$, or $\hat{h}_0=2.296$). The density  vanishes at the edge of the star indicated by the arrow ($\hat{r}=\hat{r}_0$).
}
\label{Fig:criticrho}
\end{figure}

\begin{figure}[htbp]
\centerline{
 \includegraphics[height=2.2in]{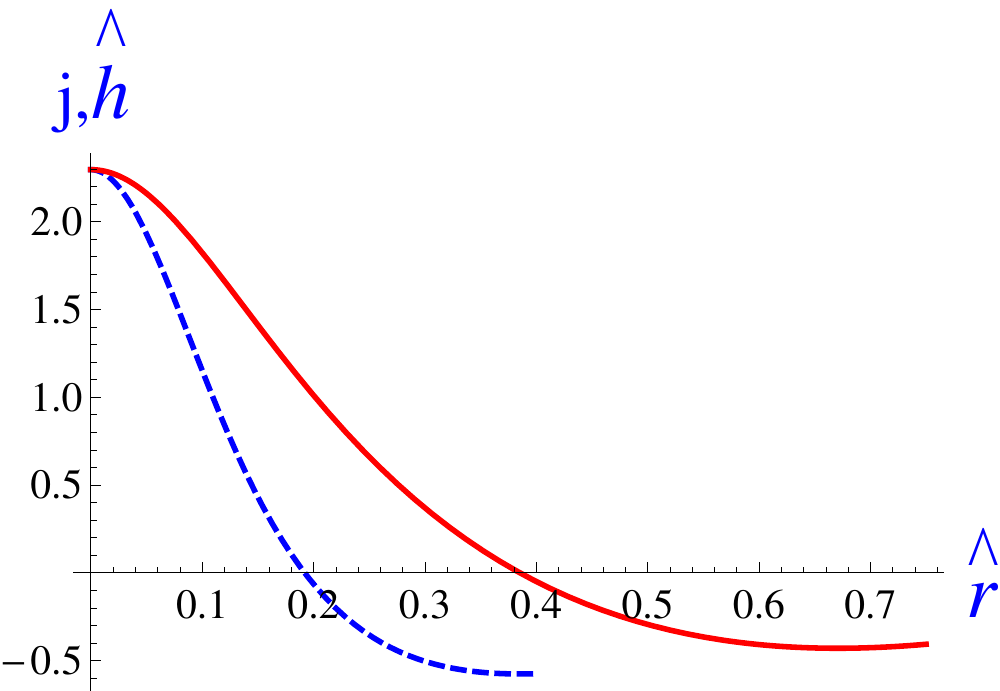}
}
\caption{Numerical solution of equations (\ref{eq:5.2}) and (\ref{eq:5.3}),
radial profile of the enthalpy $\hat{h}(\hat{r})$ (solid red curve)
and  neutral mode $j(\hat{r})$ (dashed blue curve) for $\hat{h}_0=2.296$
corresponding to the saddle-node, point $A$ of Fig. \ref{Fig:spi}.
}
\label{Fig:heq}
\end{figure}

The boundary conditions of equation (\ref{eq:5.2}) at $\hat{r}=0$ are $\hat{h}(0)= \hat{h}_0$ and
$\hat{h}_{,\hat{r}}(0)=0$.  For a given value of $\hat{h}_0$, the
smallest root of $\hat{h}(\hat{r})$, which is also the one of
$\hat{\rho}(\hat{r})$, see Figs. \ref{Fig:criticrho} and  \ref{Fig:heq}, defines the normalized radius $\hat{r}_{0}$ of
the star. The radius $r_0$ of the star is therefore
$r_0=\sqrt{T}\hat{r}_0$. On the other hand,  Gauss theorem
 applied at the surface of the star where $M=1$ (i.e. $\hat{M}_0=1/T^{3/2}$) leads to $\hat{h}_{,\hat{r}}(\hat{r}_0)=-1/(\sqrt{T}r_0^2)$. From
these equations, we obtain\footnote{We can come back to the original
(dimensional) variables by making the substitution $R\rightarrow
R/L=R\rho_*^{1/3}/M^{1/3}$ and $T\rightarrow T/(G\rho_{*}L^{2})=
T/(G\rho_{*}^{1/3}M^{2/3})$.}
\begin{equation}
r_0=\left (\frac{\hat{r}_0}{-\hat{h}_{,\hat{r}}(\hat{r}_0)}\right )^{1/3},\qquad T=\frac{1}{\left (-\hat{r}_0^2 \hat{h}_{,\hat{r}}(\hat{r}_0)\right )^{2/3}}.
\label{add1}
\end{equation}

\begin{figure}[htbp]
\centerline{
\includegraphics[height=2.2in]{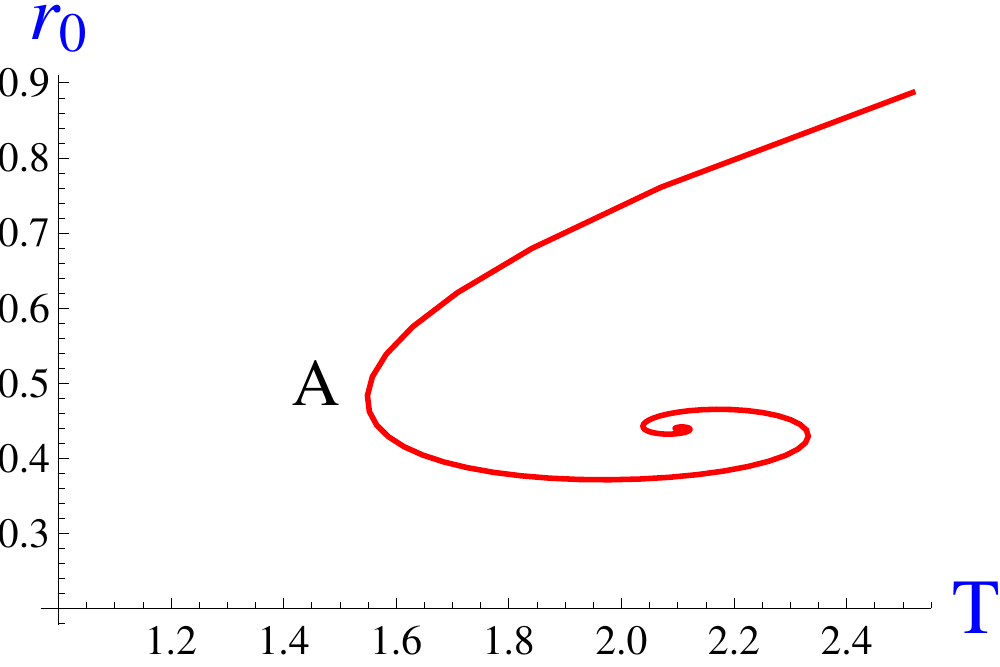}
}
\caption{Radius $r_0= \hat{r}_0 \hat{M}_0^{-{1}/{3}}$
versus temperature $T=\hat{M}_0^{-{2}/{3}}$, obtained by solving equations (\ref{eq:5.2})-(\ref{add1}) (increasing the input parameter $\hat{h}_0$).
}
\label{Fig:spi}
\end{figure}

The solution of equation (\ref{eq:5.2}), drawn in
Fig. \ref{Fig:heq} solid line, has a single free parameter $\hat{h}_0$ since
its Taylor expansion near ${\hat{r}} = 0 $ is like $\hat{h} =
\hat{h}_0 + h_2 {\hat{r}}^2 +...$ with $\hat{h}_0$ free, $h_2 = -
\frac{2\pi}{3} \hat{\rho}(\hat{h}_0)$, and so on for the higher order
coefficients. By varying $\hat{h}_0$ from $0$ to $+\infty$ we can
obtain the whole series of equilibria $r_0(T)$ giving the radius of the
star as a function of the temperature, using the quantities $\hat{h}_0$ (or
$\hat{r}_{0}$) as a parameter. The result is a spiralling curve shown in
Fig. \ref{Fig:spi} where only the upper part is stable, the
solution loosing its stability at the saddle-node (turning point A),
as studied in the next subsection\footnote{This temperature-radius relation $T(R)$
is the counterpart of the mass-radius relation $M(R)$ of boson stars in
general relativity, that also presents a spiralling behavior
\cite{ch}. The dynamical stability of the configurations
may be determined from the theory of Poincar\'e on the linear series
of equilibria as explained in \cite{aaa}. If we plot the temperature
$T$ as a function of the parameter $\hat{h}_0$, a change of stability
can occur only at a turning point of temperature. Since the system is
stable at high temperatures (or low $\hat{h}_0$) because it is
equivalent to a polytrope $n=1$ that is known to be stable, we
conclude that the upper branch in Fig. \ref{Fig:spi} is stable up to the
turning point $A$. Then, the series of equilibria loses a mode of
stability at each turning point of temperature $T$ and becomes more
and more unstable.}.  The saddle-node is found numerically to occur at
$\hat{h}_0=2.296..$, or $\hat{\rho}_0= 27.1299..$, that leads to the
following critical values for the mass, temperature and radius
respectively, $\hat{M}_c=0.52$, $T_c=1.546...$ and $\hat{r}_c=0.385..$
(hence $r_c=\sqrt{T_c}\hat{r}_c=0.479...$). The center of the spiral
is obtained for $\hat{h}_0 \rightarrow \infty$.

There is a saddle-node bifurcation when equation (\ref{eq:5.2}) linearized
about the profile $\hat{h}(\hat{r})$ determined previously has a non
trivial solution. This corresponds to the neutral mode $\delta h$ defined
by the unscaled equation (\ref{he8}). In terms of the scaled variables
this linearized equation reads
\begin{equation}
\Omega[j(\hat{r})] =  j_{,\hat{r}^2}  + \frac{2}{\hat{r}} j_{,\hat{r}} + 4 \pi  \frac{{\mathrm{d}}\hat{\rho}}{{\mathrm{d}}\hat{h}} j(\hat{r})=0
\mathrm{,}
\label{eq:5.3}
\end{equation}
where  $\Omega$ is a linear operator acting on function $j$ of
$\hat{r}$. Let us precise that we have the following boundary conditions: arbitrary $j(0)$ and $j'(0)=0$. Furthermore, we automatically have  $j'(\hat{r}_0)=0$ since $\delta M(r_0)=0$.
 The neutral mode $j(\hat{r})$, valid at the critical temperature $T_c$, is pictured in Fig. \ref{Fig:heq},
dashed blue line. We consider below the dynamics of the
function $M(r,t)$ which is the mass contained inside the sphere of
radius $r$ in the star.

\section{Dynamics close to the saddle-node: derivation of  Painlev\'e I equation}
\label{sec_dynA}

In this section we show that the dynamics close to the saddle-node reduces to Painlev\'{e} I equation. This property will be proved first by showing that  the normal form of the full Euler-Poisson system (\ref{iso1})-(\ref{iso3}) is of Painlev\'{e} I  form, secondly by comparing the normal form solutions to the full Euler-Poisson ones derived by using a numerical package for high-resolution central schemes \cite{progbalbas}.

\subsection{Simplification of the hydrodynamic
 equations close to the saddle-node}
\label{sec_simpli}

We now consider the dynamical evolution of the star, in particular
its gravitational collapse when the temperature falls below $T_c$. In this section and in the following one we use a simplified model where advection has been neglected, an approximation valid in the first stage of the collapse only.
In the following we restrict ourselves to spherically symmetric cases,
likely an approximation in all cases, and certainly not a good
starting point if rotation is present. However this allows a rather
detailed analysis without, hopefully, forgetting anything
essential. Defining $u$ as the radial component of the velocity, let
us estimate the order of magnitude of the various terms in Euler's
equations during the early stage of the collapse, namely when equation
(\ref{eq:ab}) is valid (this assuming that it can be derived from the
fluid equations, as done below). The order of magnitude of $u_{,t}$ is
the one of $\ddot{R}$, that is $\dot{R}/t_0$, with $t_0$ the
characteristic time defined by equation (\ref{eq:to}).
The order of magnitude of the advection term $u u_{,r}$ is
$\dot{R}^2/r_0$ (here $R$ is dimensional), because one assumes (and will show) that the
perturbation during this early stage extends all over the
star. Therefore $u u_{,r} \sim u_{,t} (R/r_0) $ is smaller than
$u_{,t}$ by a factor $R/r_0 $, which is the small a-dimensional
characteristic length scale defined by the relation
(\ref{eq:rs}). Neglecting the advection term in equations
(\ref{iso2}) and (\ref{iso4}) gives
\begin{equation}
\frac{\partial}{\partial t}(\rho {\bf u})= \rho\frac{\partial}{\partial t}{\bf u}=   -\nabla p-\rho\nabla\Phi.
\label{iso5}
\end{equation}
In the spherically symmetric case it becomes
\begin{equation}
u_{,t} =-\frac{1}{\rho}p_{,r} - \frac{4\pi G}{r^2} \int_0 ^r {\mathrm{d}}r' r'^2 \rho(r',t)
\mathrm{,}
\label{eq:8}
\end{equation}
where we used Gauss theorem
\begin{equation}
\Phi_{,r} = \frac{4 \pi G}{r^2} \int_0 ^r {\mathrm{d}}r' r'^2 \rho(r',t)
\mathrm{,}
\label{gauss}
\end{equation}
derived from Poisson equation (\ref{iso3}). Taking the divergence of the integro-differential dynamical equation (\ref{eq:8}) allowing to get rid of the integral term, we obtain
\begin{equation}
\left (\frac{2}{r}u+u_{,r}\right )_{,t} =-\left ( h_{,r^2}  + \frac{2}{r} h_{,r} + 4 \pi G  \rho(h) \right )
\mathrm{,}
\label{eq:8b}
\end{equation}
which is the  dynamical equation for the velocity field. This equation has been
derived from the Euler-Poisson system (\ref{iso1})-(\ref{iso3}) where the
advection has been neglected, that is  valid during the time interval of order
$t_0$ before the critical time. To derive the Painlev\'e I equation from the
dynamical equation (\ref{eq:8b}) we consider its right-hand side as a function
of $\rho$ with an equation of state of the form $p(\rho)=\rho_* T
f(\rho/\rho_*)$ depending on a slow parameter $T$, and we expand the solution
near a saddle-node bifurcation which exists when there is more than one steady
solution of equation (\ref{eq:8b})  for a given total mass $M=4\pi\int_0
^{\infty} {\mathrm{d}}r' r'^2 \rho(r')$  and temperature $T$, two  solutions
merging and disappearing as the temperature crosses a critical value
$T_c$. This occurs for the equation of state
defined by equation (\ref{eos3}), see  Fig. \ref{Fig:spi} where a saddle-node exists at point $A$. Although this formulation in
terms of the velocity field $u(r,t)$ is closely related to the heuristic
description developed in Sec. \ref{Scaling laws}, in the following we find it
more convenient to work in terms of the mass profile $M(r,t)$. Obviously the two
formulations are equivalent.

\subsection{The equation for the mass profile $M(r,t)$}
\label{msaM}

In view of studying the dynamics of the solution close to the
saddle-node, let us assume a slow decrease of the temperature versus
time, of the form $T=T_c(1-\gamma' t)$ with positive $\gamma'$ in
order to start at negative time from an equilibrium state. Taking the time
derivative of the equation of continuity (\ref{iso1}) and using equation (\ref{iso5}),
we get the two coupled equations\footnote{These equations are similar
to the Smoluchowski-Poisson system (describing self-gravitating
Brownian particles in the strong friction limit) studied in \cite{cs04}
except that it is second order in time instead of first order in
time.}
\begin{equation}
\frac{\partial^2\rho}{\partial t^2}=\nabla\cdot (\nabla p+\rho\nabla\Phi),
\label{iso6}
\end{equation}
\begin{equation}
\Delta\Phi=4\pi G\rho.
\label{iso7}
\end{equation}
According to the arguments given in Sec. \ref{sec_simpli}, these equations
are valid close to the saddle-node during the early stage of the
collapse\footnote{These equations are also valid for small
perturbations about an equilibrium state since we can
neglect the advection term ${\bf u}\cdot \nabla {\bf u}$ at linear
order.}. By contrast, when we are deep in the collapse regime (see Secs. \ref{sec:singular} and \ref{sec:beyond}) the advection term is important and we must come back to the full Euler-Poisson system (\ref{iso1})-(\ref{iso3}).

In the following, we use the dimensionless variables of Sec. \ref{sec_eos}. In the spherically symmetric case, using Gauss theorem
(\ref{gauss}), the system (\ref{iso6})-(\ref{iso7}) writes
\begin{equation}
\frac{\partial^2\rho}{\partial t^2}=\frac{1}{r^2}\left\lbrack r^2 p_{,r} +\rho  \int_0 ^r {\mathrm{d}} r' 4\pi r'^2 \rho(r')\right\rbrack_{,r}.
\label{iso6.2}
\end{equation}
It has to be completed by the boundary conditions imposing zero mass
at the center of the star, and a constant total mass
\begin{equation}
\int_0 ^{r_0} {\mathrm{d}}  r' 4\pi r'^2 \rho(r',t)=1,
\label{mass0}
\end{equation}
where $r_0$ is the star radius (practically the smallest root of $\rho(r) = 0$).
Let us define the variable
\begin{equation}
M(r,t)= \int_0 ^r {\mathrm{d}}  r'4\pi r'^2 \rho(r', t)
\label{mass}
\end{equation}
which represents the mass of fluid contained inside a sphere of radius $r$ at
time $t$. Multiplying the  two sides of equation (\ref{iso6.2}) by $ 4\pi r^2$, and
integrating them with respect to the radius, we obtain the dynamical equation
for the mass profile $M(r,t)$,
\begin{equation}
\frac{\partial^2 M(r,t)}{\partial t^2}= 4 \pi r^2 p_{,r} +  \frac{1}{r^2}M_{,r}  M,
\label{eq:d2M}
\end{equation}
where the term $p_{,r}=  p'(\rho) \rho_{,r}$ has to be expressed as a function of $\rho(r,t)=\frac{1}{4 \pi r^2}M_{,r}$ and $\rho_{,r}(r,t)=\frac{1}{4 \pi r^2}(M_{,r^2}-\frac{2}{r}M_{,r})$. Using the relation (\ref{eos3}), one has
\begin{equation}
p'(\rho)=  T\left (1-\frac{1}{\sqrt{1+{\rho}}}\right ).
\label{eq:m1}
\end{equation}
The first term of equation (\ref{eq:d2M}) becomes
\begin{equation}
4\pi r^2 p_{,r}=  T \mathcal{L}(M) g(M_{,r})
\end{equation}
with
\begin{equation}
\left \{ \begin{array}{l}
\mathcal{L}(M)= M_{,r^2}-\frac{2}{r}M_{,r}\\
  g(M_{,r})=1-\frac{1}{\sqrt{1+\frac{1}{4 \pi r^2}M_{,r}}}
  \mathrm{.}
\end{array}
\right.
\label{eq:m11}
\end{equation}
Introducing this expression into equation (\ref{eq:d2M}), the dynamical equation for $M(r,t)$ writes
\begin{equation}
\frac{\partial^2 M(r,t)}{\partial t^2}=  T \mathcal{L}(M)g(M_{,r}) + \frac{1}{r^2}M_{,r}  M.
\label{eq:d2Mf}
\end{equation}
The boundary conditions to be satisfied  are
\begin{equation}
 \left \{ \begin{array}{l}
M(0,t)=0  \\
M(r_0(t),t)= 1 = 4 \pi\int_0 ^{r_0(t)} {\mathrm{d}} r' r'^2 \rho(r',t)
\mathrm{.}
\end{array}
\right. \label{eq:scales}
\end{equation}
In the latter relation  the radius of the star $r_0(t)$ depends on time. However
this dependance will be  neglected below, see equation (\ref{eq:bcMn}), because
we ultimately find that the star collapses, therefore its radius will decrease,
leading to $r_0(t) < r_c$, or $M(r_0(t),t)= M(r_c) $ as time goes on.

\subsection{Equilibrium state and neutral mode}
\label{eqmsa}

A steady solution of equation (\ref{eq:d2Mf}) is determined by
\begin{equation}
T \mathcal{L}(M)g(M_{,r}) + \frac{1}{r^2}M_{,r}  M=0.
\label{mm1}
\end{equation}
Using Gauss theorem $\Phi_{,r}={M(r)}/{r^2}$,
 and the equilibrium relation $\Phi_{,r}=-h_{,r}$, we can easily check that equation (\ref{mm1}) is equivalent to equation (\ref{eq:5.2}). We now consider a small perturbation about a steady state and write $M(r,t)=M(r)+\delta M(r,t)$ with $\delta M(r,t)\ll M(r)$. Linearizing equation (\ref{eq:d2Mf}) about this steady state and writing the perturbation as $\delta M(r,t)\propto e^{\lambda t}$, we obtain the eigenvalue equation
\begin{eqnarray}
\lambda^2\delta M=T\left\lbrack {\cal L}(\delta M)g(M_{,r})+{\cal L}(M)g'(M_{,r})\delta M_{,r}\right\rbrack\nonumber\\
+\frac{1}{r^2}(M\delta M)_{,r}.
\label{mm3}
\end{eqnarray}
The neutral mode, corresponding to $\lambda=0$, is determined by the differential equation
\begin{equation}
T\left\lbrack {\cal L}(\delta M)g(M_{,r})+{\cal L}(M)g'(M_{,r})\delta M_{,r}\right\rbrack +\frac{1}{r^2}(M\delta M)_{,r}=0.
\label{mm4}
\end{equation}
Using Gauss theorem
$\delta\Phi_{,r}={\delta M(r)}/{r^2}$, and the relation $\delta\Phi_{,r}=-\delta h_{,r}$ satisfied at the neutral point (see Sec. \ref{sec_henm}), we can check that equation (\ref{mm4}) is equivalent to equation (\ref{eq:5.3}). This implies that the neutral mass profile is given by
\begin{equation}
\delta M(r)=-r^2 j_{,r}.
\label{mm5}
\end{equation}

\subsection{Normal form of the mass profile $M(r,t)$}

\begin{figure}[htbp]
\centerline{
\includegraphics[height=1.7in]{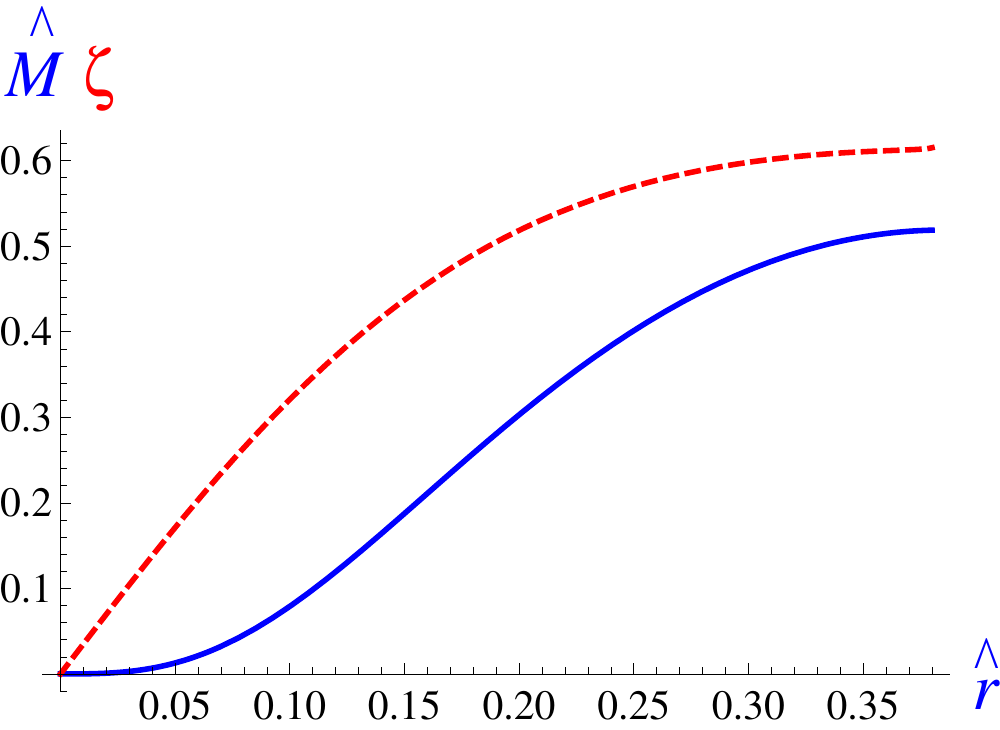}
}
\caption{ Mass ${\hat{M}}$ (solid blue line) inside the star versus the radial variable ${\hat{r}}$ at the
saddle-node, solution of equations (\ref{eq:5.2})-(\ref{eq:massr})  for $T=T_c$, i.e. $\hat{h}_0=2.296$. The dashed red line is for $\zeta(\hat{r})$, solution of equation (\ref{eq:zetadiff1}) with appropriate initial conditions for solving the adjoint problem (in this caption, we have restored the ``hat'' on the variables).
}
\label{Fig:criticM}
\end{figure}

The derivation of the normal form close to the saddle-node proceeds mainly along the lines of \cite{cs04}\footnote{The authors of \cite{cs04} study the dynamics of Smoluchowski-Poisson equations close to a saddle-node but for a {\it fixed} value of the temperature $T\rightarrow T_c^{-}$.}. The mass profile is expanded as
\begin{equation}
 M(r,t)= M^{(c)}(r) + \epsilon M^{(1)}(r,t)+\epsilon^2 M^{(2)}(r,t)+...
\label{eq:serM}
\end{equation}
where $M^{(c)}(r)$ is the equilibrium profile at $T=T_c$ (see  above) drawn in solid line in Fig. \ref{Fig:criticM}, and $\epsilon$ is a small parameter which characterizes a variation of the temperature with respect to its value at the collapse. We set
\begin{equation}
T=T_c(1-\epsilon^2 T^{(2)}),
\label{eq:epsilon}
\end{equation}
which amounts to defining $\epsilon^2 T^{(2)}=\gamma' t$, and rescaling
the time as $ t=t'/\epsilon ^{{1}/{2}}$ (this implies that $\gamma'\sim \epsilon^{5/2}$ is a small quantity). Substituting the
expansion (\ref{eq:serM}) into equation (\ref{eq:d2Mf}), we get at
leading order the equilibrium relation
\begin{equation}
T_c \mathcal{L}^{(c)} g^{(c)} + \frac{1}{r^2}M_{,r}^{(c)}  M^{(c)}=0,
\label{eq:M0}
\end{equation}
which  has to satisfy the boundary conditions
\begin{equation}
M^{(c)}(0)= M^{(c)}_{,r}(0)=0;\qquad M^{(c)}(r_c)= 1
.
\end{equation}
To order $1$ we have
\begin{equation}
T_c\left( \mathcal{L}^{(1)}g^{(c)} + \mathcal{L}^{(c)}g'^{(c)} M^{(1)}_{,r} \right)+  \frac{1}{r^2}(M^{(1)}  M^{(c)})_{,r}=0,
\label{eq:M1}
\end{equation}
and to order $2$
\begin{equation}
\frac{\partial^2 M^{(1)}(r,t')}{\partial t'^2} = T_c \mathcal{F}^{(2)}+  \frac{1}{r^2}\left\lbrack (M^{(2)}  M^{(c)})_{,r}+ M^{(1)}  M^{(1)}_{,r} \right\rbrack,
\label{eq:M2}
\end{equation}
where
\begin{equation}
\mathcal{F}^{(2)}=\mathcal{F}^{(2)}_1+\mathcal{F}^{(2)}_2+\mathcal{F}^{(2)}_3
\label{eq:F2}
\end{equation}
with
\begin{equation}
\mathcal{F}^{(2)}_1=\left(\mathcal{L}^{(2)}-T^{(2)}\mathcal{L}^{(c)} \right) g^{(c)},
\end{equation}
\begin{equation}
\mathcal{F}^{(2)}_2=\mathcal{L}^{(1)}g'^{(c)} M^{(1)}_{,r},
\end{equation}
\begin{equation}
\mathcal{F}^{(2)}_3=  \mathcal{L}^{(c)}\left\lbrack g'^{(c)} M_{,r}^{(2)}+\frac{g''^{(c)}}{2}(M_{,r}^{(1)})^2 \right\rbrack,
\end{equation}
where  $\mathcal{L}^{(c)}=\mathcal{L}(M^{(c)})$, $\mathcal{L}^{(n)}=\mathcal{L}(M^{(n)})$, $g^{(c)}=g(M_{,r}^{(c)})$, $g'^{(c)}=(\frac{dg}{dM_{,r}})^{(c)}$ and $g''^{(c)}=(\frac{d^2g}{dM_{,r}^2})^{(c)}$.
The $r$-dependent quantities can be written in terms of the equilibrium density function $\rho^{(c)}(r)$ as
\begin{equation}
 \left \{ \begin{array}{l}
\mathcal{L}^{(c)}=4\pi r^2\rho_{,r}^{(c)}, \\
g^{(c)} = 1-\frac{1}{\sqrt{1+\rho^{(c)}}}, \\
g'^{(c)}  =\frac{1}{8\pi r^2(1+\rho^{(c)})^{3/2}}, \\
g''^{(c)} = -\frac{3}{4(4\pi r^2)^2(1+\rho^{(c)})^{5/2}}
\mathrm{.}
\end{array}
\right. \label{eq:gg'}
\end{equation}
The boundary conditions are
 \begin{equation}
 \left \{ \begin{array}{l}
M^{(n)}(0,t')=0  ; M^{(n)}_{,r}(0,t')=0; \\
 M^{(n)}(r_c,t')=0
\mathrm{.}
\end{array}
\right. \label{eq:bcMn}
\end{equation}

Let us rescale the quantities in equations  (\ref{iso6.2})-(\ref{eq:bcMn}) by using the critical value $T_c$ for the temperature in
the rescaled variables. We thus define
$\hat{T}=T/T_c$, $\hat{r} =r/\sqrt{T_c}$, $\hat{t}=t$, $\hat{M}=M/T_c^{3/2}$, $\hat{h}={h}/{T_c}$, and $\hat{\rho}=\rho$. This rescaling leads to the same
expressions as the unscaled ones in equations
(\ref{iso6.2})-(\ref{eq:bcMn}), except that $T_c$ is set to
one. Furthermore, at the critical point, the rescaled variables coincide with those introduced in Sec. \ref{sec_tr}. In the following, we drop the superscripts to simplify the notations.

\begin{figure}[htbp]
\centerline{
(a) \includegraphics[height=1.7in]{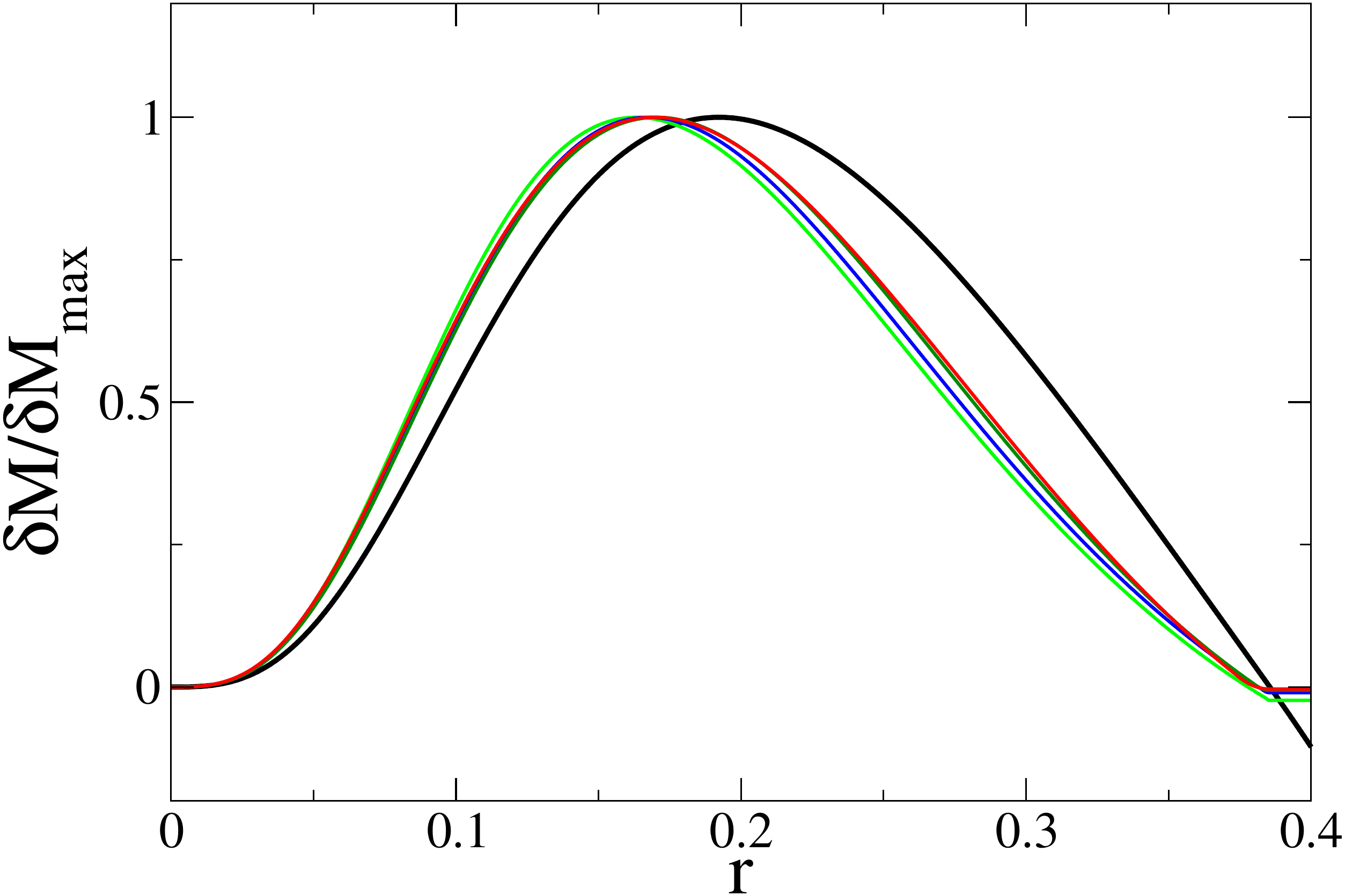}
}
\centerline{
(b)\includegraphics[height=1.7in]{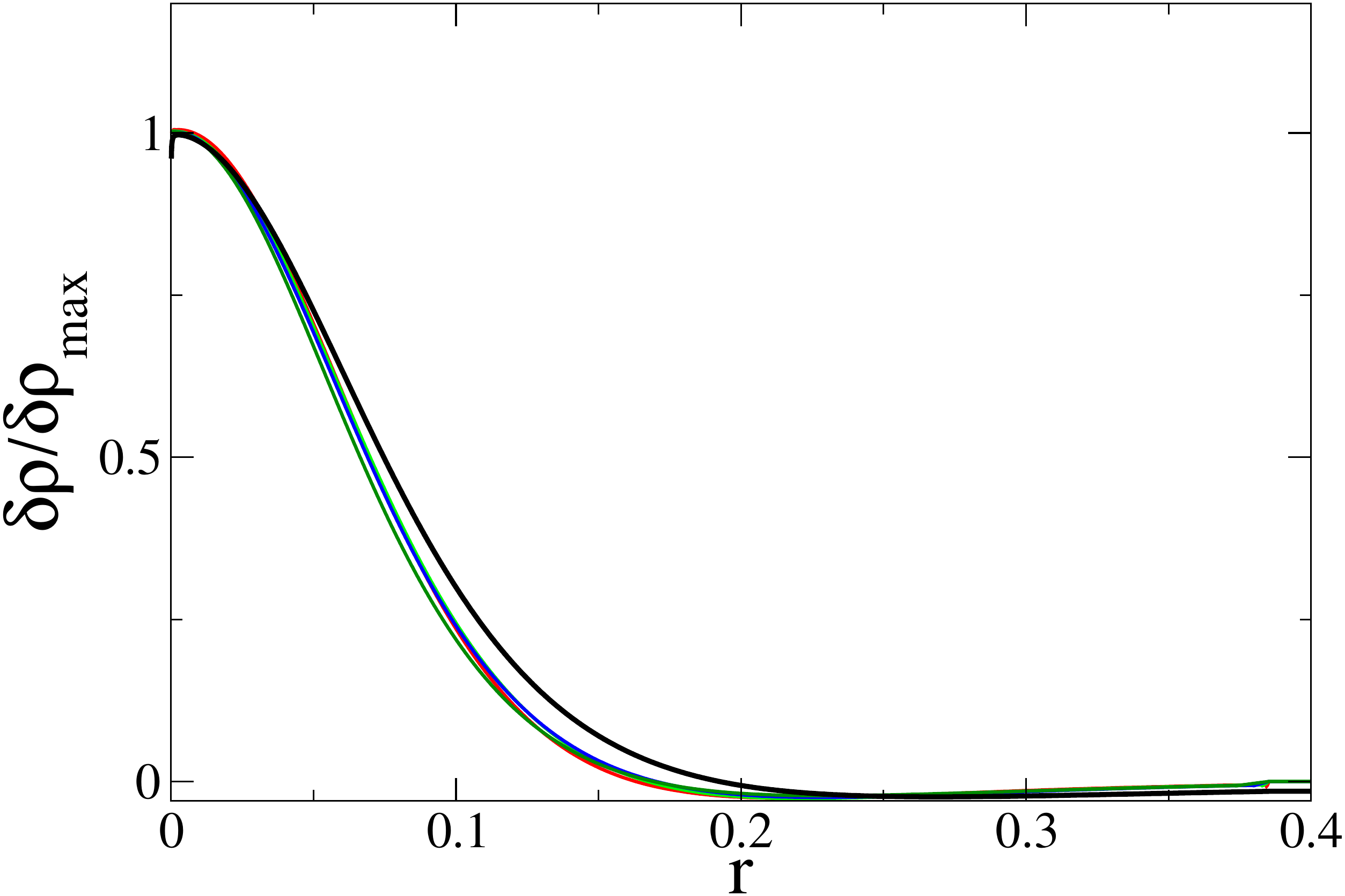}
}
\caption{Comparison between theory (thick black curves) and the numerics (thin colored curves) for the first order terms: (a) mass $M^{(1)}({r,t})$
 (b) density $\rho^{(1)}(r,t)$, in scaled variables. The numerical curves  correspond to the times
$t=0.2$ to $0.6$ in Fig. \ref{Fig:Painleve}.
}
\label{Fig:criticMR}
\end{figure}

The foregoing  equations have a clear interpretation. At zeroth order, equation (\ref{eq:M0}) corresponds to the equilibrium state (\ref{mm1}), equivalent to equation (\ref{eq:5.2}), at the critical point $T_c$. The critical mass profile is drawn in Fig. \ref{Fig:criticM} solid line. At order $1$, equation (\ref{eq:M1}) has the same form as the differential equation (\ref{mm4}), equivalent to equation (\ref{eq:5.3}), determining the neutral mode (corresponding to the critical point).  Because equation (\ref{eq:M1}) is linear, its solution is
\begin{equation}
M^{(1)}(r,t')= A^{(1)}(t')F(r),
\label{eq:Maf1}
\end{equation}
where
\begin{equation}
F(r) = \delta M(r)= - r^2 j_{,r},
\label{eq:Maf}
\end{equation}
according to equation (\ref{mm5}). This solution, drawn  in Fig. \ref{Fig:criticMR}-(a), thick black line, fulfills the boundary conditions (\ref{eq:bcMn}). The corresponding density profile $\rho^{(1)}(r,t')= A^{(1)}(t')\delta \rho(r)$ is drawn in Fig. \ref{Fig:criticMR}-(b), where
\begin{equation}
 \delta \rho(r)= \frac{F_{,r}}{4\pi r^2}= j(r)\left (\frac{d\rho}{ dh}\right )_{(c)}.
\label{eq:rho1}
\end{equation}
At order $2$, equation (\ref{eq:M2}) becomes
\begin{equation}
F(r)\ddot{A}^{(1)}(t') = -T^{(2)}\mathcal{L}^{(c)} g^{(c)} +\mathcal{D}(F) A^{(1)2} + \mathcal{C}(M^{(2)}),
\label{eq:AFeq}
\end{equation}
where
\begin{equation}
 \mathcal{D}(F)=\frac{1}{r^2}FF_{,r} +\frac{1}{2}\mathcal{L}^{(c)} g''^{(c)} F_{,r}^2 +g'^{(c)}{\cal L}(F)F_{,r},
 \label{eq:Deq}
\end{equation}
and
\begin{equation}
\mathcal{C}(M^{(2)})= \mathcal{L}^{(2)} g^{(c)} +\frac{1}{r^2}(M^{(2)}M^{(c)} )_{,r}+\mathcal{L}^{(c)} g'^{(c)}  M^{(2)}_{,r}.
\label{eq:Ceq0}
\end{equation}

To write the dynamical equation for $A(t)^{(1)}$ in a normal form, we  multiply equation (\ref{eq:AFeq}) by a function $\zeta(r)$ and integrate over $r$ for $0 <r <r_c$,  where $r_c$ is the radius of the star at $T=T_c$. We are going to derive the function $\zeta(r)$ so that the term ${\cal C}(M^{(2)})$  disappears after integration (see Appendix \ref{sec_cl} for details about the boundary conditions). Introducing the slow decrease of the temperature versus time, $T^{(2)}\sim \gamma' t/\epsilon^2$, and making the rescaling $A=\epsilon A^{(1)}$ to eliminate $\epsilon$ (we note that $A(t)$ is the true amplitude of the mass profile $\delta M(r,t)$), the result writes
\begin{equation}
\ddot{A}(t)= \tilde{\gamma} t+K A^2,
\label{eq:Ceq}
\end{equation}
where
\begin{equation}
\tilde{\gamma}= -\gamma' \frac{\int_0^{r_c}{\mathrm{d}}r {\cal L}^{(c)}(r) g^{(c)}(r)\zeta(r)}{\int_0^{r_c}{\mathrm{d}}rF(r)\zeta(r)}
\label{eq:coefft}
\end{equation}
is found equal to $\tilde{\gamma}= 120.2...\gamma'$ and
\begin{equation}
K= \frac{\int_0^{r_c}{\mathrm{d}}r \mathcal{G}(r) \zeta(r)}{\int_0^{r_c}{\mathrm{d}}rF(r)\zeta(r)},
\label{eq:coeffA2}
\end{equation}
with
\begin{eqnarray}
\mathcal{G}(r)=\frac{1}{2}{\cal L}^{(c)}(r) g''^{(c)}(r)F_{,r}^2 +g'^{(c)}(r) F_{,r}(F_{,r^2}-\frac{2}{r}F_{,r})\nonumber\\
+\frac{1}{r^2} F(r)F_{,r}\qquad\qquad
\end{eqnarray}
is found  to have the numerical value $K=12.32...$. We have therefore established that the amplitude $A(t)$ of the mass profile $\delta M(r,t)$ satisfies
 Painlev\'e I equation.

By definition the function $\zeta$ must satisfy, for any function $M^{(2)}(r)$, the integral relation
\begin{equation}
\label{ire}
\int_0 ^{r_c} {\mathrm{d}}r\, \mathcal{C}(M^{(2)})(r) \zeta(r) =0.
\end{equation}
Let us expand  $\mathcal{C}$ as
\begin{equation}
\mathcal{C}(M^{(2)})=
g^{(c)} M^{(2)}_{,r^2} +b M^{(2)}_{,r}+c M^{(2)}
\label{eq:integC2}
\end{equation}
with  $b(r) = -{2g^{(c)}}/{r}+{M^{(c)}}/{r^2}+\mathcal{L}^{(c)}g'^{(c)}$  and $c(r) = {M^{(c)}_{,r}}/{r^2}$, or in terms of the equilibrium values of the density and potential functions at the saddle-node
\begin{equation}
 \left \{ \begin{array}{l}
 g^{(c)}(r) = 1-\frac{1}{\sqrt{1+\rho^{(c)}}}, \\
b(r) = -\frac{2g^{(c)}}{r} -h_{,r}^{(c)}+\frac{\rho_{,r}^{(c)}}{2(1+\rho^{(c)})^{3/2}}, \\
c(r) = 4\pi \rho^{(c)}
\mathrm{.}
\end{array}
\right.
\label{eq:abc}
\end{equation}

Integrating equation (\ref{ire}) by parts, and using  $M^{(2)}=0$ on the boundaries $r=0$ and $r=r_c$ (see Appendix \ref{sec_cl}),
 we find that $\zeta(r)$ must be a solution of the second order differential equation
\begin{equation}
(g^{(c)} \zeta)_{,r^2}-(b \zeta)_{,r}+ c\zeta=0,
\label{eq:zetadiff}
\end{equation}
with the initial condition $\zeta(0)=0$ (the radial
derivative $\zeta_{,r}(0)$ is a free parameter since the differential equation
is of second order). At the edge of the star we do not have $\zeta(r_c) = 0$, see below, but rather
$\zeta_{,r}(r_c)=0$: the radial derivative of
$\zeta$ vanishes because the second order differential equation
(\ref{eq:zetadiff1}) becomes a first order one (since $g^{(c)}(r_c) =0$, see
equation (\ref{eq:gg'})). This does not happen in the case studied in \cite{cs04}
where the pressure-density relation was $p = \rho T$, that leads to similar
relations as here,  but $g^{(c)}(r_c)=1$.
The differential equation for the unknown function $\zeta(r)$ writes
\begin{equation}
 g^{(c)}(r)\zeta_{,r^2}+ a_1(r) \zeta_{,r}+  a_0(r)\zeta=0,
\label{eq:zetadiff1}
\end{equation}
where the coefficients
\begin{equation}
 \left \{ \begin{array}{l}
a_1(r) = 2g^{(c)}_{,r} -b(r), \\
a_0(r) = c(r)+ g^{(c)}_{,r^2}(r) -b_{,r}(r),
\end{array}
\right.
\label{eq:a1a0}
\end{equation}
may be expressed in terms of the radial density using equations
(\ref{eq:gg'}) and (\ref{eq:abc}). It turns out that for $r=r_c$ we have $g^{(c)}=
a_0=0$, but $a_1(r_c) \neq0$, that gives the boundary
relation $\zeta _{,r}(r_c)=0$.

\begin{figure}[htbp]
\centerline{
\includegraphics[height=1.7in]{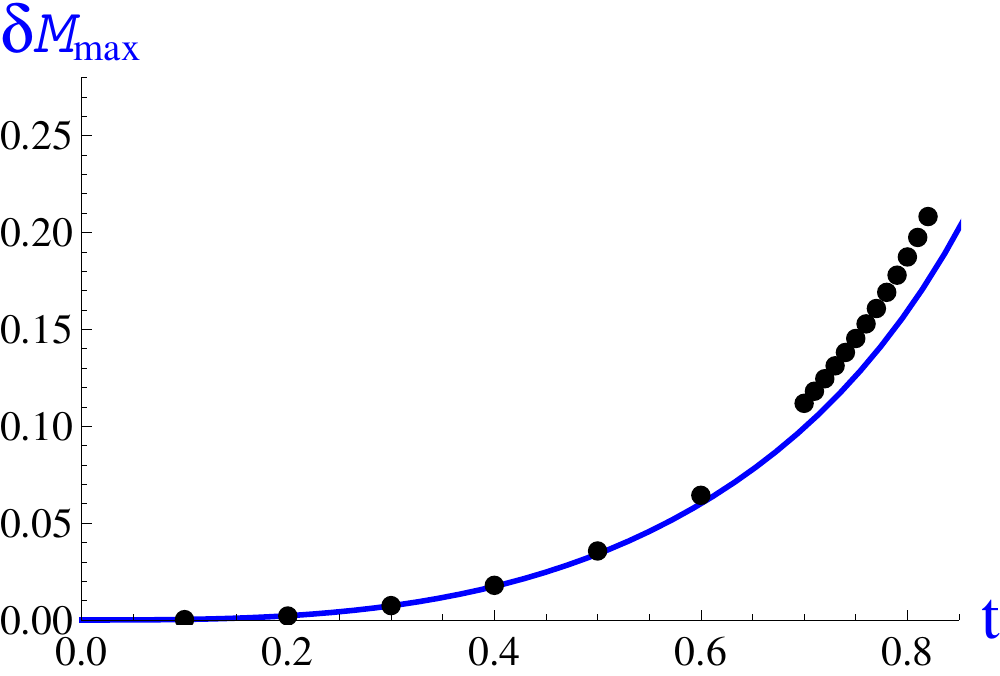}
}
\caption{Comparison between normal form (solid blue  curve) and numerical solution (dots) for the maximum of $M^{(1)}(r,t)$ versus time. In the numerical simulations of the Euler-Poisson system we start from the critical profile $M_c(r)$ at $t=0$ and decrease the temperature as $T(t)=1-\gamma' t$ with $\gamma'=0.1$.
}
\label{Fig:Painleve}
\end{figure}

The solution of equation (\ref{eq:zetadiff1}) with the condition
$\zeta(0)=0$ is shown in Fig. \ref{Fig:criticM}, red dashed line,
where $\zeta_{,r}(r_c)=0$.  Figure \ref{Fig:Painleve} shows the
evolution of the maximum value $M^{(1)}_{max} (t)$ of the profile
$M^{(1)}(r,t)$ with time (solid line). This quantity
is proportional to the function $A(t)$ that is the solution of
Painlev\'e equation (\ref{eq:Ceq}). It is compared with the numerical
solution of the full Euler-Poisson equations (dots). We see that the
results agree for small amplitudes but that the agreement ceases to be
correct at large amplitudes where our perturbative approach loses its
validity. It particular, the real amplitude increases more rapidly,
and the singularity occurs sooner, than what is predicted by
Painlev\'e equation.


{\it Remark:} According to the results of Sec. \ref{Scaling laws}, and coming back to the original (but still dimensionless) variables, we find that the collapse time in the framework of Painlev\'e equation is $t_*=\hat{t}_*/(K\tilde\gamma)^{1/5}$ with $\hat{t}_*\simeq 3.4$, i.e.
\begin{equation}
t_*=0.79... \left |\frac{T_c}{\dot T}\right |^{1/5}.
\label{collp}
\end{equation}
On the other hand, close to the collapse time, the amplitude of the mass profile diverges as $A(t)\sim (6/K)(t_*-t)^{-2}$ i.e.
\begin{equation}
A(t)\sim 0.487 \frac{1}{(t_*-t)^2}.
\label{collpf}
\end{equation}

\subsection{Discussion}
\label{subsec:disc}

This section was devoted to an explicit derivation of the ``universal" Painlev\'e I equation for the beginning of the collapse following the slow crossing of the saddle-node bifurcation for the equilibrium problem. We have chosen to expose this detailed derivation in a simple model of equation of state and without taking into account exchange of energy in the fluid equations. Of course this makes our analysis  qualitatively correct (hopefully!) but surely not quantitatively so for real supernovae, an elusive project anyway. We have shown that the Painlev\'e I equation represents the actual solution of the full Euler-Poisson system until the changes out of the solution at the saddle-node equilibrium are too large to maintain the validity of a perturbative approach. Our analysis explains well that the collapse of the star can be a very fast process following a very long evolution toward a saddle-node bifurcation.
As we shall explain in the next section, after the crossing of the saddle-node bifurcation, the solution of the Euler-Poisson equations have a finite time singularity at the center. We point out that this happens when the radius of the star has the order of magnitude it had at the time of the saddle-node bifurcation.
Therefore the size of the core should remain orders of magnitude smaller than the star radius, as  found for the Penston-Larson solution which predicts a core containing a very small portion of the total star mass. If the saddle-node bifurcation is the key of the implosion mechanism, this result should  not depend on the equation of state.
However the question of how massive is the self-collapsing core has received various answers. For supernovae in massive stars, starting from the hypothesis that pressure and gravity forces are of the same order during the collapse, Yahil \cite{Yahil} considered  equations of state of the form $p =K \rho^\Gamma$ with adiabatic indices in the range $ 6/5 < \Gamma \le 4/3$. He found that the ratio of the mass inside the core and the Chandrasekhar mass is almost constant, between $1.1$  and unity in this range of $\Gamma$. Moreover he found that the core moves at less than the sound speed, that was  considered as essential for all its parts to move in unison \cite{Bethe}. In the next section we show that the hypothesis that pressure and gravity forces are of the same order is not relevant to describe the collapse. Our derivation leads to a drastically different velocity field, which is supersonic in the core  and subsonic outside, tending to zero at the edge of the star.

\section{Finite time singularity of solutions of Euler-Poisson equations: pre-collapse}
\label{sec:singular}

The perturbation analysis presented so far can deal only with perturbations of small amplitude, that is corresponding to a displacement small compared to the radius of the star. We have seen that, at least up to moderate values of the amplitude of perturbations to the equilibrium solution, the analysis derived from Painlev\'e equation yields correct results, not only for the exponents, but also for all the numerical prefactors. This defines somehow completely the starting point of the ``explosion of the star". But there is still a long way toward the understanding of supernovae. As a next step forward, we shall look at the dynamics of the solution of the Euler-Poisson equations with radial symmetry, starting with a quasi-equilibrium  numerical solution of the equations of motion. We emphasize the importance of the initial conditions for solving the dynamics, a delicate problem which could lead to various solutions as discussed and illustrated in \cite{Brenner} for instance. The most noticeable feature of our numerical study is the occurrence of a singularity at the center after a finite time. To describe the numerical results, we must invoke a singularity of the second kind, in the sense of Zel'dovich \cite{Zel}. Contrary to the singularity of the first kind where the various exponents occurring in the self-similar solution are derived by a simple balance of all terms present in the equations, a singularity of the second kind has to be derived from relevant asymptotic matching, that may require to neglect some terms, as described in the present section.

The occurrence of a finite time singularity in the collapse of a
self-gravitating sphere has long been a topic of investigations.  An
early reference is the paper by Mestel \cite{mestel} who found the
exact trajectory of a particle during the free-fall\footnote{By
free-fall, we mean a situation where the collapse is due only to the
gravitational attraction, i.e. in which pressure forces are
neglected. This corresponds to the Euler-Poisson system
(\ref{iso1})-(\ref{iso3}) with $p=0$.} of a molecular cloud
(neglecting the pressure forces), assuming spherically symmetry. The
exact Mestel solution displays a self-similar solution of the
pressureless Euler-Poisson system as shown later on by Penston
\cite{Penston}, that leads to a finite time singularity with an
asymptotic density as $\rho(r) \sim r^{-\alpha}$ with $\alpha=12/7$,
smaller than $2$ (an important remark, as will be shown in the next
subsection).  Taking account of the pressure forces, another
self-similar solution was found independently by Penston
\cite{Penston} and Larson \cite{larson} which is usually called the
Penston-Larson solution. It is characterized by $\alpha=2$. This
solution was proposed to describe the gravitational collapse of an
isothermal gas assuming that pressure and gravitational forces scale
the same way. This corresponds to a self-similarity of the first kind
(the exponent being defined simply by balancing all the terms in the
original equations) by contrast to self-similarity of the second kind,
or in the sense of Zel'dovich, that we are considering below. In the
Penston-Larson solution, the magnitude of the velocity remains finite,
something in contradiction with our numerical findings. Moreover this
solution has a rather unpleasant feature, noticed by Shu \cite{Shu}:
it implies a finite constant inward supersonic velocity far from the
center, although one would expect a solution tending to zero far from
the center, as observed numerically. We present below another class of
singular solution which better fits the numerical observations than
the one of Penston \cite{Penston} and Larson \cite{larson}. In the
numerics we start from a physically relevant situation which consists
in approaching slowly the saddle-node bifurcation in a
quasi-equilibrium state.  As time approaches the collapse, we observe
that the numerical velocity tends to infinity in the core of the
singularity and decays to zero far from the center, in agreement with
the theoretical solution proposed, equations
(\ref{eq:rhoa})-(\ref{eq:ua}) below with $\alpha$ larger than $2$.
The equations we start from are the Euler-Poisson equations for the
mass density $\rho(r, t)$ and radial speed $u(r, t)$,
\begin{equation}
\rho_{,t} + \frac{1}{r^2} \left( r^2 \rho u \right)_{,r} = 0
\mathrm{.}
\label{eq:Euler.1}
\end{equation}
\begin{equation}
\rho\left(u_{,t} + u u_{,r}\right) = - T \rho_{,r} - \frac{G M(r,t) \rho}{r^2}
\mathrm{,}
\label{eq:Euler.2}
\end{equation}
with
 \begin{equation}
M(r,t)  = 4 \pi \int_0^r {\mathrm{d}}r' r'{^2} \rho(r',t)
\mathrm{.}
\label{eq:Euler.2.1}
\end{equation}
In the equations above, we consider the case of an isothermal equation
of state, $p=\rho T$, which amounts to considering the equation of
state (\ref{eos3}) in the limit of large density, that is the case in
the central part of the star.  The temperature $T$ has the physical
dimension of a square velocity, as noticed first by Newton, and $G$ is
Newton's constant.  The formal derivation of self-similar solutions
for the above set of equations is fairly standard. Below we focus on
the matching of the local singularity with the outside and on its
behavior at $r = 0$. A solution blowing-up locally can do it only if
its asymptotic behavior can be matched with a solution behaving
smoothly outside of the core.  More precisely, one expects that
outside of the singular domain (in the outer part of the core) the
solution continues its slow and smooth evolution during the blow-up,
characterized in particular by the fact that the velocity should
decrease to zero at the edge of the star meanwhile the local solution
(near $r=0$) evolves infinitely fast to become singular.

In summary, contrary the Penston-Larson derivation which imposes the
value $\alpha=2$ by balancing the terms in the equations and leads to
a free parameter value $R(0)$, our derivation starts with an unknown
$\alpha$ value (larger than $2$), but leads to a given value of
$R(0)$. In our case the unknown $\alpha$ value is found after
expanding the solution in the vicinity of the center of the star.
This yields a nonlinear eigenvalue problem of the second kind in the
sense of Zel'dovich \cite{Zel}, as was found, for instance, in the
case of the Bose-Einstein condensation \cite{BoseE,bosesopik} while
the Penston-Larson singular solution is of the first kind (again
because it is obtained by balancing all terms in the equations).

\subsection{General form of self-similar solutions}
\label{subsec5A}

The solution we are looking after is of the type for the density $\rho$,
\begin{equation}
 \rho(r, t) = (- t)^{\beta} R\left(r (-t)^{\beta/\alpha}\right) \mathrm{,}
\label{selfrho}
\end{equation}
and for the radial velocity $u$,
\begin{equation}
u(r, t) = (- t)^{\gamma} U\left(r (-t)^{\beta/\alpha}\right) \mathrm{,}
\label{selfu}
\end{equation}
where $\alpha$, $\beta$ and $\gamma$ are real exponents to be found. The
functions $R(.)$ (different from the function $R(t)$ introduced at the beginning
of this paper. We keep this letter to remind that it is the
scaled density $\rho$) and $U(.)$ are numerical functions with values of order one when their
argument  is of order one as well. They have to satisfy coupled differential
equations without small or large parameter (this also concerns the boundary
conditions). To represent a solution blowing up at time $t = 0$ (this time 0 is
not the time zero where the saddle-node bifurcation takes place; we have kept
the same notation to make the mathematical expressions lighter), one expects
that the density at the core diverges. This implies $\beta$ negative. Moreover
this divergence happens in a region of radius tending to zero at $t =0$.
Therefore $\alpha$ must be positive. Finally, at large distances of the collapsing core the solution must become independent on time.
This implies that $R(.)$ and $U(.)$ must  behave with
\begin{equation}
\xi= r(-t)^{\beta/\alpha}
\mathrm{,}
\label{Rasgh}
\end{equation}
as power laws when $\xi\gg 1$ such that the final result
obtained by combining this power law behavior with the pre-factor $(-t)^{\beta}$ for $R$ and $(-t)^{\gamma}$ for $U$ yields functions $\rho$ and $u$ depending on $r$ only, not on time.
  Therefore one must have
\begin{equation}
R(\xi)\sim \xi^{-\alpha}\mathrm{,}
   \label{Ras}
\end{equation}
and
 \begin{equation}
U(\xi)\sim \xi^{-\gamma\alpha/\beta}\mathrm{.}
   \label{Uas}
\end{equation}
In that case,
\begin{equation}
 \left \{ \begin{array}{l}
\rho(r,t)\propto r^{-\alpha}, \\
u(r,t)\propto r^{-\gamma\alpha/\beta}
\mathrm{,}
\end{array}
\right. \label{eq:asrhou}
\end{equation}
for $r\rightarrow
+\infty$ where the proportionality constants are independent on time.

Inserting those scaling assumptions in the dynamical equations, one finds that equation (\ref{eq:Euler.1}) imposes the relation
 \begin{equation}
  \frac{\beta}{\alpha}+\gamma + 1=0.
   \label{betaalpha}
\end{equation}
This relation is also the one that yields the same order of magnitude to the two terms $u_{,t}$ and $u u_{,r}$ on the left-hand side of equation (\ref{eq:Euler.2}). If one assumes, as usually done, that all terms on the right-hand side of equation (\ref{eq:Euler.2}) are of the same order of magnitude at $t$ tending to zero, this imposes $\alpha = -\beta= 2$ and $\gamma=0$. This scaling corresponds to the Penston-Larson solution. However, let us leave $\alpha$ free (again contrary to what is usually done where $\alpha = 2$ is selected) and consider the relative importance of the two terms in the right-hand side of equation (\ref{eq:Euler.2}), one for the pressure and the other for gravity. The ratio pressure to gravity is of order $t^{2\beta/\alpha-\beta}$. Therefore the pressure becomes dominant for $t$ tending to zero if $\alpha < 2$, of the same order as gravity if $\alpha = 2$ and negligible compared to gravity if $\alpha > 2$ (in all cases for $\beta$ negative). For pressure dominating gravity (a case where very likely there is no collapse because the growth of the density in the core yields a large centrifugal force acting against the collapse toward the center), the balance of left and right-hand sides of equation  (\ref{eq:Euler.2}) gives $\gamma=0$ and $\beta=-\alpha$, while in the opposite case, i.e. for $\alpha>2$,  it gives
 \begin{equation}
  \beta=-2
   \mathrm{,}
   \label{beta}
\end{equation}
 and
\begin{equation}
 \gamma= 2/\alpha - 1\mathrm{.}
  \label{gamma}
\end{equation}
   Therefore the velocity in the collapse region where $r \sim (-t)^{-\beta/\alpha}$ diverges only in the case of gravity dominating pressure ($\alpha>2$).

Our numerical study shows clearly that velocity diverges in the
collapse region. We believe that the early numerical work by Larson
\cite{larson} does not contradict our observation that $\alpha$ is
larger than 2: looking at his Figure 1, page 276, in log scale, one
sees rather clearly that the slope of the density as a function of $r$
in the outer part of the core is close to $-2$, but slightly smaller
than $(-2)$. The author himself writes that this curve ``approaches
the form $r^{-2}$" without stating that its slope is exactly $(-2)$,
and the difference is significant, without being very large. The slope
$- \alpha = - {24}/{11}$ derived below fits better the asymptotic
behavior in Figure 1 of Larson \cite{larson} than the slope $(-2)$
does (the same remarks apply to
Figure 1 of Penston \cite{Penston}). Therefore we look for a solution with
$\alpha >2$ for which the gravitational term dominates the pressure in
equation (\ref{eq:Euler.2}). As shown below, the existence of a
solution of the similarity equations requires that $\alpha$ has a well
defined value, one of the roots of a second degree polynomial, and the
constraint $\alpha>2$ allows us to have a velocity field decaying to
zero far from the singularity region, as observed in our numerics,
although $\alpha < 2$ yields a velocity field growing to infinity far
from the collapse region, something that forbids to match the collapse
solution with an outer solution remaining smooth far from the
collapse.  The case $\alpha = 2$ imposes a finite velocity at
infinity, also something in contradiction with the numerical results.

\subsection{A new self-similar solution where gravity dominates over pressure}
\label{subsec5B}

\subsubsection{Eigenvalue problem of the second kind}

In the following, we assume that gravity dominates over pressure forces, i.e. $\alpha>2$. The set of two integro-differential equations (\ref{eq:Euler.1}) and (\ref{eq:Euler.2}) becomes a set of coupled equations for the two numerical functions $R(\xi)$ and
$U(\xi)$  such that
 \begin{equation}
\rho(r,t) = (-t)^{-2} R(r(-t)^{-2/\alpha})
\mathrm{,}
\label{eq:rhoa}
\end{equation}
and
 \begin{equation}
 u(r,t) = (-t)^{-1+\frac{2}{\alpha}} U(r(-t)^{-2/\alpha})
 \mathrm{,}
\label{eq:ua}
\end{equation}
where
 $\xi=r(-t)^{-2/\alpha}$ is the scaled radius. As explained previously, we
must have
\begin{equation}
R(\xi)\sim \xi^{-\alpha}, \qquad {\rm and}\qquad  U(\xi)\sim \xi^{-(\alpha/2-1)},
\label{sunm}
\end{equation}
for
$\xi\rightarrow +\infty$ in order to have a steady profile at large distances.
The equations of conservation of mass and momentum become in scaled variables
 \begin{equation}
2 R + \frac{2 \xi}{\alpha} R_{,\xi} + \frac{2}{\xi} R U + (R U)_{,\xi} = 0
\mathrm{,}
\label{eq:Euler.4}
\end{equation}
 \begin{equation}
\left (1 - \frac{2}{\alpha}\right ) U + \frac{2}{\alpha} \xi U_{,\xi}  + U U_{,\xi} = - \frac{4 \pi G}{\xi^2} \int_0^{\xi} {\mathrm{d}}\xi'\,  \xi'^2 R(\xi')
\mathrm{.}
\label{eq:Euler.5}
\end{equation}
The integro-differential equation (\ref{eq:Euler.5}) can be transformed into a
differential equation, resulting into the following second order differential
equation for $U(.)$, supposing $R(.)$ known,
\begin{eqnarray}
U_{,\xi^2}\left (U+\frac{2}{\alpha}\xi\right ) + U_{,\xi}\left[1+ \frac{4}{\alpha}+\frac{2}{\xi}U + U_{,\xi}\right]\nonumber\\
-\frac{2\gamma}{\xi}U +4 \pi G R=0
\mathrm{.}
\label{eq:Euler.5bis}
\end{eqnarray}
From now on, we use the dimensionless
variables defined in Sec. \ref{sec_eos}. Concerning the initial conditions
(namely the conditions at $\xi =0$), they are derived from the possible Taylor
expansion of $U$ and $R$ near $\xi = 0$, like
\begin{equation}
R = R_0 + R_2 \xi^2  + R_4 \xi^4
+...
\end{equation}
and
\begin{equation}
U = U_1 \xi  + U_3 \xi^3+ U_5 \xi^5 + ...
 \end{equation}
Putting those
expansions in equations (\ref{eq:Euler.4}) and (\ref{eq:Euler.5}), one finds
$U_1 = -{2}/{3}$ and $R_0 = {1}/({6 \pi })$. Note that $R = R_0 $ and $ U
= \xi U_1$ is an exact solution of the equations (\ref{eq:Euler.4}) and
(\ref{eq:Euler.5}), that is not the usual case for such Taylor expansions.
This corresponds to the well-known free-fall solution of a
homogeneous sphere \cite{Penston}.   It follows from this peculiarity that, at
next order, we obtain  a linear homogeneous algebraic relation because the zero
value of $R_2$ and $U_3$ must be a solution. Inserting the above values of $R_0$
and $U_1$ at this order, we obtain the homogeneous relations
\begin{equation}
  \frac{3-\alpha}{3\alpha} R_2 + \frac{5}{24 \pi} U_3 = 0
  \mathrm{,}
  \label{eq:R2-U3}
 \end{equation}
and
\begin{equation}
 4 \pi R_2+ 5 \frac{12-5\alpha}{3\alpha} U_3  = 0 \mathrm{.}
 \label{eq:R2-U3b}
 \end{equation}
This has a non trivial solution (defined up to a global multiplying factor - see below for an explanation) if the determinant of the matrix of the coefficients is zero, namely if $\alpha $ is a root of the second degree polynomial
\begin{equation}
\frac{7}{3}\alpha^2 -18\alpha +24=0
 \label{eq:poly}
 \mathrm{.}
 \end{equation}
This shows that $\alpha$ cannot be left free and has to have a well defined value. However, it may happen that none of these two values of $\alpha$ is acceptable for the solution $R(\xi),U(\xi)$ we are looking for, so that we should take $R_2=U_3=0$ and pursue the expansion at next order. This is the case for our problem because one solution of equation (\ref{eq:poly}) is $\alpha=12/7$ which does not belong to  the domain $\alpha >2$ we are considering (because we assume that the gravity effects are stronger than the pressure effects)\footnote{We note that the exponent $\alpha=12/7$ was previously found by Penston \cite{Penston} for the free-fall of a pressureless gas ($T=0$) by assuming a regular Taylor expansion $\rho=\rho_0+\rho_2 r^2+...$ close to the origin. This solution is valid if $T$ is exactly zero but, when $T>0$, as it is in reality, this solution cannot describe a situation where gravity dominates over pressure (the situation that we are considering) since $\alpha=12/7<2$.
This is why Penston \cite{Penston} and Larson \cite{larson} considered a self-similar solution of the isothermal Euler-Poisson system
(\ref{eq:Euler.1})-(\ref{eq:Euler.2.1}) where both pressure and gravity terms
scale the same way. Alternatively, by assuming a more general expansion
$\rho=\rho_0+\rho_k r^k+...$ with $k>2$ close to the origin, we find a new
self-similar solution where gravity dominates over pressure.},
 and the other
solution $\alpha=6$ is excluded by the argument in section
\ref{subsec:upperbound} below.

Therefore we have to choose $R_2=U_3=0$ and consider the next order terms of the expansion, which also provides a homogeneous linear system for the two unknown coefficients $R_4$ and $U_5$. It is
\begin{equation}
4 \frac{3-\alpha}{3\alpha} R_4 + \frac{7}{12 \pi} U_5 = 0 \mathrm{,}
\end{equation}
and
\begin{equation}
4 \pi R_4+ 7 \frac{8-3\alpha}{\alpha} U_5  = 0 \mathrm{,}
\end{equation}
which has non trivial solutions if $\alpha$ is a root of the secular equation
\begin{equation}
\frac{11}{4}\alpha^2 -17\alpha +24=0
 \label{eq:polybis}
 \mathrm{,}
 \end{equation}
  whose solutions are $\alpha=4$ or $\alpha={24}/{11}$. The value $\alpha=4$ is excluded by the argument in section  \ref{subsec:upperbound}
  whereas the solution
  \begin{equation}
  \alpha=\frac{24}{11}
   \label{eq:alpha}
\end{equation}
could be the relevant one for our problem. In that case, we get $\beta=-2$ and $\gamma=-1/12$. The density decreases at large distances as $r^{-24/11}$ and the velocity as $r^{-1/11}$ (while in the Penston-Larson solution, the density decreases at large distances as $r^{-2}$ and the velocity tends to a constant value). Of course, we
can carry this analysis by beginning the expansion with an arbitrary power $k$ bigger than $2$ like $R=R_0+R_k \xi^k+...$ and
$U=U_1\xi+U_k\xi^{k+1}+...$ with arbitrary $k$ (actually, $k$ must be even for
reasons of regularity of the solution). In that case, we find the two exponents
 \begin{equation}
\alpha(k)=\frac{6k}{2k+3}
 \label{eq:alphak}
\end{equation}
 and $\alpha=3k/(k-1)$. We note that the first exponent varies
between $0$ (homogeneous sphere) and $3$, while the second exponent is larger
than $3$ for $k>1$ which is unphysical by the argument in section
\ref{subsec:upperbound}.

\begin{figure}[htbp]
\centerline{
(a) \includegraphics[height=1.5in]{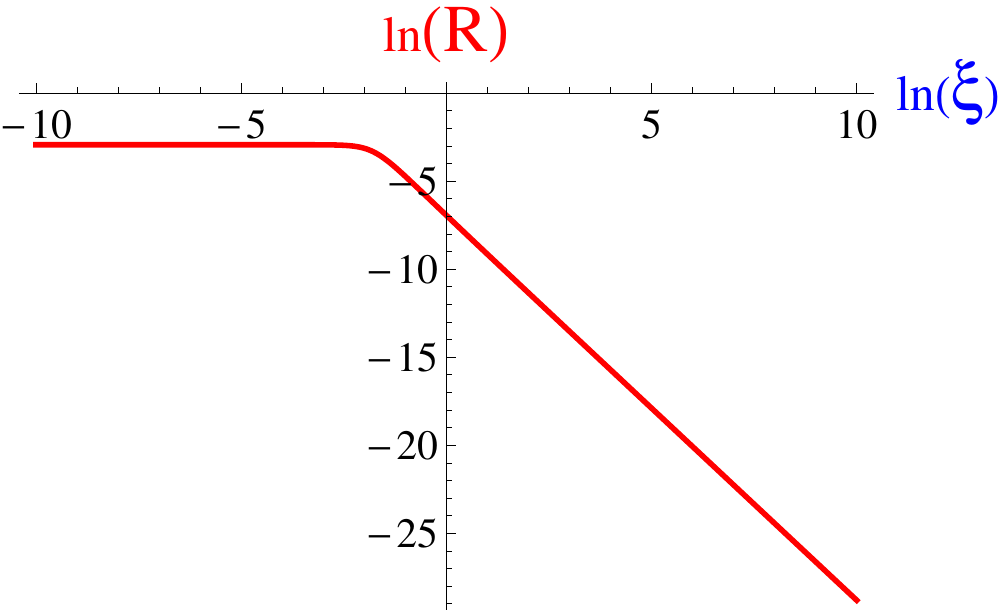}
}
\centerline{
(b) \includegraphics[height=1.5in]{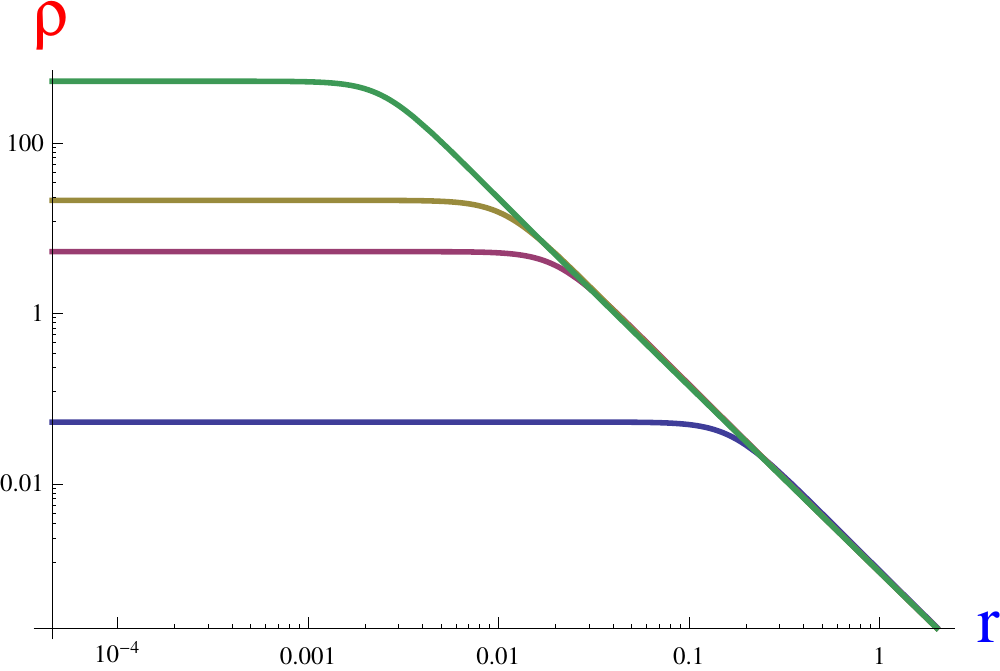}
}
\caption{Density of the self-similar problem obtained by solving
equations (\ref{eq:Euler.6})-(\ref{eq:Euler.7})  with
$\alpha=24/11$. (a) $R(\xi)$; (b) $\rho(r,t)$ versus $r$ at times $1,0.1,0.05,0.01,0.001$. The initial
conditions are $R(y_i)=R_0+R_4 \exp(4 y_i)$, $V(y_i)=U_1  +U_5 \exp(4 y_i)$,
$V_{,y}(y_i)=U_1+4 U_5 \exp(4 y_i)$ at $y_i=-10$, with $R_0=\frac{1}{6\pi}$,
$U_1=-\frac{2}{3}$ , $R_4 = -\frac{7(8 - 3\alpha)}{4\pi\alpha}U_5$ and
$U_5=10^2$.
}
\label{Fig:Rfig}
\end{figure}

In the case considered above, we note that the exponent $\alpha(4)=24/11$ is close
to $2$ so that it is not in contradiction with previous numerical simulations
analyzed in terms of the Penston-Larson solution (which has $\alpha=2$).
Moreover there is obviously a freedom in the solution  because, even with
$\alpha$ root of the secular equation, $R_4$ and $U_5$ are determined up to a
multiplicative constant. This is the consequence of a property of symmetry of
the equations  (\ref{eq:Euler.4}) and  (\ref{eq:Euler.5}): if $\left(R(\xi),
U(\xi)\right)$ is a solution, then $\left(R(\xi/\lambda), \lambda^{-1}
U(\xi/\lambda)\right)$ is also a solution with $\lambda$ an arbitrary positive
number. This freedom translates into the fact that $U_5$ and $R_4$ are defined
up to a multiplication by the same arbitrary (positive) constant. If $U_5$ and
$R_4$ are multiplied by $\lambda$, the next order coefficients of the Taylor
expansion, like $U_9$ and $R_8$ ($U_7$ and $R_6$  being set to zero) should be
multiplied by $\lambda^2$, and more generally the coefficients $U_{4n+1}$ and
$R_{4n}$, $n$ integer, by $\lambda^{2n}$, the coefficients $U_{2n}$ and
$R_{2n+1}$ being all zero.

 \begin{figure}[htbp]
\centerline{
 (a)\includegraphics[height=1.5in]{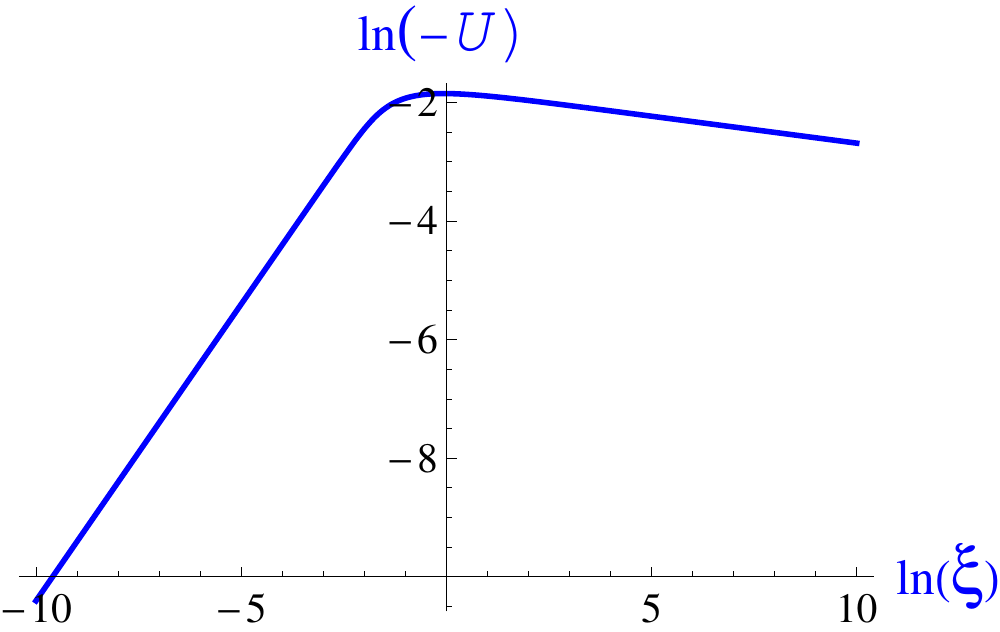}
 }
 \centerline{
 (b)\includegraphics[height=1.5in]{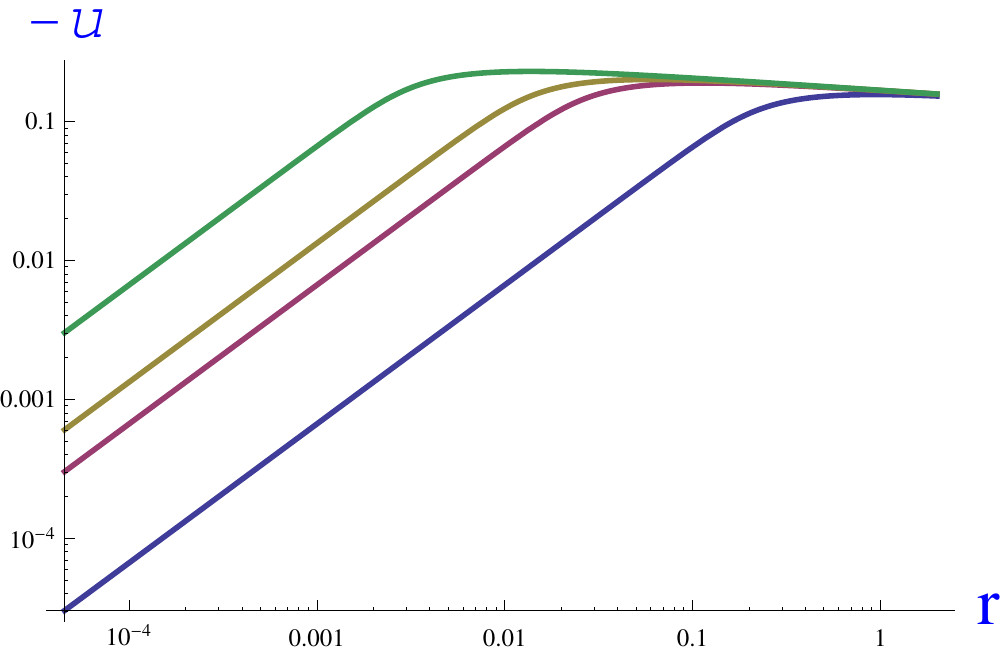}
}
\caption{Velocity of the self-similar problem, obtained by solving
equations (\ref{eq:Euler.6})-(\ref{eq:Euler.7})  with
$\alpha=24/11$.  (a) $-U(\xi)$, (b) $-u(r,t)$ versus $r$ at same times  and with same initial conditions as in Fig. \ref{Fig:Rfig}.
}
\label{Fig:Ufig}
\end{figure}

The behavior of $U(\xi)$ and $R(\xi)$ at $\xi \to \infty$ was derived in equation (\ref{sunm}). As one can see, the power law behavior for $R$ at $\xi$ infinity follows from the assumption that terms linear with respect to $R$ in equation (\ref{eq:Euler.4}) become dominant at large $\xi$. Keeping the terms linear with respect to $U$ in equation (\ref{eq:Euler.5}) and canceling them yields $U(\xi) \sim \xi^{1 - \alpha/2}$.
This shows that both the perturbation to $u$ and $\rho$ described by the self-similar solution have first a constant amplitude far from the core (defined as the range of radiuses $r \sim (-t)^{2/\alpha}$) and then an amplitude tending to zero as the distance to the core increases, which justifies that the linear part of the original equation has been kept to derive this asymptotic behavior of the similarity solution. As already said, this large distance behavior of the self-similar solution makes possible the matching of this collapsing solution with an outer solution behaving smoothly with respect to time.

The numerical solution of equations (\ref{eq:Euler.4})-(\ref{eq:Euler.5}) was actually obtained by using the system (\ref{eq:Euler.6})-(\ref{eq:Euler.7}) for the coupled variables  $R, V={U}/{\xi}$, then changing the variable $\xi$ into   $y=\ln(\xi)$. It writes
 \begin{equation}
  2 R + \frac{2}{\alpha}R_{,y} + 3 R V + (R V)_{,y} = 0
\mathrm{,}
\label{eq:Euler.6}
\end{equation}
and
\begin{equation}
\mathcal{A}_{,y}(V)+ 3  \mathcal{A}(V) + 4 \pi R(y) = 0
\mathrm{,}
\label{eq:Euler.7}
\end{equation}
where $\mathcal{A}(V)= V + \frac{2}{\alpha} V_{,y} + V^2 + V V_{,y}$. The self-similar solutions $R(\xi)$ and $-U(\xi)$  are drawn in log scale in Figs.  \ref{Fig:Rfig} and  \ref{Fig:Ufig} respectively together with the corresponding time dependent density and velocity $\rho(r,t)$ and $-u(r,t)$. In
Appendix \ref{sec:freefall}, by proceeding differently, we obtain the
self-similar solution of the free-fall analytically, in parametric form. As shown later, the analytical solution is  equivalent to the numerical solution of equations
(\ref{eq:Euler.6})-(\ref{eq:Euler.7}), see Fig. \ref{Fig:compar}.

\subsubsection{An upper bound for $\alpha$}
\label{subsec:upperbound}

We have seen that $\alpha$ must be larger than $2$. It is interesting to look at a possible upper bound. Such a bound can be derived as follows. At the end of the collapse, the density and radial velocity follow simple power laws near $r =0$, derived from the asymptotics of the self-similar solution. As said below, at the end of the collapse one has precisely $\rho(r) \sim r^{-\alpha}$. Therefore, from elementary estimates, the total mass converges if $\alpha$ is less than 3, which gives an upper bound for $\alpha$. In summary, the exponent $\alpha$ has to be in the range
\begin{equation}
2 < \alpha < 3
\label{rangealpha}
\end{equation}
in order for a physically self-similar solution to fulfill the condition that gravity is dominant over pressure.

\subsubsection{ Homologous solution for general polytropic equations of state}
\label{Gammaqcq}

The self-similar solution that we have found is independent on the pressure term
in the original equation for momentum. Therefore, it is natural to ask the
question of its dependence on the equation of state (namely the
pressure-density relation). Because the density diverges at $r = 0$ in the
similarity solution, it is reasonable to expect that, if the pressure grows too
much at large densities, it will become impossible to neglect the pressure term
compared to gravity. Let us consider a pressure depending on $\rho$ with a power
law of the form $p = K \rho^{\Gamma}$ with $\Gamma\equiv 1+1/n$ a real exponent
and $K$ a positive constant.
 We know already that, if $\Gamma = 1$, the pressure
term can be neglected in the collapsing core, and
the collapsing
solution is characterized by the exponent
$\alpha=24/11$.
The same system of
equations (\ref{eq:Euler.4})-(\ref{eq:Euler.5}) for the self-similar
solution will be found whenever the pressure can be neglected. Therefore we
expect that the above solution is valid, with the same $\alpha$, as long
as the power $\Gamma$ in the pressure-density relation leads to negligible
pressure effects in the collapsing region.
Putting the power law estimate derived from the similarity solution without pressure, one finds that the
marginal exponent $\Gamma$ is $\Gamma_c = 2  - {2}/{\alpha}$ which for
$\alpha=24/11$ is equal to
\begin{equation}
\Gamma_c = \frac{13}{12}, \qquad (n_c=12) \mathrm{.}  \label{Gammac}
\end{equation}
For $\Gamma > \Gamma_c$, the pressure becomes formally dominant
compared to gravity in the collapse domain (still assuming
$\alpha=24/11$), although if $\Gamma$ is less than $ \Gamma_c$ the
pressure is negligible compared to gravity in the same collapse
domain. When the pressure is dominant, either there is no collapse
because the outward force it generates cannot physically produce an
inward collapse, or other scaling laws with a different $\alpha$ yield
a collapsing solution different from the one that we have derived (see
below). If $ \Gamma$ is less than $\Gamma_c = {13}/{12}$ the collapse
is driven by dominant gravity forces and the scaling laws derived
above apply and are independent on the value of $\Gamma$. This occurs
because the values of the exponents $\alpha={24}/{11}$, $\beta=-2$, and
$\gamma= -{1}/{12}$ were deduced from the Euler-Poisson equations after
canceling the pressure term in the right-hand side of equation
(\ref{eq:Euler.2}).

Let us be more general and consider other possible values of $\alpha$.

If we assume that pressure and gravity forces are of the same order, the exponents are
\begin{equation}
\alpha=\frac{2}{2-\Gamma}, \qquad \beta=-2, \qquad  \gamma=1-\Gamma.
\label{press-et-grav}
\end{equation}
The condition $\alpha<3$ (see Section \ref{subsec:upperbound}) implies that $\Gamma<4/3$.  It is well-known that a polytropic star with index  $\Gamma>4/3$ is dynamically stable, so there is no collapse.  The critical index $\Gamma=4/3$ corresponds to ultra-relativistic fermion stars such as white dwarfs and neutron stars. In that case, the system collapses and forms a core of mass of the order of the Chandrasekhar mass as studied by Goldreich and Weber \cite{gw}. The collapse of polytropic spheres with ${6}/{5} \le  \Gamma \le {4}/{3}$ described by Euler-Poisson equations has been studied by Yahil \cite{Yahil}. For $\Gamma<4/3$, the star collapses in a finite time but since $\alpha<3$ the mass at $r=0$ at the collapse time $t=0$ is zero (in other words, the density profile is integrable at $r=0$ and there is no Dirac peak).

We can also consider the case where gravity forces overcome pressure forces so that the system experiences a free fall. If we compare the magnitude of the pressure and gravity terms in the Euler-Poisson system when the homologous solutions (\ref{selfrho})-(\ref{selfu}) are introduced, we find that the pressure is negligible if $\alpha>2/(2-\Gamma)$. Therefore, for a given polytropic index $\Gamma$, the pressureless homologous solutions are characterized by the exponents
\begin{equation}
\frac{2}{2-\Gamma} < \alpha \le  3,
\label{rangealphaGamma}
\end{equation}
and
\begin{equation}
\beta=-2, \quad \gamma=2/\alpha-1.
\label{BetaGamma}
\end{equation}
The collapse exponent $\alpha$ is selected by considering the behavior
of the solution close to the center. Setting $R(\xi)=R_0 + R_k \xi^k$
and $U(\xi)=U_1 + U_{k+1} \xi^{k+1}$, the relation (\ref{eq:alphak})
between $\alpha$ and $k$ leads to the following choice: $\alpha$ will
be the smallest value of $\alpha(k)$ satisfying both relations
(\ref{rangealphaGamma}) and (\ref{eq:alphak}) for $k$ even. If follows
that
\begin{equation}
\alpha=\frac{12}{7} \quad {\rm for}\quad  \Gamma \le  \frac{5}{6},
\label{alphaPenston}
\end{equation}
which is the exponent derived by Penston \cite{Penston} for zero pressure or $T=0$ assuming  $k=2$. Next, we find
\begin{equation}
\alpha=\frac{24}{11}, \quad {\rm for}\quad  \frac{5}{6} <\Gamma \le  \frac{13}{12},
\label{alphanous}
\end{equation}
as obtained above assuming $k=4$. Finally, we find that
\begin{equation}
\alpha=\frac{6k}{2k+3}, \quad {\rm for} \quad  \frac{4k-3}{3k} <\Gamma \le \frac{4k+5}{3k+6},
\end{equation}
for any $k\ge 4$ even. We note that there is no solution for $\Gamma\ge 4/3$ since the polytropic stars with such indices are stable as recalled above.

Finally, when pressure forces dominate gravity forces, the scaling exponents are obtained by introducing the self-similar form (\ref{selfrho})-(\ref{selfu}) into the Euler-Poisson system without gravity forces, yielding
 \begin{equation}
\beta= -\frac{2}{{2}/{\alpha}+\Gamma-1}, \qquad  \gamma=-\frac{\Gamma-1}{{2}/{\alpha}+\Gamma-1}.
\label{pressuredom}
\end{equation}
However, this situation is not of physical relevance to our problem since it describes a slow ``evaporation'' of the system instead of a collapse.

\subsection{Comparison of the self-similar solution with the numerical results}

\subsubsection{Invariant profiles and scaling laws}

\begin{figure}[htbp]
\centerline{
\includegraphics[height=2.0in]{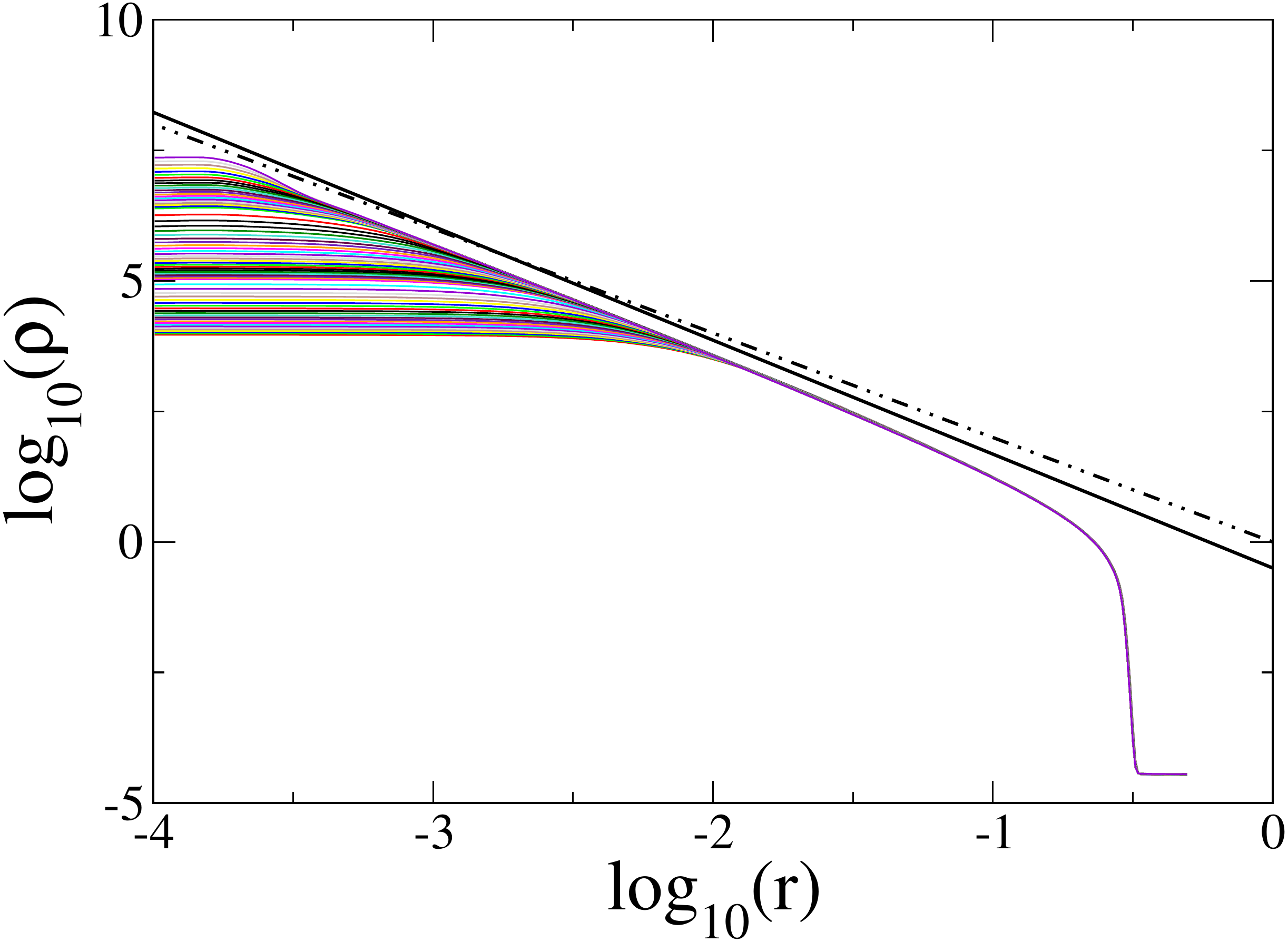}
}
\caption{Density $\rho(r,t)$ versus the radial variable $r$ in ${\rm log}_{10}$ scale: numerical solutions of the full Euler-Poisson system, equations (\ref{eq:Euler.1})-(\ref{eq:Euler.2}) at different times before the collapse. The solid line with slope $-24/11$ fits better the asymptotic behavior (large $r$) of the curves than the dotted-dashed line with slope $-2$.
}
\label{Fig:RBruno1}
\end{figure}

\begin{figure}[htbp]
\centerline{
\includegraphics[height=2.0in]{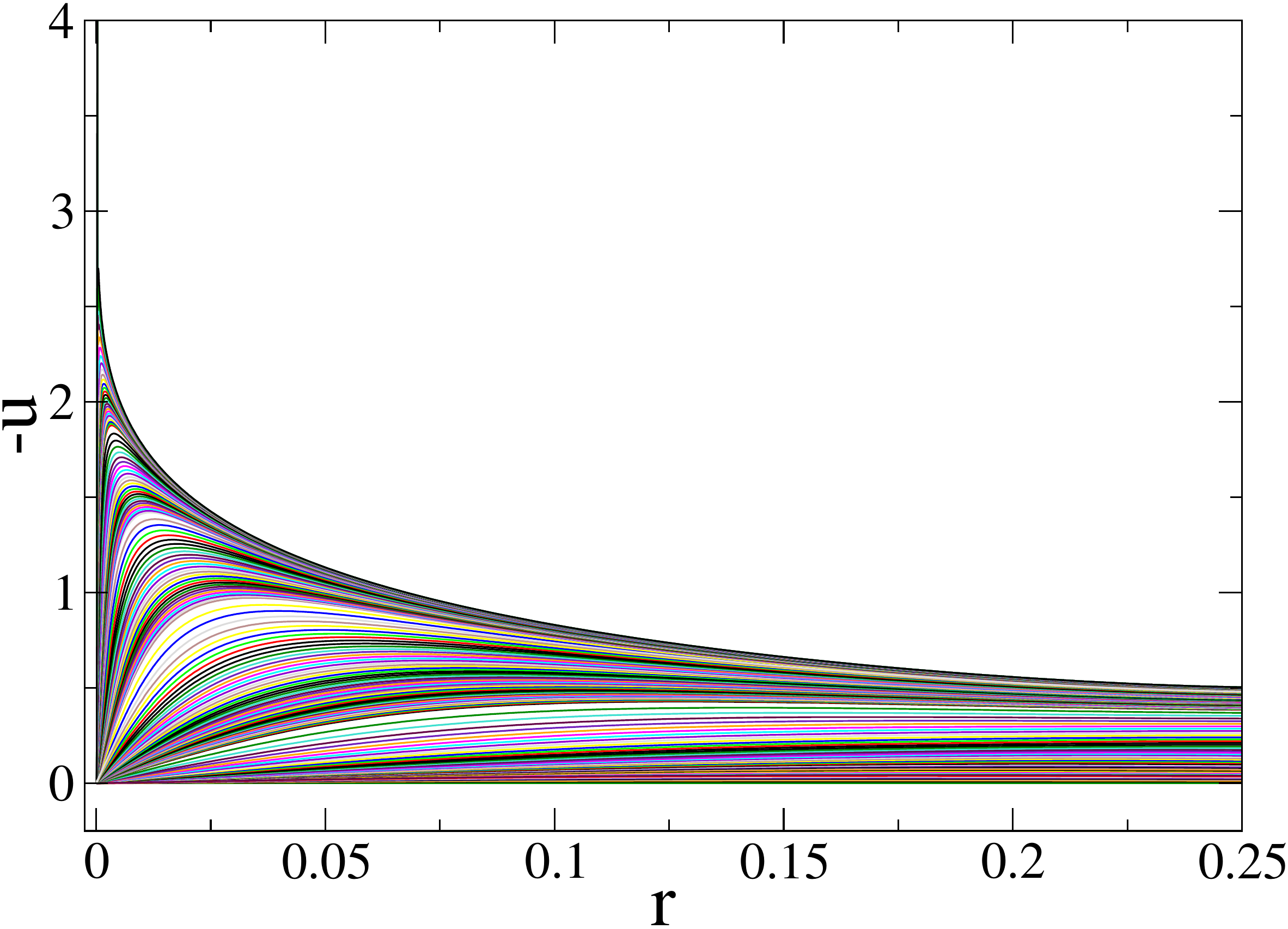}
}
\caption{Velocity $-u(r,t)$ versus the radial variable $r$: numerical solutions of the full Euler-Poisson system, equations (\ref{eq:Euler.1})-(\ref{eq:Euler.2}) at different times before the collapse.
}
\label{Fig:UBruno1}
\end{figure}

The numerical solutions of the full Euler-Poisson system were obtained using a variant of the centpack program \cite{progbalbas} by Balbas and Tadmor.
Comparing our theoretical predictions of the self-similar solution just before collapse with the numerical solution of the full Euler-Poisson system, we find that both lead to the same result, namely they give a value of the  exponent $\alpha$ slightly larger than two. The numerical solutions of $\rho(r,t)$ and $u(r,t)$ versus the radial variable $r$ at different times before the collapse are shown in Figs. \ref{Fig:RBruno1}  and \ref{Fig:UBruno1}  respectively.

\begin{figure}[htbp]
\centerline{
\includegraphics[height=2.0in]{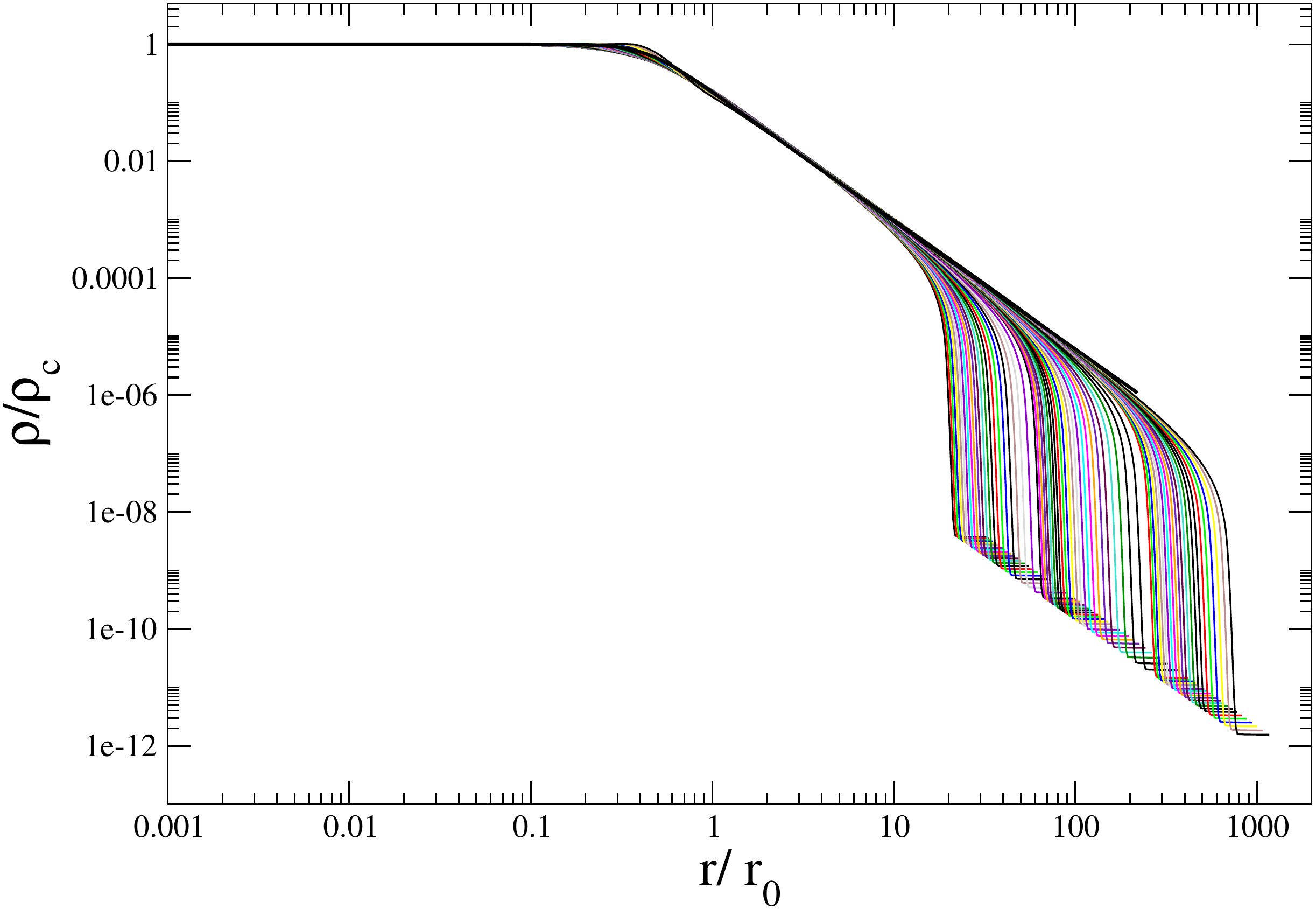}
}
\caption{ Self-similar density curves $\rho(r,t)/\rho(0,t)$  versus $r/r_0(t)$ in $\log$ scale with $r_0(t)$ defined in the text and $\alpha=24/11$.
}
\label{Fig:numrhoBruno}
\end{figure}

\begin{figure}[htbp]
\centerline{
\includegraphics[height=2.0in]{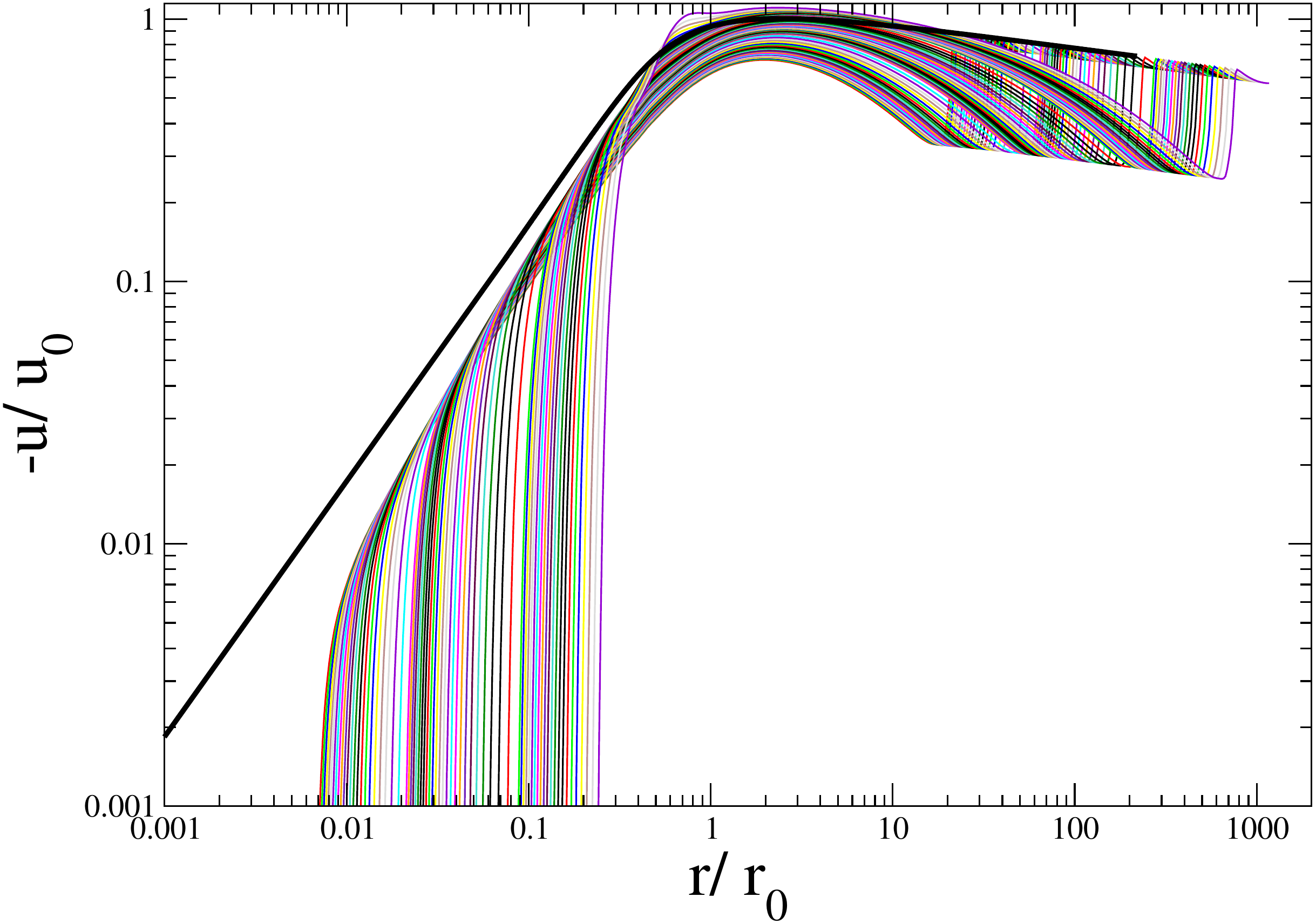}
}
\caption{ Velocity ratio $-u(r,t)/u_0(r,t)$ versus $r/r_0(t)$ in $\log$ scale deduced from the curves of Fig. \ref{Fig:UBruno1} with the definition of $r_0(t)$ given in the text and $\alpha=24/11$. A shock wave is visible at the edge of the star, see the oscillations of the velocity.
}
\label{Fig:UBruno}
\end{figure}

To draw the self-similar curves, we may get around the difficult task of the exact determination of the collapse time by  proceeding as follows. We define a core radius $r_0(t)$ such that $ \rho(0,t) r_0(t)^\alpha =1$ (or any constant value), then we draw $\rho(r,t)/\rho(0,t)$ and $u(r,t)/u(r_0,t)$ versus $r/r_0(t)$. The merging of the successive curves should be a signature of the self-similar behavior. The result is shown in Figs. \ref{Fig:numrhoBruno}  and \ref{Fig:UBruno} for the density and velocity respectively. The $\log$ scale of the density curve illustrates the expected asymptotic behavior (large $\xi$ values)  $R \sim \xi^{-\alpha}$ or $\rho(r,t)/\rho(0,t) \sim (r/r_0(t))^{-\alpha}$. The asymptotic behavior of the velocity, $U \sim \xi^{1-\alpha/2}$ is less clear on Fig. \ref{Fig:UBruno} where the curves display an oscillating behavior below the line with slope $1-\alpha/2$. We attribute the progressive decrease of the curves below the expected asymptote to the shock wave clearly visible in the outer part of the velocity curves (in addition, as discussed by Larson \cite{larson} p. 294, the velocity profile approaches the self-similar solution much slower than the density). In Figs. \ref{Fig:numrhoBruno} and \ref{Fig:UBruno} the black curves display the theoretical self-similar solution shown in Figs. \ref{Fig:Rfig}-(a) and \ref{Fig:Ufig}-(a), which has analytical parametric expression  given in Appendix \ref{App:precoll}.

\begin{figure}[htbp]
\centerline{
\includegraphics[height=2.0in]{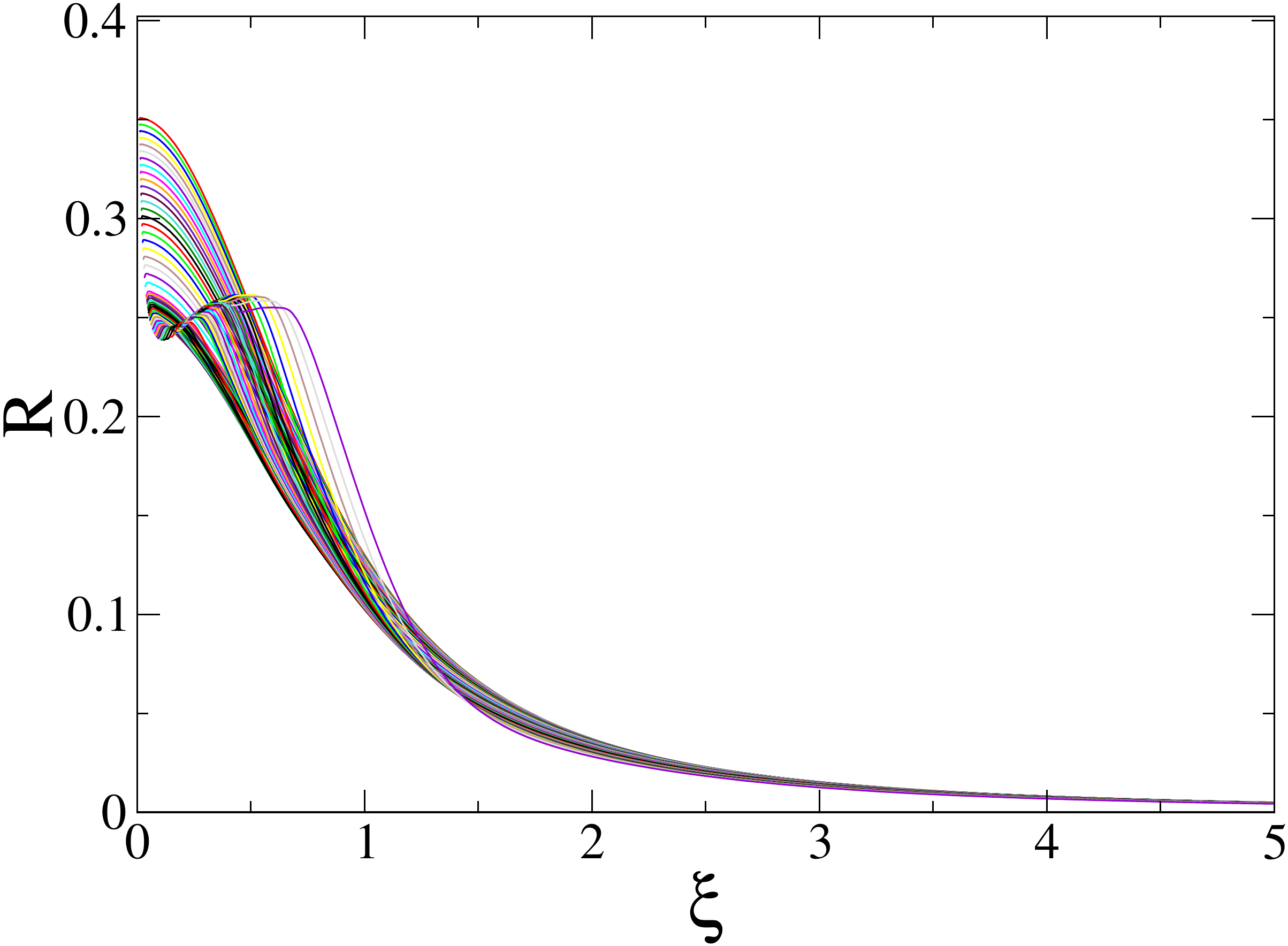}
}
\caption{
 Numerical self-similar density  curves $\rho(r,t)(-t)^2$ versus $\xi=r(-t)^{-{2}/{\alpha}}$ for $\alpha=24/11$.
}
\label{Fig:RBruno}
\end{figure}

In Fig. \ref{Fig:numrhoBruno} the merging density curves have all the
same ordinate at the origin, since we have plotted
$\rho(r,t)/\rho(0,t)$. To complete the comparison between the theory
and the simulation for the self-similar stage, we have also drawn the
series of self-similar density curves $R(\xi)$ in order to check
whether the central behavior of the numerical curves agrees with the
expected value $R(0)=1/(6\pi)$. To do this we have first to define the
collapse time as precisely as possible, then to plot the quantity $
(t_*-t)^2 \rho(r,t)$ versus $r/(t_*-t)^{{2}/{\alpha}}$. These curves are
shown in Fig. \ref{Fig:RBruno}. They clearly merge except in a close
domain around the center. We observe that the numerical value at
$\xi=0$ is noticeably larger than the expected value
$R(0)=1/(6\pi)\simeq 0.05$ (it is also substantially
larger than the value $0.133...$ corresponding to the Penston-Larson
solution). This shows that the system has not entered
yet deep into the self-similar regime. Therefore, our numerical
results should be considered with this limitation in mind.   However,
a precise study displays a clear decrease of the value of
$(t_*-t)^2\rho(0,t)$ during the approach to collapse, as illustrated
in Fig. \ref{Fig:rhoc}, which shows a good trend of the evolution (see below).

 \begin{figure}[htbp]
\centerline{
\includegraphics[height=2.0in]{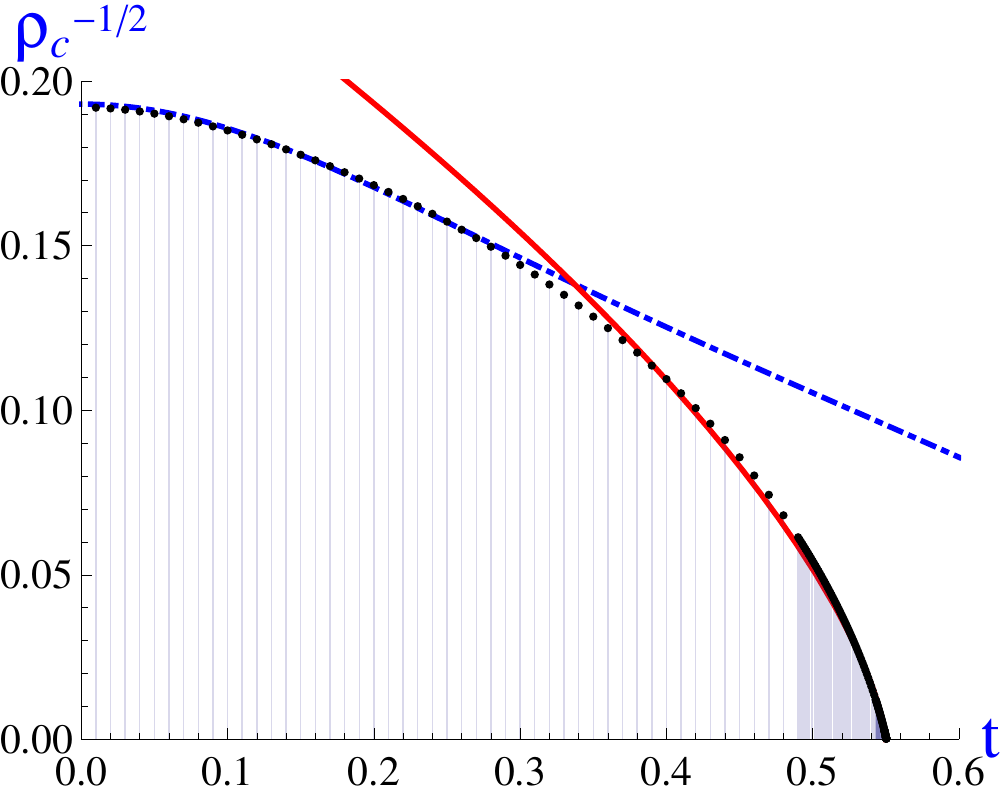}
}
\caption{ Behavior of the density at the center of the star. We plot  $\rho_c^{-1/2}$ versus $t$ to show a quasi-linear time dependence of the numerical solution in the Painlev\'{e} and in the pre-collapse regimes. The dots are the numerical results, the blue dotted-dashed curve is the Painlev\'{e} solution, the red curve the self-similar one which includes an additional second order term, see text.
}
\label{Fig:rhoc}
\end{figure}

In Fig. \ref{Fig:rhoc} we compare the numerics with the theory, both in the Painev\'{e} regime described in section \ref{sec_dynA} and in the self-similar regime described here.  In these two regimes, the central density is expected to behave as $(t_* - t)^{-2}$, see equations (\ref{eq:t-2})  and  (\ref{eq:rhoa}) for the Painlev\'{e} and the homologous regime respectively. Therefore we  draw $\rho(0,t)^{-1/2}$  which should decrease linearly with time (with different slopes).  The dots result from the numerical integration of the full Euler-Poisson equations at constant temperature (actually similar results are obtained with a temperature decreasing with time), with initial condition at temperature $T=0.9 T_c$ (out of equilibrium). At the beginning of the integration, in the Painlev\'{e} regime, the density $\rho_c(t)=\rho(0,t)$ is expected to evolves as $\rho_c(0) + \rho_1 A(t)$, where $A(t)$ is the solution of the modified version of equation (\ref{eq:Ceq}), valid for constant temperature, which writes
\begin{equation}
\ddot{A }=\left (1-\frac{T}{T_c}\right )\gamma + K A^2,
\end{equation}
with $\gamma=120.2$, and $K=12.32$, as in section \ref{sec_dynA}. The dotted-dashed blue line displays the function $\rho_p(t)^{-1/2}$ with  $\rho_p(t)=36 A(t)+26.85$, where the coefficients are fitted to the numerical Euler-Poisson solution, and the initial conditions for the Painlev\'{e} equation are $A(0)=\dot{A}(0)=0$.

Close to the collapse time ($t_*=0.55$ in the numerics), the numerical solution $\rho(0,t)$ is expected to behave as $\frac{1}{6 \pi}(t_*-t)^{-2}$, up to an additional second order term. A term of order $(t_*-t)^{-4/3}$ was chosen because it is the perturbation associated to the eigenvalue $\lambda=-2/3$ 
 of the linear analysis around the fixed point $\mathcal{C}=[\overline{R}_0=1/(6\pi); \overline{U}_1=-2/3]$\footnote{See the next subsection where the change of variable in equation (\ref{eq:appell}) is a trick converting a problem with algebraic decay into  exponential decay permitting spectral analysis.} and fits well the numerical results. The red curve displays  the function $ \rho_f(t)^{-1/2}$ with $\rho_f(t)=\frac{1}{6\pi} (t_*-t)^{-2} +6.5(t_*-t)^{-4/3}$, which agrees well with the numerical dots, indicating that the Euler-Poisson solution tends to converge towards the self-similar form close to the center, whereas with some delay.  In the following subsection we show that  the fixed point $\mathcal{C}$ is a saddle point, with one stable direction but another unstable. It follows that the numerical solution has a priori no reason to reach $\mathcal{C}$. However we observe that it clearly tends towards this fixed point as the collapse is approached.

\subsubsection{Dynamical behavior close to the center}
\label{sec:fixcenter}

Recall that we have derived the theoretical value of the exponent $\alpha =24/11$ by expanding the density as $R(\xi)=R_0 + R_4 \xi^4 +...$ close to $\xi =0$, with $R_0=1/(6\pi)$ and $U=U_1\xi + U_5 \xi^5+...$ with $U_1=-2/3$ ($R_4 $ and $U_5$ being defined up to a multiplicative coefficient).
 In order to explain the discrepancy between the numerics and the theoretical value $R(0)=1/(6\pi)$,  we  look at the stability of the self-similar solution close to $\xi=0$. Let us assume here that $R$ and $U$ are functions of $\xi$ and time, with $\xi= r(-t)^{-2/\alpha}$ and define the time dependent variable  \cite{Apple}:
\begin{equation}
s=-\ln{(-t)}.
\label{eq:appell}
\end{equation}
We set
 \begin{equation}
\rho(r,t) = (-t)^{-2} R(\xi,s)
  \mathrm{,}
\label{eq:rhoXs}
\end{equation}
and
  \begin{equation}
  u(r,t) = (-t)^{-1+\frac{2}{\alpha}} U(\xi,s)
   \mathrm{.}
\label{eq:uXs}
\end{equation}
where  the variable $s$ is positive for small $t$, increasing up to infinity as collapse is  approached. Substituting  this ansatz in equations
(\ref{eq:Euler.1})-(\ref{eq:Euler.2}) which include the terms due to pressure and gravity, yields the dynamical equations for $R$ and $U$:
\begin{equation}
 R_{,s} + R_{,\xi}\left (U+\frac{2}{\alpha}\xi\right )+RU_{,\xi}+  \frac{2}{\xi} R (U +\xi) = 0
\mathrm{,}
\label{eq:Euler.4s}
\end{equation}
and
 \begin{eqnarray}
  U_{,s}+ U_{,\xi}\left (U+\frac{2}{\alpha}\xi\right )- \gamma U + T\frac{R_{,\xi}}{R}e^{2\gamma s} \nonumber\\
  +  \frac{4 \pi }{\xi^2} \int_0^{\xi} {\mathrm{d}}\xi' {\xi'}^2 R(\xi',s)=0
\mathrm{,}
\label{eq:Euler.5s}
\end{eqnarray}
where $\gamma$ is negative, see equation (\ref{gamma}).

These two coupled equations generalize the self-similar study of Larson \cite{larson}, Penston \cite{Penston}  and Brenner-Witelski \cite{Brenner} to the case of an exponent $\alpha$ different from $2$. Besides the fact that in equations (\ref{eq:Euler.4s})-(\ref{eq:Euler.5s}) the $\alpha$-dependent coefficients are slightly different from theirs, the main difference  with  previous works is that here the prefactor $e^{2\gamma s}$ of the pressure term decreases as $s$ increases (as the collapse is approached), while this factor was unity in their case.

The self-similar functions $U$ and $R$  can be expanded as $R(\xi,s)=R_0(s) + R_2(s) \xi^2+ R_4(s) \xi^4 +...$  and $U=U_1(s)\xi + U_3(s)\xi^3+...$ close to $\xi =0$.  Writing $R_i(s)=\overline{R}_i +r_i(s)$ and $U_i(s)=\overline{U}_i +u_i(s)$, for $i=0,1,2...$ one gets the asymptotic relations $ \overline{R}_0=1/(6\pi)$ and $\overline{U}_1=-2/3$ at lowest order, which is strictly the steady-state values found above in the equations without pressure, because asymptotically the pressure term vanishes. However these asymptotic values are not stable, as we shall prove now.

Because we are interested in what happens just before the collapse time,
we can neglect the pressure term in equation (\ref{eq:Euler.5s}). It
becomes
 \begin{equation}
  U_{,s}+ U_{,\xi}\left (U+\frac{2}{\alpha}\xi\right )- \gamma U +  \frac{4 \pi
}{\xi^2} \int_0^{\xi} {\mathrm{d}}\xi' {\xi'}^2 R(\xi',s)=0
\mathrm{.}
\label{eq:Euler.5s.wp}
\end{equation}
The autonomous system (\ref{eq:Euler.4s}) and (\ref{eq:Euler.5s.wp}) has the
useful property to reduce itself to a closed set of ODE's for $R_0(s)$ and
$U_1(s)$. This set reads
 \begin{equation}
  U_{1,s}+ U_{1} \left (U_1 +\frac{2}{\alpha}\right ) +\left (1 - \frac{2}{\alpha}\right ) U_{1} +
\frac{4 \pi }{3} R_0=0
\mathrm{,}
\label{eq:Euler.5s.wp.or}
\end{equation}
and
 \begin{equation}
 R_{0,s} + 3 R_{0} U_1 + 2 R_0 = 0
\mathrm{.}
\label{eq:Euler.Ros}
\end{equation}
This system has three fixed points (namely solutions independent on $s$): (i)
the point $\mathcal{C}$=$[\overline{ R}_0 = {1}/{(6\pi)}; \overline{U}_1 = -{2}/{3}]$ defined in the previous subsection (the values at $\xi =0$
of $R$ and $U$, solution of the similarity equations already derived); (ii)
also $[\overline{R}_0 =\overline{ U}_1 = 0]$; (iii) and finally $[\overline{R}_0 = 0;\overline{ U}_1 = -1]$.

Writing $R=\overline{R}_0+\delta r e^{\lambda s}$, and $U=\xi(\overline{U}_1+\delta u e^{\lambda s})$, the linear stability analysis of equations (\ref{eq:Euler.5s.wp.or})-(\ref{eq:Euler.Ros}) in the vicinity of the fixed point $[\overline{R}_0, \overline{U}_1]$ gives the eigenvalues equation
 \begin{equation}
 \lambda^2+ (5 \overline{U}_1 + 3)\lambda + (2\overline{U}_1+1)(3\overline{U}_1+2)-4 \pi \overline{R}_0=0
\mathrm{.}
\label{eq:eigenv}
\end{equation}
It follows that the fixed point $\mathcal{C}$ has one unstable and one stable direction in the phase
plane, with eigenvalues $+1$ and $-2/3$, independently of the $\alpha$ value.

The fixed point $\overline{R}_0 = 0$ and $\overline{U}_1 = -1$ has two unstable
directions with a degenerate eigenvalue $+1$, although $\overline{R}_0
= \overline{U}_1 = 0$ is stable in all directions, with eigenvalues
$-1$ and $-2$.  The consequences for the whole solution are not
completely clear. This could explain why in the numerical work it
seems so hard to get the right value of $\overline{R}_0$.  This could
be either because the initial condition for this set of ODE's does not
allow to reach the fixed point $\overline{ U}_1 = -{2}/{3}$ and
$\overline{R}_0= {1}/({6\pi})$ or because the numerics does not have
the accuracy necessary to reach in logarithmic times the fixed
point. Moreover, this fixed point, because it is stable in only one
direction and unstable in the other, is reached from special initial
conditions, on its stable manifold. Otherwise the solution are
attracted either to infinity or to $\overline{R}_0 = \overline{ U}_1 =
0$, depending on the initial condition.

\subsubsection{Near  the stable fixed point}

Assuming that the solution approaches  the stable fixed point $\overline{R}_0 = \overline{ U}_1 = 0$, one may write $R(s,\xi)=\delta r(s,\xi)$  and $ U(s,\xi)=\xi \delta u(s,\xi)$, where $\delta r$ and $\delta u$ are smaller than unity. Setting $x=-\ln(\xi)$, the functions  $\delta r(s,x)$ and $\delta u(s,x)$ are solutions of a linear autonomous system derived from equations (\ref{eq:Euler.4s}) and (\ref{eq:Euler.5s.wp}). We obtain
\begin{equation}
\delta r_{,s}(s,x) -\frac{2}{\alpha}\delta r_{,x}(s,x)+ 2 \delta r(s,x)=0
  \mathrm{,}
\label{eq:deltar}
\end{equation}
and
  \begin{equation}
 \delta  u_{,s}(s,x) +\delta u(s,x)+ \frac{4\pi}{3} \delta r(s,x)=0
   \mathrm{,}
\label{eq:deltau}
\end{equation}
where both variables $s$ and $x$ are positive and go to infinity as the collapse is approached.

The solution of the linear homogeneous equation (\ref{eq:deltar}) is
\begin{equation}
\delta r(s,x)=e^{-2s}  \tilde{r}\left (\frac{2}{\alpha}s+x\right )
  \mathrm{,}
\label{eq:soldeltar}
\end{equation}
where $ \tilde{r}=\delta r(s,0)$ is the profile of the density at the initial time $t_0$  of the collapse regime, with  $s=-\ln(t_0-t_*) $ by definition. It follows that the solution of the linear equation (\ref{eq:deltar}) decreases exponentially to zero as the collapse is approached.

\section{Beyond the singularity: post-collapse}
\label{sec:beyond}

The question of the post-collapse was considered by Yahil \cite{Yahil} in his study of Euler-Poisson equations with a polytropic equation of state $p=K\rho^\Gamma$ with ${6}/{5} \le \Gamma \le {4}/{3}$. For the critical index $\Gamma=4/3$, corresponding to ultra-relativistic neutron stars, during the homologous collapse all the mass in the core contracts towards the
center, such that at the singularity time there is a non-zero mass, of the order of the Chandrasekhar mass, at $r=0$ \cite{gw}. In that case, the post-collapse regime begins with a non-zero mass at $r= 0$,
represented in the equations by a Dirac peak at $r=0$.  This is not
what happens for polytropic equations of state with $\Gamma<4/3$ when pressure and gravity are of the same order \cite{Penston,larson,Yahil}, or in our description of the self-similar collapse where gravity overcomes pressure forces (free fall), because, at the singularity time $t=0$, as we have seen, the density does not
write as a Dirac distribution but as a power law $\rho(r,0) \propto
r^{-\alpha}$ which yields for $\alpha< 3$ a mass converging at $r = 0$ (the large distance behavior
is to be matched with an outer solution to make the total mass
finite).  Because we do not expect a Dirac peak of finite mass at $r = 0$ at
the time of the singularity, our post-collapse situation looks
(mathematically) like the one of the dynamics of the Bose-Einstein
condensation where the mass of the condensate begins to grow from zero
{\it{after}} the time of the singularity
\cite{BoseE,bosesopik}\footnote{Some analogies between the
post-collapse dynamics  of self-gravitating Brownian particles \cite{post} and the
Bose-Einstein condensation have been discussed in \cite{bosesopik}.}.

Let us derive the equations for the self-similar dynamics after the collapse. As
in the case of the post-collapse dynamics of self-gravitating
Brownian particles \cite{post} and of the  Bose-Einstein condensation
\cite{BoseE,bosesopik}, we have to add to the equations of density and momentum
conservation an equation for the mass at the center. Let $M_c(t)$ be this mass.
It is such that $M_c(0) = 0$. We need an equation for its growth. The mass flux
across a sphere of radius $r$ is $ J = 4 \pi r^2 \rho(r) u(r)$. Therefore the
equation for $M_c(t)$ is
\begin{equation}
{M}_{c,t} = \left[-4 \pi r^2 \rho(r) u(r)\right]_{r \to 0}
\mathrm{.}
\label{eq:Mc}
\end{equation}
To have a non zero limit of $ \left[-4 \pi r^2 \rho(r)
u(r)\right]$ as $r$ tends to zero constrains the behavior of $u(r)$ and
$\rho(r)$ near $r =0$. The velocity near $r = 0$ is a free-fall velocity. At $r$
very small, it is completely dominated by the attraction of the mass at $r = 0$.
Therefore it can be estimated by taking the relation of energy conservation in
free-fall, with a zero total energy, because at such short distances the initial
velocity is negligible compared to the velocity of free-fall. This yields $ u
\approx - \left({2 M_c}/{r}\right)^{1/2}$, which shall define the limit behavior
of  $u(r, t)$ near $r = 0$. Because $r^2 \rho(r) u(r)$ must tend to a finite
value at $r = 0$, one must have $\rho(r) \sim r^{-3/2}$. Note that this gives an infinite density at $r=0$ for $t>0$ while  $\rho(0)$ was finite before the collapse time; but close to $r=0$ the density $\rho(r)$ decreases (versus $r$) less rapidly for positive $t$ than it did
 for negative $t$.

The equations one has to solve now are the same as before plus the attraction by
the
mass $M_c(t)$ at $r = 0$ included (the pressure being again considered as
negligible, which is to be checked at the end),
 \begin{equation}
\rho_{,t} + \frac{1}{r^2} \left( r^2 \rho u \right)_{,r} = 0
\mathrm{,}
\label{eq:Euler.1+}
\end{equation}
 \begin{equation}
u_{,t} + u u_{,r} = - \frac{G M(r,t)}{r^2}
\mathrm{,}
\label{eq:Euler.2+}
\end{equation}
and
 \begin{equation}
M(r,t)  = 4 \pi \int_0^r {\mathrm{d}}r' r'^2 \rho(r',t) + M_c(t)
\mathrm{.}
\label{eq:Euler.2.1+}
\end{equation}
The equation (\ref{eq:Mc}) for $M_c(t)$ with the initial condition $M_c (0) = 0$ has to be added to the set of equations of motion.  The scaling laws of this system are derived as was done for the self-similar dynamics before the singularity. Because the equations after singularity include the whole set of equations leading to the singularity, the scaling laws are the same as before, with a free exponent like the one denoted as $\alpha$ (this assuming, as we shall check it, that the scaling laws have as much freedom as they had before collapse, which is not necessarily true because one has another equation (\ref{eq:Mc}) for another unknown function, $M_c(t)$). But the free exponent has to be the same as before collapse, because the asymptotic behavior of the solution remains the same before and after collapse: at very short times after collapse only the solution very close to $r = 0$ is changed by the occurrence of a finite mass at $r = 0$, a mass which is very small at short positive time. Therefore we look for a self-similar solution of the equations above with the same scaling laws as before collapse for $\rho(r, t)$ and $u(r,t)$ plus another scaling for $M_c(t)$:
\begin{equation}
\rho(r,t) = t ^{-2} R_+(r t^{-2/\alpha}) \mathrm{,}
\end{equation}
\begin{equation}
u(r,t) =
t^{-1+\frac{2}{\alpha}} U_+(r t^{-2/\alpha}) \mathrm{,}
\end{equation}
and
\begin{equation}
M_c(t) = K_M t^b
\mathrm{,}
\end{equation}
where $\alpha=24/11$ and $b$ is a positive
exponent to be found.

Moreover there has been a change of sign
from $(-t)$ to $t$ in the scaled functions, which is obviously due to the fact
that we are looking for positive times after the singularity, this one taking
place at $t =0$.  To have the two terms on the right-hand side of equation
(\ref{eq:Euler.2.1+}) of the same order of magnitude with respect to $t$ imposes
\begin{equation}
b = \frac{6}{\alpha} - 2,
\end{equation}
a positive exponent as it should be (recall the condition
that $\alpha$ is less than 3). For $\alpha=24/11$, we get $b=3/4$. This yields the following set of definitions of
the self similar unknowns after collapse,
 \begin{equation}
\rho (r, t) = t^{ - 2} R_+ (\xi_+)
\mathrm{,}
\label{eq:Euler+rho}
\end{equation}
 \begin{equation}
u (r, t) = t^{2/\alpha - 1} U_{+}(\xi_+)
\mathrm{,}
\label{eq:Euler+u}
\end{equation}
and
 \begin{equation}
M_c (t) = K_{M}t^{6/\alpha - 2}
\mathrm{.}
\label{eq:Euler+M}
\end{equation}
The stretched radius is $\xi_+ = r t^{-2/\alpha}$.
The equations to be satisfied by the scaled functions are
 \begin{equation}
- 2 R_+ - \frac{2 \xi_+}{\alpha} R_{+,\xi_+} + \frac{2}{\xi_+} R_+ U_+ + (R_+ U_+)_{,\xi_+} = 0
\mathrm{,}
\label{eq:Euler.4+}
\end{equation}
and
 \begin{eqnarray}
\left (1 - \frac{2}{\alpha}\right ) U_+ + \frac{2}{\alpha} \xi_+ U_{+,\xi_+}   - U_+ U_{+,\xi_+} \nonumber\\
=\frac{G}{\xi_+^2} \left(4 \pi \int_0^{\xi_+} {\mathrm{d}}\xi'_+ {\xi'}_+^2 R_+(\xi'_+) + K_M \right)
\mathrm{.}
\label{eq:Euler.5+}
\end{eqnarray}
The coefficient $K_M$ in equation (\ref{eq:Euler.5+}) is related to the limit values of $R_+$ and $U_+$ near $\xi_+ = 0$. The solution of the two equations near $\xi_+ = 0$ are
\begin{equation}
R_+ \approx K_R \xi_+^{-3/2} \mathrm{,}
\end{equation}
and
\begin{equation}
U_+ \approx K_U \xi_+^{-1/2} \mathrm{.}
\end{equation}
Equation (\ref{eq:Euler.4+}) does not constrain the coefficients $K$'s. By setting to zero the coefficient of the leading order term, of order $\xi_+^{-5/2}$ near $\xi_+ = 0$,  in equation (\ref{eq:Euler.5+}) yields a relationship between the $K$'s,
\begin{equation}
K_U = -(2 G K_M )^{1/2} \mathrm{.}
\end{equation}
Another relation comes from  equation (\ref{eq:Mc}).
It yields
\begin{equation}
K _M = - \frac{2\pi}{3/\alpha - 1} K_U K_R \mathrm{.}
\end{equation}
Therefore there is only one free parameter among the three coefficients $K$'s.
This free parameter is fixed by the matching with the large distance behavior of
$R_+$ and $U_+$, which is defined itself by the matching with the outside of the
collapse domain.

  \begin{figure}[htbp]
\centerline{
 \includegraphics[height=1.6in]{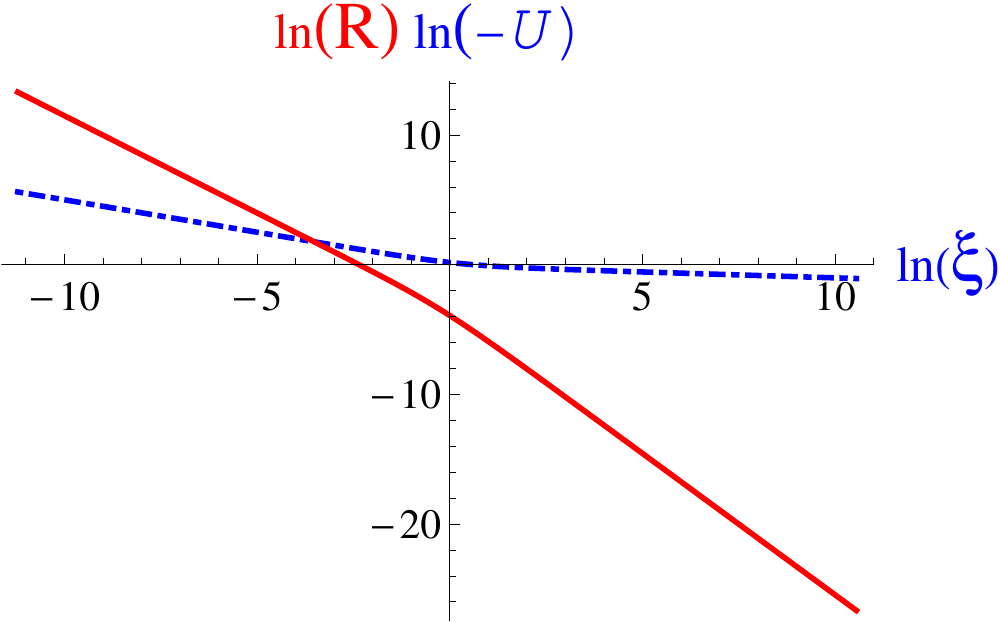}
}
\caption{ Self-similar density $R_{+}(\xi_+)$ and velocity $U_{+}(\xi_+)$ in the post-collapse regime, solution of  equations (\ref{eq:numRV}), in $\ln$ scale, to be compared with solutions in the pre-collapse regime drawn in Figs. \ref{Fig:Rfig}-\ref{Fig:Ufig} also in $\ln$ scale.
}
\label{Fig:comparpost}
\end{figure}

The system (\ref{eq:Euler.4+})-(\ref{eq:Euler.5+}) was solved numerically by using the coupled variables $R,V={U}/{\xi}$ and $y=\ln(\xi)$ (dropping the $+$ indices) as in subsection \ref{subsec5B}, that gives the coupled equations analogous to equations (\ref{eq:Euler.6})-(\ref{eq:Euler.7}),
 \begin{equation}
 \left \{ \begin{array}{l}
  -2 R - \frac{2}{\alpha}R_{,y} + 3 R V + (R V)_{,y} = 0 \\
\mathcal{A}_{,y}(V)+ 3  \mathcal{A}(V) - 4 \pi R(y) = 0
\end{array}
\right. \label{eq:numRV}
\end{equation}
with $\mathcal{A}(V) = V+\frac{2}{\alpha}V_{,y}- V^2 - V  V_{,y} $, which are free of the inner core mass term.
In Appendix \ref{sec:freefall}, by proceeding differently, we obtain an analytical solution of the post-collapse dynamics which agrees with  the numerical solution of equations (\ref{eq:numRV}), see Fig. \ref{Fig:comparpost} for comparison.

\section{Conclusion and perspectives}
\label{sec:conc}

This contribution introduced a theory of the early stage of supernova
explosion which assumes that this belongs to the wide class of
saddle-node bifurcations with a slow sweeping of the parameters across
the bifurcation range. This explains well the suddenness of the
explosion occurring after aeons of slow evolution. The hugely
different time scales combine into a single intermediate time scale
for the slow-to-fast transition which could be of the order of several
hours. This transition is described by a ``universal" dynamical
equation, the Painlev\'e I equation. Comparing this prediction with a
model of star presenting a saddle-node bifurcation shows a
quantitative agreement with the predictions based on general arguments
of bifurcation theory.

This shows at least one thing, namely that the collapse of the star by
the loss of equilibrium between pressure and gravitational forces is a
global phenomenon depending on the full structure of the star and
cannot be ascribed, for instance, to an instability of the core
reaching the Landau-Chandrasekhar limit mass, as often assumed. We
also looked at the evolution of the star following the onset of
instability, namely when the amplitude of the perturbations grows to
finite values and cannot be described by the Painlev\'e I equation
anymore. In our equation of state model, the pressure becomes
proportional to the density in the large density limit. The pressure
increase is likely less steep than what is expected for the inner core
of stars, even though there are big uncertainties on the interior of
stars, particularly the ones yielding supernovae: showing no early
warning on the incoming explosion they are not scrutinized
spectroscopically.  Nevertheless, an analysis of this situation
teaches us a few interesting lessons. First,
we do not consider self-similar (or homologous) collapse in the usual
sense (where pressure and gravity scale the same way) because our
numerical results and our analysis lead us to claim that the pressure
becomes negligible in the core. Secondly, we find a new self-similar
free-fall solution toward the center.

Our numerical results together
with physical considerations about the velocity field make us argue
that besides the mathematically correct Penston-Larson solution, our
new self-similar (free-fall) solution is relevant to describe the
collapse. In other words, writing self-similar equations is not enough
to guaranty their relevance for a given problem because there can be
more than one such kind of solution, like in the present case, where
Zel'dovich type 2 solution corresponds to the numerical results,
although a type 1 solution also exists, but is not relevant.

The numerical results presented here were obtained by starting from the equilibrium state of the star at the saddle-node, then decreasing slowly the temperature. However we notice that the same conclusions are obtained when starting slightly away from the saddle-node point and performing the numerical integration at constant temperature. We point out that the previous numerical studies of gravitational collapse by Penston \cite{Penston}, Larson \cite{larson} and later by
others
\cite{Brenner}  were performed starting from a uniform density initial state (and finite radius), that represents initial conditions which are very far from ours and from any physical situation; nevertheless these authors did find a density behaving asymptotically (at large distance) as $r^{-\alpha}$, with $\alpha$ larger than $2$, as we find here.

 The free-fall
solution we found is \textit{not} the free-fall solution studied for
many years, because the exponents of our self-similar solution are not
the ones usually found. This conclusion is based upon a detailed
comparison between the direct numerical solution of the evolution
equations and the solution of the simpler equations for the
self-similar problem. As far as we are aware, although the
self-similar paradigm is often invoked in this field, such a detailed
comparison between dynamical solutions of the full Euler-Poisson
system and the full self-similar solution has not yet been done (the
merging of the curves before the collapse time was not shown). We show
that it is a relatively non trivial endeavor to perform such a
comparison.  Moreover we point out that our self-similar pressure-free
solution is more tricky to derive than the standard Penston-Larson
homologous solution including the pressure for which standard scaling
laws (Zel'dovich first kind) can be derived formally without any
difficulty. Finally we have mentioned that the center is a saddle point for our self-similar solution. Numerically this property is manifested in the behavior towards $r=0$ of the density profile $\rho(r,t)-\rho(0,t)$ which should pass from $r^2$  to $r^4$ in the self-similar regime (for generic initial conditions). The mechanism of this change of exponent, if it really occurs, has not been clearly identified and requires a deeper study.

This work leaves open many questions. One central issue is how the
scenario we outlined, namely slow starting in the universality class
Painlev\'e I, and later finite time collapse toward the central core,
is dependent on the pressure/density relation. We suspect that, if the
pressure increases more rapidly with the density than linearly at
large densities, there will be no finite time singularity. Likely,
because shock waves will form, irreversible transformations will take
place in those shock waves and another equation of state will become
relevant for the star.

\begin{acknowledgement}
We greatly acknowledge the ``Fondation des Treilles" where this work was initiated, and Paul Clavin for many very stimulating discussions.

\end{acknowledgement}


\appendix

\section{Boundary conditions to derive the normal form}
\label{sec_cl}

Let us derive the  boundary conditions to solve the integral equation
(\ref{ire}) by transforming it into the differential equation
(\ref{eq:zetadiff}).
We have to cancel the terms $$\left[g^{(c)} \zeta M_{,r}^{(2)}\right]_{0}^{r_c}, \quad \left[(g^{(c)} \zeta)_{,r} M^{(2)}\right]_{0}^{r_c},\quad  {\rm and} \quad \left[b \zeta M^{(2)}\right]_{0}^{r_c}.$$

$(\textit{i})$ At $r_c$  we have  $g^{(c)}(r_c)=0$ and $M^{(2)}(r_c)=0$
 that ensure the cancelation of the terms $g^{(c)} \zeta M_{,r}^{(2)}$, $g^{(c)} \zeta_{,r} M ^{(2)}$,  $b \zeta M ^{(2)}$, and $g^{(c)}_{,r} \zeta M ^{(2)}$
  at $r=r_c$ (while $g^{(c)}_{,r}$ and $\zeta$ are both non zero at $r=r_c$, see Fig. \ref{Fig:criticM}). This suppresses all the  terms taken at $r=r_c$.

$(\textit{ii})$ At $r=0$ we impose $\zeta=0$ that cancels the terms $g^{(c)} \zeta M_{,r}^{(2)}$,  $g^{(c)}_{,r}\zeta M ^{(2)}$,  and $ b \zeta M^{(2)}$. The last term $g^{(c)} \zeta_{,r} M ^{(2)}$ vanishes under the condition $M^{(2)}(0)=0$ (because $g^{(c)}$ and $\zeta_{,r}$ are both non zero at $r=0$). This suppresses all the  terms taken at $r=0$.

\section{Analytical self-similar solutions for the free-fall}
\label{sec:freefall}

Penston \cite{Penston} has given an exact solution of the free-fall problem without
thermodynamic pressure ($p=0$). It could seem that, because of the absence of
thermodynamic pressure, this is irrelevant for the problem of singularity in the
evolution of the collapsing core of models of stars. However, this is not quite
true because we have shown that during the collapse this thermodynamic pressure
becomes negligible, and so the evolution of the system is essentially like a
free-fall. By analyzing the equations for this pressureless collapse we have
shown that, actually, a discrete set of solutions exists, with different
singularity exponents. The free-fall solution found by Penston corresponds to
the exponent $\alpha =12/7$. Since this exponent is smaller than $2$ pressure
effects become important at a certain point of the evolution (Penston obtains
the estimate $\delta t/t_f\sim 10^{-4}$) and this is why he  considers in a
second step the case where pressure and gravity forces are of the same order
leading to another self-similar solution (the Penston-Larson solution) with
$\alpha=2$. Actually, we propose another possibility which is in agreement with our numerical results (and actually with many others). We show below that other
exponents than $12/7$ are possible for the free-fall, some of them being larger
than $2$ and providing therefore a possible solution of the initial problem in
which gravity always dominates over pressure forces\footnote{
 It does not mean that the Penston-Larson solution is incorrect. It represents a mathematically exact (type I) self-similar solution of the isothermal Euler-Poisson equations. However, we argue that {\it other} (type II) self-similar solutions exist in which gravity overcomes pressure. They are characterized by $\alpha>2$ and by a density behaving as $\rho_0+\rho_k r^{k}$ with $k>3$ close to the origin (see below), while the Penston-Larson solution has $\alpha=2$ and the density behaves as $\rho_0+\rho_2 r^{2}$ close to the origin. Our numerical work (despite its limitations because we follow the collapse only over a few decades in density) together with important physical considerations (e.g. the fact that the velocity profile in our solution decreases to zero instead of tending to a constant value) suggest that these new solutions
are
relevant to describe the collapse.}.
Our solutions are based on
the choice of initial conditions for the radial dependence of the density
$\overline{\rho}(a)=\rho_0(1-a^k/A^k)$ where $a$ is the radial variable (same notations as in Penston \cite{Penston}). The exponent $k$
is left free, although Penston takes $k=2$ with the comment:
``we are 'almost always' correct in taking the form
$\overline{\rho}(a)=\rho_0(1-a^2/A^2)$''.

We consider a sphere of gas initially at rest and call $M(a,0)$ the mass of gas contained within the sphere of radius $a$ and $\overline{\rho}(a)=3M(a,0)/4\pi a^3$ the average density of that sphere. Using Gauss theorem, the Euler equation (\ref{iso2}) with the pressure neglected is equivalent to
\begin{eqnarray}
\frac{d^2 r}{dt^2}=\frac{du}{dt}=-\frac{GM(a,0)}{r^2},
\label{ff1}
\end{eqnarray}
where $r$ and $u$ are the position and the velocity at time $t$  of a
fluid particle  located at $r=a$ at $t=0$.  This equation can be solved
analytically \cite{mestel} and the solution can be expressed in parametric form
as
\begin{eqnarray}
r=a\cos^2\theta,
\label{ff2}
\end{eqnarray}
\begin{eqnarray}
t=\left (\frac{3}{8\pi G\overline{\rho}(a)}\right )^{1/2}\left (\theta+\frac{1}{2}\sin(2\theta)\right ),
\label{ff3}
\end{eqnarray}
where $\theta$ runs between $0$ and $\pi/2$. Taking $\theta=\pi/2$, we find that a particle initially at $r=a$ arrives at $r=0$ at a time $t(a)=(3\pi/32G\overline{\rho}(a))^{1/2}$. Setting $a=0^+$ in the foregoing expression, we find that the first particle reaches the center at the time
\begin{eqnarray}
t_{f}=\left (\frac{3\pi}{32 G \rho_0}\right )^{1/2},
\label{ff4}
\end{eqnarray}
where $\rho_0=\overline{\rho}(0)$. This is called the free-fall time. At $t=t_f$, the central density becomes infinite ($\rho_c=+\infty$).

Using the equation of motion (\ref{ff2})-(\ref{ff3})  giving $r=r(a)$ and the
relation $\rho(r,t)r^2\, dr=\rho(a,0) a^2\, da$, which is equivalent to the
equation of continuity (\ref{iso1}), we can determine the evolution of the
density profile $\rho(r,t)$ and of the velocity profile $u(r,t)$ in the pre-
and post-collapse regimes. For $t\rightarrow t_f$ and $r$ not too large, they
have a self-similar form. The derivation of this self-similar solution follows
rather closely the one by Penston with the only difference that his assumption
$\overline{\rho}(a)=\rho_0(1-a^2/A^2)$ is replaced by
$\overline{\rho}(a)=\rho_0(1-a^k/A^k)$. Therefore, we skip the details of the
derivation and directly give the final results.

\subsection{The pre-collapse regime}
\label{App:precoll}

\begin{figure}[htbp]
\centerline{
 (a)\includegraphics[height=1.25in]{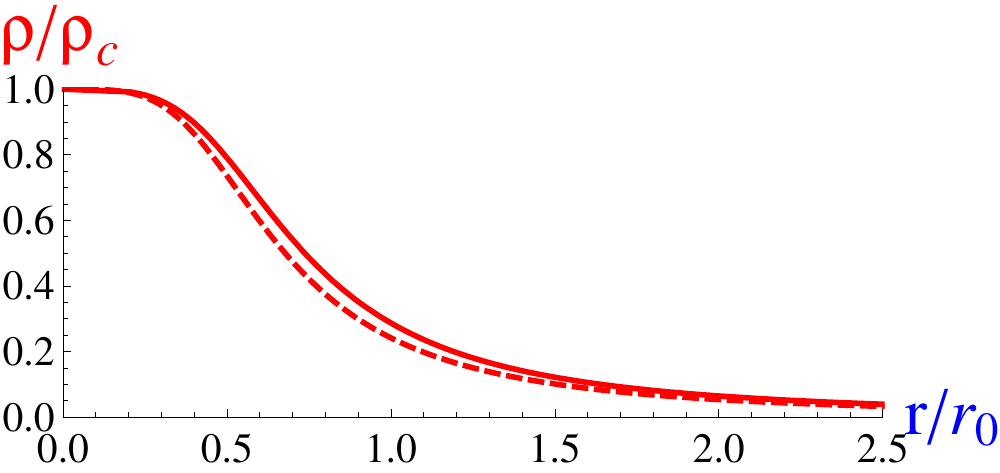}
}
\centerline{
 \includegraphics[height=1.25in]{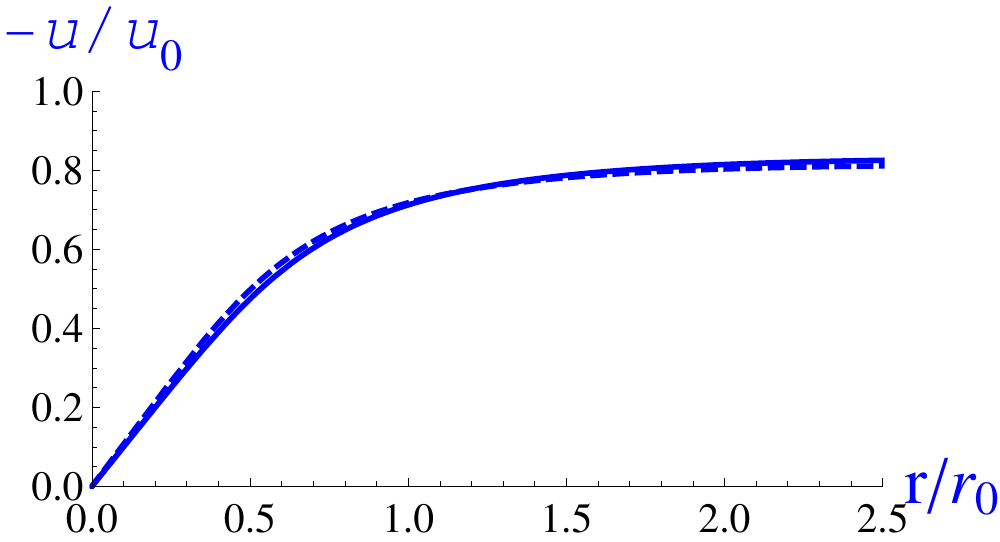}
}
(b) \caption{Parametric solutions (\ref{ff5})-(\ref{ff7}) compared with the self-similar solutions of section \ref{subsec5B} for $\alpha=24/11$. (a) Density $\rho(r,t)/\rho_c(t)$  versus $r/r_0(t)$ for $k=4$ in solid line,  $R(\xi)/R(0)$ versus $\xi=r/r_0(t)$  for $R_4= - 4$ in dashed line. (b) Velocity  $u(r,t)/u_0(t)$ versus $r/r_0(t)$ in solid line; $-1.6 U(\xi)$ in dashed line.
}
\label{Fig:compar}
\end{figure}

In the pre-collapse regime ($t<t_{f}$), the self-similar density and velocity profiles are given in parametric form by
\begin{eqnarray}
\frac{\rho(r,t)}{\rho_c(t)}=\frac{3}{3+2(3+k)y+(3+2k)y^2},
\label{ff5}
\end{eqnarray}
\begin{eqnarray}
\frac{r}{r_0(t)}=y^{1/k}(1+y)^{2/3},
\label{ff6}
\end{eqnarray}
\begin{eqnarray}
\frac{u(r,t)}{u_0(t)}=-\frac{y^{1/k}}{(1+y)^{1/3}},
\label{ff7}
\end{eqnarray}
where $y=\frac{1}{2}(\frac{a}{A})^k\frac{t_f}{\delta t}$ goes from $0$ to $+\infty$ (here $\delta t=t_f-t$). For $k=4$, the curves ${\rho(r,t)}/{\rho_c(t)}$ and $-{u(r,t)}/{u_0(t)}$ drawn in Fig. \ref{Fig:compar}, solid lines, coincide with the self-similar numerical solution (dashed line) of equations (\ref{eq:Euler.4})-(\ref{eq:Euler.5}) derived in section \ref{subsec5B}.

In the above parametric representation the central density is given by the relation
\begin{eqnarray}
\rho_c(t)=\left (\frac{4}{3\pi}\right )^2\rho_0\left (\frac{t_{f}}{t_{f}-t}\right )^2.
\label{ff8}
\end{eqnarray}
Using equation (\ref{ff4}) it can be rewritten as
\begin{eqnarray}
\rho_c(t)=\frac{1}{6\pi G}\frac{1}{(t_f-t)^2},
\label{ff8b}
\end{eqnarray}
which agrees with the result of Sec. \ref{subsec5B}. Moreover, we have
\begin{eqnarray}
r_0(t)=\left (\frac{3\pi}{4}\right )^{2/3}2^{1/k}A\left |\frac{t_{f}-t}{t_{f}}\right |^{(2k+3)/3k},
\label{ff9}
\end{eqnarray}
\begin{eqnarray}
u_0(t)=\frac{\pi}{2^{(k-1)/k}}\left (\frac{4}{3\pi}\right )^{1/3}\frac{A}{t_{f}}\left |\frac{t_{f}-t}{t_{f}}\right |^{(3-k)/3k}.
\label{ff10}
\end{eqnarray}
For $r\rightarrow 0$, we get
\begin{eqnarray}
\rho(r,t)\sim \rho_c(t) \left\lbrack 1-\frac{2}{3}(3+k)\left (\frac{r}{r_0(t)}\right )^{k}\right\rbrack,
\label{ff11}
\end{eqnarray}
\begin{eqnarray}
u(r,t)\sim -u_0(t) \frac{r}{r_0(t)}.
\label{ff12}
\end{eqnarray}
For $r\rightarrow +\infty$, we get
\begin{eqnarray}
\rho(r)\sim \rho_0 \frac{3}{2k+3}\left (\frac{8}{3\pi}A^k\right )^{6/(2k+3)}\frac{1}{r^{6k/(3+2k)}},
\label{ff13}
\end{eqnarray}
\begin{eqnarray}
u(r)\sim -\left (\frac{8\pi\rho_0 G}{3}\right )^{1/2} \left (\frac{8}{3\pi}A^k\right )^{3/(2k+3)}r^{(3-k)/(3+2k)},\nonumber\\
\label{ff14}
\end{eqnarray}
which are independent on time as it should.  We have $\rho\sim
r^{-\alpha_k}$ and $u\sim r^{\nu_k}$ with
\begin{eqnarray}
\alpha_k=\frac{6k}{2k+3},\qquad \nu_k=\frac{3-k}{2k+3}.
\label{ff15}
\end{eqnarray}
The expressions (\ref{ff13}) and (\ref{ff14}) also give the density and  velocity
profiles for all $r$ at $t=t_{f}$.  For $k=2$, we get $\alpha_2=12/7$ and $\nu_2=1/7$; for $k\rightarrow +\infty$, we get $\alpha_{\infty}=3$ and $\nu_{\infty}=-1/2$; for
$k=4$, we get $\alpha_4=24/11$ and $\nu_4=-1/11$. The exponent
$\alpha$ achieves the critical value $2$ for $k=3$. For $k<3$, i.e. $\alpha<2$, the pressure
wins over gravity as we approach the collapse time $t_{f}$, and
the free-fall solution is not valid anymore.  For $k>3$, i.e. $\alpha>2$, the gravity always wins over
pressure so the free-fall solution may be valid for all times.


Let us discuss the form of the density and velocity profiles depending on $k$.

For any $k$, the density profile $\rho(r,t)$ starts from a finite value (for $t<t_f$) and decreases with the distance $r$. The central density $\rho_c(t)$ increases with time and diverges at the collapse time $t_f$. At $t=t_f$, the density profile is singular at the origin.

For $k<3$, i.e. $\alpha<2$, the velocity profile $-u(r,t)$ starts from zero at $r=0$ and increases with the distance $r$. The magnitude of the velocity $u_0(t)$ decreases with time and tends to zero at the collapse time $t_f$. At $t=t_f$, the velocity is still zero at the origin.

For $k=3$, i.e. $\alpha=2$, the velocity profile $-u(r,t)$ starts from zero at $r=0$ (for $t<t_f$), increases with the distance $r$, and reaches an asymptotic value $u_0$ (the prefactor $u_0(t)$ is constant). At $t=t_f$, the velocity profile has a constant non-zero value $u_0$.

For $k>3$, i.e. $\alpha>2$, the velocity profile $-u(r,t)$ starts from zero at $r=0$, increases with the distance $r$, reaches a maximum, and decreases towards zero at large distances. The  prefactor $u_0(t)$ increases with time and diverges at the collapse time $t_f$. At $t=t_f$, the velocity profile is singular at the origin.


\subsection{The post-collapse regime}

\begin{figure}[htbp]
\centerline{
(a) \includegraphics[height=1.75in]{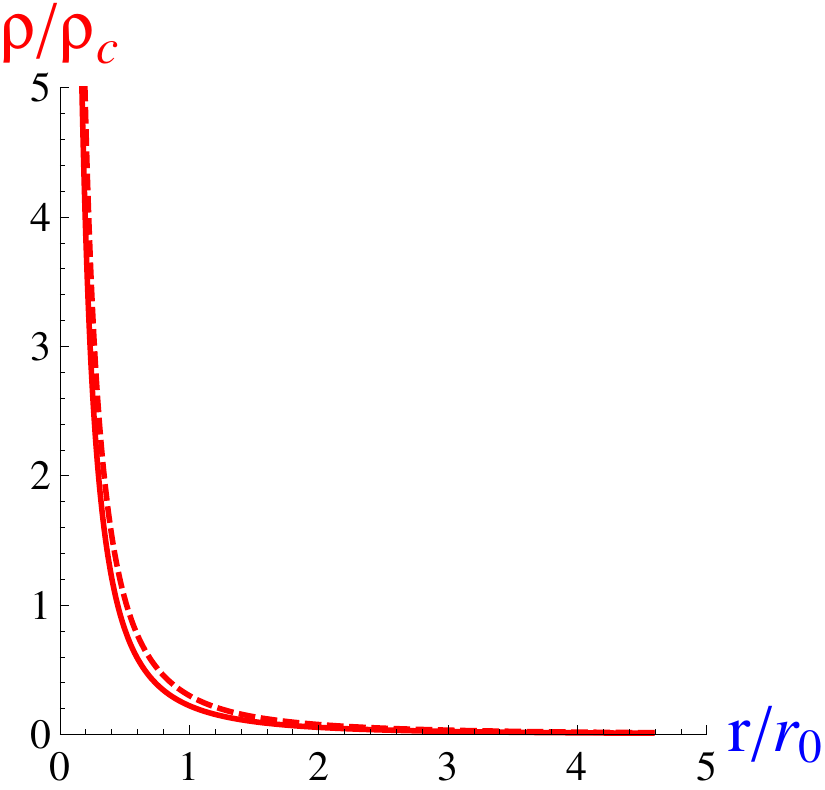}
}
\centerline{
(b)  \includegraphics[height=1.75in]{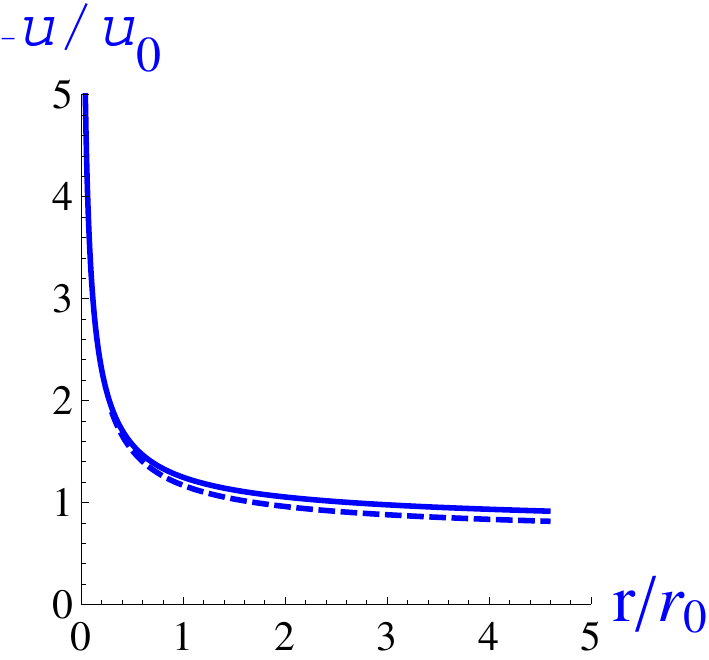}
}
\caption{Parametric solutions of equations (\ref{ff16})-(\ref{ff18}) and self-similar solutions of equations (\ref{eq:numRV}) in the post-collapse regime. (a) density $\rho(r,t)/\rho_c(t)$ versus $r/r_0(t)$ for $k=4$, or $\alpha=24/11$,  in solid line, $15 R(\xi)$ versus $\xi=r/r_0(t)$ for $K_U=-1$ in dashed line, (b) velocity $-u(r,t)/u_0(t)$ versus $r/r_0(t)$  in solid line, $-U(\xi)$ in dashed line.
}
\label{Fig:postcoll}
\end{figure}

In the post-collapse regime ($t>t_{f}$), the self-similar density and velocity profiles are given in parametric form by
\begin{eqnarray}
\frac{\rho(r,t)}{\rho_c(t)}=\frac{3}{3+2(3+k)y+(3+2k)y^2},
\label{ff16}
\end{eqnarray}
\begin{eqnarray}
\frac{r}{r_0(t)}=|y|^{1/k}|1+y|^{2/3},
\label{ff17}
\end{eqnarray}
\begin{eqnarray}
\frac{u(r,t)}{u_0(t)}=-\frac{|y|^{1/k}}{|1+y|^{1/3}},
\label{ff18}
\end{eqnarray}
where $y$ goes from $-\infty$ to $-1$, and $\rho_c(t)$, $r_0(t)$ and $u_0(t)$ are defined by equations (\ref{ff8})-(\ref{ff10}) as in the pre-collapse regime.
For $r\rightarrow +\infty$, the behavior is the same as in the pre-collapse regime, but for $t>t_f$ and $r\rightarrow 0$, we get
\begin{eqnarray}
\rho(r,t)\sim \rho_c(t) \frac{3}{2k}\left (\frac{r_0(t)}{r}\right )^{3/2},
\label{ff19}
\end{eqnarray}

\begin{eqnarray}
u(r,t)\sim -u_0(t) \left (\frac{r_0(t)}{r}\right )^{1/2}.
\label{ff20}
\end{eqnarray}
We note that the density and the velocity are always singular at $r=0$. For any $k$, the density profile $\rho(r,t)$ is decreasing, as illustrated in Fig. \ref{Fig:postcoll}-(a) . For $k<3$, the velocity profile $-u(r,t)$ decreases, reaches a minimum value, and increases. For $k=3$ it decreases towards an asymptotic value $u_0$ and for $k>3$ it decreases towards zero, see Fig. \ref{Fig:postcoll}-(b).

Finally, the mass contained in the Dirac peak $\rho_{D}({\bf r},t)=M_{D}(t)\delta({\bf r})$ at time $t>t_{f}$ is
\begin{eqnarray}
M_{D}(t)=\frac{8\pi}{3}2^{(3-k)/k}\rho_0 A^{3}\left (\frac{t-t_{f}}{t_{f}}\right )^{3/k}.
\label{ff21}
\end{eqnarray}
The mass in the core grows algebraically with an exponent $b_k=3/k$. For $k=2$, we get $b_2=3/2$; for $k\rightarrow +\infty$, we get $b_{\infty}=0$; for $k=3$, we get $b_3=1$; for
$k=4$, we get $b_4=3/4$.

\subsection{The homogeneous sphere}

Finally, for completeness, we recall the solution corresponding to the collapse of a homogeneous sphere with mass $M$, initial density $\rho_0$ and initial radius $R_0$. Since $\overline{\rho}(a)=\rho_0$, we find from equations (\ref{ff2})-(\ref{ff4}) that all the particles collapse at $r=0$ at the same time $t_f$. Therefore, a Dirac peak $\rho_{D}({\bf r})=M\delta({\bf r})$ is formed at $t=t_f$. The evolution of the radius $R(t)$ of the homogeneous sphere is given by
\begin{eqnarray}
R(t)=R_0\cos^2\theta,\qquad \frac{t}{t_f}=\frac{2}{\pi}\left (\theta+\frac{1}{2}\sin(2\theta)\right ),
\label{ff22}
\end{eqnarray}
where $\theta$ runs between $0$ and $\pi/2$. For $t\rightarrow t_f$, we get
\begin{eqnarray}
R(t)=R_0\left (\frac{3\pi}{4}\right )^{2/3}\left (1-\frac{t}{t_f}\right )^{2/3}.
\label{ff23}
\end{eqnarray}
The density $\rho_c(t)=3M/4\pi R(t)^3$ increases as
\begin{eqnarray}
\rho_c(t)=\rho_0 \left (\frac{4}{3\pi}\right )^{2}\left (1-\frac{t}{t_f}\right )^{-2}.
\label{ff24}
\end{eqnarray}
The velocity field is $u(r,t)=-H(t)r$ with
\begin{eqnarray}
H=-\frac{\dot R}{R}=\frac{2}{3}(t_f-t)^{-1}.
\label{ff25}
\end{eqnarray}

 \end{document}